\documentclass[twocolumn,superscriptaddress,amsmath,amssymb,floatfix,aps,notitlepage,pre,tightenlines,longbibliography]{revtex4-1}
\let\Omega\varOmega

\usepackage[font=small,format=plain,labelfont=bf,justification=raggedright,singlelinecheck=false]{caption}
\usepackage[T1]{fontenc}
\usepackage[utf8]{inputenc}
\usepackage{helvet}

\usepackage{soul}
\usepackage[usernames,dvipsnames]{xcolor}
\usepackage{mathrsfs, hyperref}
\usepackage{amssymb, amsbsy, amsmath, latexsym, dsfont, array, layout, graphics,mathrsfs,braket,amsfonts,amsthm,
  amssymb,graphicx,subfigure, youngtab,color,bm,mathtools,braket,verbatim,url,cleveref,natbib,hypernat,extarrows,inputenc}

\makeatletter
\newcommand\newsubcommand[3]{\newcommand#1{#2\sc@sub{#3}}}
\def\sc@sub#1{\def\sc@thesub{#1}\@ifnextchar_{\sc@mergesubs}{_{\sc@thesub}}}
\def\sc@mergesubs_#1{_{\sc@thesub#1}}
\makeatother

\newcommand{\bs}[1]{\boldsymbol{#1}}

\newcommand{\vc}{{\it V. cholerae} }

\newcommand{\Jep}{J_\mathrm{e,\parallel}}

\newcommand{\lamg}{\lambda_\mathrm{g}}

\newcommand{\kg}{k_\mathrm{g}}
\newcommand{\kr}{k_\mathrm{r}}

\newcommand{\degreeC}{$^\circ$C}

\newsubcommand{\Fg}{\mathbf{F}}{\mathrm{g}}
\newsubcommand{\Fe}{\mathbf{F}}{\mathrm{e}}
\newsubcommand{\Fep}{\mathbf{F}}{\mathrm{e,\parallel}}
\newcommand{\Ft}{\mathbf{F}}

\newsubcommand{\Fd}{\mathbf{F}}{\mathrm{d}}
\newcommand{\Few}{\mathbf{F}_\mathrm{e}^\mathrm{w}}

\newcommand{\Feo}{\mathbf{F}_\mathrm{e}^0}
\newcommand{\Feop}{\mathbf{F}_\mathrm{e,\parallel}^0}
\newcommand{\gammao}{\gamma^0}

\newcommand{\Gb}{G_\mathrm{b}}
\newcommand{\Gs}{G_\mathrm{s}}

\newsubcommand{\Bo}{\mathbf{B}}{\parallel}

\newsubcommand{\Ainit}{\mathbf{F}^0}{\mathrm{e,\parallel}}

\begin{document}
\title{Non-uniform growth and surface friction determine\\ bacterial biofilm morphology on soft substrates}

\author{Chenyi Fei}
\affiliation{Department of Molecular Biology, Princeton University, Princeton, New Jersey 08544, USA}
\affiliation{Lewis-Sigler Institute for Integrative Genomics, Princeton University, Princeton, New Jersey 08544, USA}
\author{Sheng Mao}
\affiliation{Department of Mechanical and Aerospace Engineering, Princeton University, Princeton, New Jersey 08544, USA}
\author{Jing Yan}
\affiliation{Department of Molecular Biology, Princeton University, Princeton, New Jersey 08544, USA}
\affiliation{Department of Mechanical and Aerospace Engineering, Princeton University, Princeton, New Jersey 08544, USA}
\author{Ricard Alert}
\affiliation{Lewis-Sigler Institute for Integrative Genomics, Princeton University, Princeton, New Jersey 08544, USA}
\affiliation{Princeton Center for Theoretical Science, Princeton University, Princeton, New Jersey 08544, USA}
\author{Howard A. Stone}
\affiliation{Department of Mechanical and Aerospace Engineering, Princeton University, Princeton, New Jersey 08544, USA}
\author{Bonnie L. Bassler}
\affiliation{Department of Molecular Biology, Princeton University, Princeton, New Jersey 08544, USA}
\affiliation{The Howard Hughes Medical Institute, Chevy Chase, MD 20815, USA}
\author{Ned S. Wingreen}
\email{wingreen@princeton.edu}
\affiliation{Department of Molecular Biology, Princeton University, Princeton, New Jersey 08544, USA}
\affiliation{Lewis-Sigler Institute for Integrative Genomics, Princeton University, Princeton, New Jersey 08544, USA}
\affiliation{Princeton Center for Theoretical Science, Princeton University, Princeton, New Jersey 08544, USA}
\author{Andrej Ko\v{s}mrlj}
\email{andrej@princeton.edu}
\affiliation{Department of Mechanical and Aerospace Engineering, Princeton University, Princeton, New Jersey 08544, USA}
\affiliation{Princeton Institute for the Science and Technology of Materials (PRISM),
Princeton University, Princeton, New Jersey 08544, USA}

\begin{abstract}
During development, organisms acquire three-dimensional shapes with important physiological consequences. While the basic mechanisms underlying morphogenesis are known in eukaryotes, it is often difficult to manipulate them \emph{in vivo}. To circumvent this issue, here we present a study of developing \emph{Vibrio cholerae} biofilms grown on agar substrates in which the spatiotemporal morphological patterns were altered by varying the agar concentration. Expanding biofilms are initially flat, but later experience a mechanical instability and become wrinkled. Whereas the peripheral region develops ordered radial stripes, the central region acquires a zigzag herringbone-like wrinkle pattern. Depending on the agar concentration, the wrinkles initially appear either in the peripheral region and propagate inward (low agar concentration) or in the central region and propagate outward (high agar concentration). To understand these experimental observations, we developed a model that considers diffusion of nutrients and their uptake by bacteria, bacterial growth/biofilm matrix production, mechanical deformation of both the biofilm and the agar, and the friction between them. Our model demonstrates that depletion of nutrients beneath the central region of the biofilm results in radially-dependent growth profiles, which in turn, produce anisotropic stresses that dictate the morphology of wrinkles. Furthermore, we predict that increasing surface friction (agar concentration) reduces stress anisotropy and shifts the location of the maximum compressive stress, where the wrinkling instability first occurs, toward the center of the biofilm, in agreement with our experimental observations.
Our results are broadly applicable to bacterial biofilms with similar morphologies and also provide insight into how other bacterial biofilms form distinct wrinkle patterns.
\end{abstract}

\maketitle

The intricate shapes of organisms are determined by the spatiotemporal patterns of growth as well as the mechanical properties of their underlying biological components~\cite{goriely2017mathematics,de2018physical, genet2015heterogeneous}. Three-dimensional (3D) shape transformations in developing organisms often arise via differential growth of connected tissues~\cite{thompson1917growth, goriely2017mathematics}. Such asymmetric growth patterns generate compressive stresses within the faster growing tissues, which may cause mechanical instabilities \cite{klein2007shaping, huxley1993problems, sachs1875text}. Growth-induced mechanical instabilities drive the formation of many convoluted morphologies, such as the gyrification of brains~\cite{de2018physical,tallinen2016growth,budday2014role}, the vilification and looping of guts~\cite{shyer2013villification,savin2011growth}, and the branching of lungs~\cite{kim2015localized} as well as 3D structures of synthetic systems with patterned swelling \cite{klein2007shaping,kim2012designing,modes2016shape, gladman2016biomimetic,yang2010harnessing}. 

Biofilms, which are surface-associated bacterial communities encapsulated by a self-produced extracellular matrix~\cite{o2000biofilm,teschler2015living},  also display a variety of 3D developmental morphologies ranging from radial stripes, to concentric rings, to disordered labyrinth and herringbone patterns~\cite{serra2013microanatomy,serra2013cellulose,dietrich2013bacterial,romero2010amyloid,yan2017extracellular}. In the case of {\it Vibrio cholerae}, a model biofilm former, quantitative imaging revealed a 3D undulating topography with an intrinsic wavelength that depends on the stiffnesses of both the substrate and the biofilm~\cite{yan2019mechanical}. Over the course of growth on an agar substrate, an initially flat \vc biofilm expands and forms a 3D pattern in which a disordered core is surrounded by radial stripes extending to the edge~\cite{beyhan2007smooth}. These morphological transitions in \vc biofilms are proposed to be caused by mechanical instabilities. 

The major components of the \vc biofilm matrix and their roles in defining the biofilm's bulk and interfacial mechanical properties have been well explored~\cite{yildiz1999vibrio,yildiz2014structural,fong2007rbmbcdef,berk2012molecular,fong2017structural,teschler2015living}. \vc biofilms behave as soft viscoelastic solids similar to hydrogels, and possess finite adhesion to the agar surface on which they are grown ~\cite{yan2018bacterial}.
Thus, as the biofilm expands, it is mechanically constrained by the friction with the agar substrate. Mechanical compression due to constrained biofilm expansion ultimately triggers instabilities that result in out-of-plane deformation and the 3D biofilm morphology~\cite{zhang2016morphomechanics,yan2019mechanical}.

A key to understanding the full 3D morphodynamics of \vc biofilms involves the cells' spatially heterogeneous physiology \cite{stewart2008physiological}. Soon after the initial expansion of the biofilm, growth occurs primarily at the edge of the biofilm due to nutrient limitation near the center \cite{anderl2003role,seminara2012osmotic,yan2017extracellular,yan2019mechanical}. However, little is known about how this non-uniform growth profile, combined with the mechanical interaction between the biofilm and substrate, lead to the observed morphodynamics. While a mechanical basis for instability-induced pattern formation in biofilms has been suggested previously~\cite{zhang2016morphomechanics, zhang2017experimental}, the dynamics of stress accumulation during biofilm expansion and the consequences for global pattern formation remain largely unknown.

Here, we determine the biophysical mechanisms controlling \vc biofilm expansion and pattern formation. We show that the observed kinematic and morphodynamic features of growing biofilms are well captured by a reduced two-dimensional (2D) chemo-mechanical model. Consistent with experimentally measured velocity profiles, our model predicts three distinct kinematic stages of biofilm expansion, even before the formation of wrinkles.
We also demonstrate that non-uniform growth due to nutrient depletion generates anisotropic compressive stresses in the outer biofilm region leading to radial stripes; by contrast, in the interior of the biofilm, the compressive stresses are more isotropic, leading to zigzag herringbone-like patterns. We conclude that the spatiotemporal distribution of mechanical stresses dictates the morphodynamics of experimental biofilms grown on substrates of different agar concentrations.
Our model thus illustrates the mechanical principles underlying how growth drives the emergent 3D morphologies of biofilms.

\section*{Results}
\subsection*{Biofilm morphodynamics depends on substrate stiffness}
\begin{figure*}[t!]
\centering
\includegraphics[width=\linewidth]{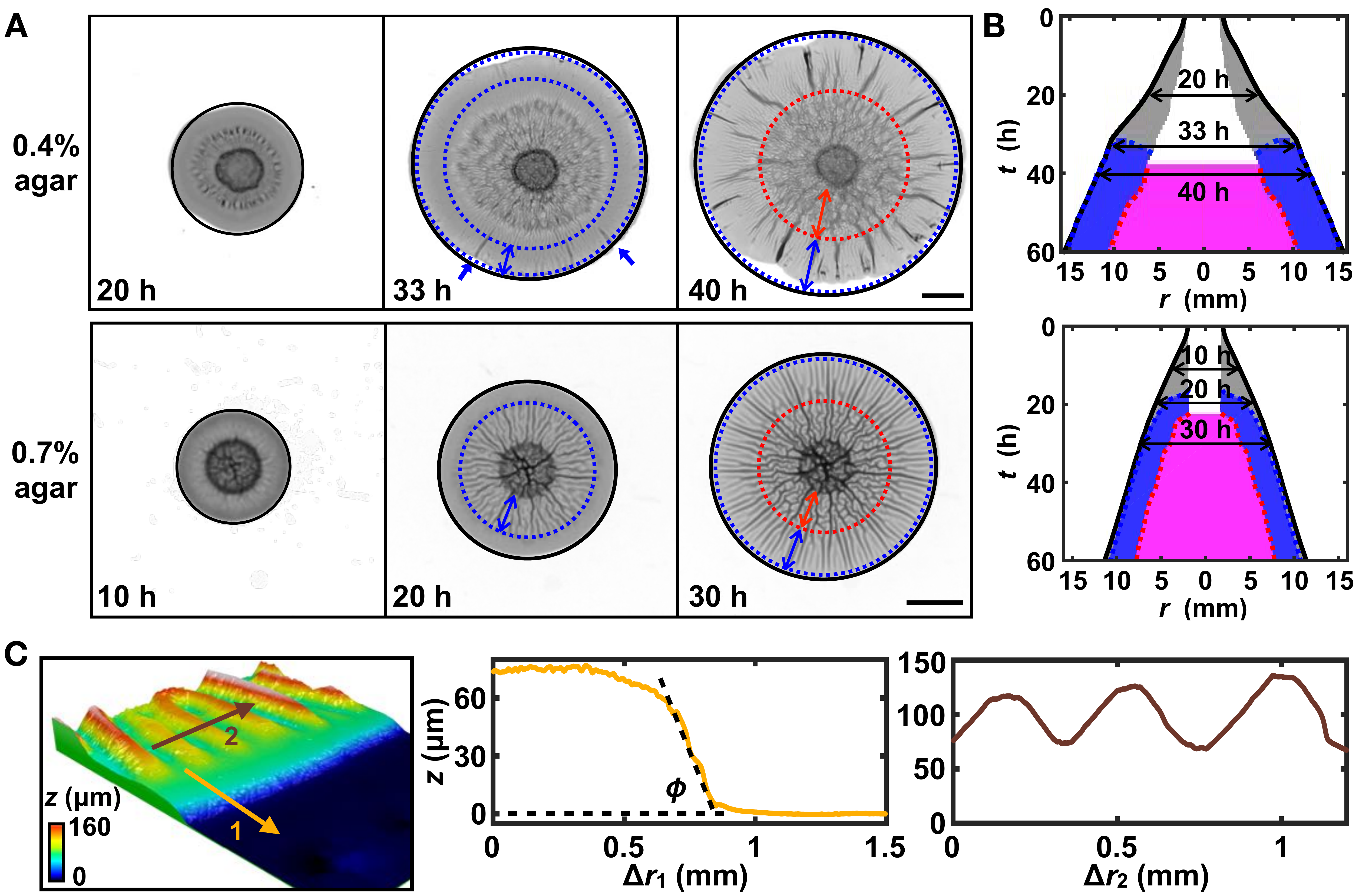}
\vspace{1em}
\caption{
{\bf Morphogenesis of \textit{V. cholerae} biofilms grown at an air-solid interface.} 
({\bf A}) Transmission images of biofilms grown on 0.4\% ({\it top}) and 0.7\% ({\it bottom}) agar substrates at the designated times, where time is measured relative to the time when a biofilm starts expanding radially. Black solid circles mark the boundaries of entire biofilms, blue dotted circles mark the boundaries of regions with radial patterns, and red dotted circles mark the boundaries of regions with zigzag patterns. Blue single-headed arrows indicate radial morphological features near the edge. Blue double-headed arrows span the regions with radial patterns. Red double-headed arrows span the regions with zigzag patterns. Scale bars: 3 mm. 
({\bf B}) Kymograph representation of the pattern formation dynamics of experimental biofilms grown on 0.4\% ({\it top}) and 0.7\% ({\it bottom}) agar substrates, where $r$ measures the distance from the center of the biofilm and $t$ is time.  Gray, blue, and magenta colors indicate, respectively, regions without patterns, with radial patterns and with zigzag patterns. The internal white zone indicates the region possessing patterns related to the initial biofilm at $t=0$. Biofilms shown in {\bf A} are marked by horizontal double-headed arrows arrows at the designated times. The boundaries of different regions (as outlined in {\bf A}) were obtained based on the intensity of transmitted light (see Methods). 
({\bf C})  {\it Left}: height map of a 3.2 mm $\times$ 2.4 mm region of the edge of a biofilm grown on 0.6\% agar ($t$ = 22 h). {\it Middle} and {\it right}: height profiles corresponding to the positions spanned by the yellow (denoted by 1) and brown (denoted by 2) arrows in the {\it left} panel. Intersecting dashed lines denote the biofilm leading angle $\phi$. The zero value for $z$ was chosen to coincide with the average height of a line profile on the agar surface.
}
\label{Fig1}
\end{figure*}

After a liquid drop inoculates \vc on an agar substrate, a biofilm initially expands radially and remains flat with no recognizable morphological features except at the center where inoculation occured (Fig.~\ref{Fig1}A and B). Expansion occurs because bacteria consume nutrients from the agar substrate, proliferate, and produce extracellular matrix.
Growing biofilms adhere to the non-growing agar substrate, and the sliding friction between biofilm and agar mechanically constrains biofilm expansion. Thus, growing biofilms become compressed and build up mechanical stresses. When the compressive stress reaches a critical value, a mechanical instability generates wrinkles (Fig.~\ref{Fig1}A). Wrinkles are vertical deformations of the biofilm together with the adhered substrate with a characteristic wavelength (Fig.~\ref{Fig1}C) that depends on the thickness of the biofilm and on the mechanical properties of the biofilm and the agar substrate~\cite{yan2019mechanical, chen2004herringbone, huang2005nonlinear}. Subsequently, as compressive stresses continue to build up, a biofilm can partially detach from the agar substrate, forming delaminated blisters \cite{yan2019mechanical}. In this manuscript, however, we restrict our focus to exploring the original wrinkle patterns outside the inoculation core -- localized cell death has been shown to facilitate pattern formation inside the inoculation core~\cite{asally2012localized}.
  
Notably, the development of wrinkle patterns depends on the stiffness of the agar substrate. For {\it V. cholerae}, after about 30 h of growth on soft substrates (low agar concentration), a pattern of radial wrinkles initially appears at the outer edge of the biofilm and subsequently propagates towards the center (Fig.~\ref{Fig1}A and B, top). By contrast, on stiff substrates (high agar concentration), radial wrinkles initially form near the center and propagate outward (Fig.~\ref{Fig1}A and B, bottom). After about 40~h of growth, herringbone-like zigzag patterns emerge in the central region, surrounded by the outer region of radial stripes. Both of these regions expand outward at approximately the same speed as the expanding edge of the biofilm (Fig.~\ref{Fig1}B). In this steadily expanding state, surface profiling by confocal microscopy reveals a wedge-shaped rim ($\sim 200~\mu$m in width) with a constant leading angle $\phi$, followed by a narrow region ($\sim 500~\mu$m in width) of nearly constant height, followed, in turn, by the region of radial stripe patterns (Fig.~\ref{Fig1}C).

\subsection*{Chemo-mechanical model of biofilm development}
\label{Sec:ChemoMechanics}

\begin{figure*}[t!]
\centering
\includegraphics[width=\textwidth]{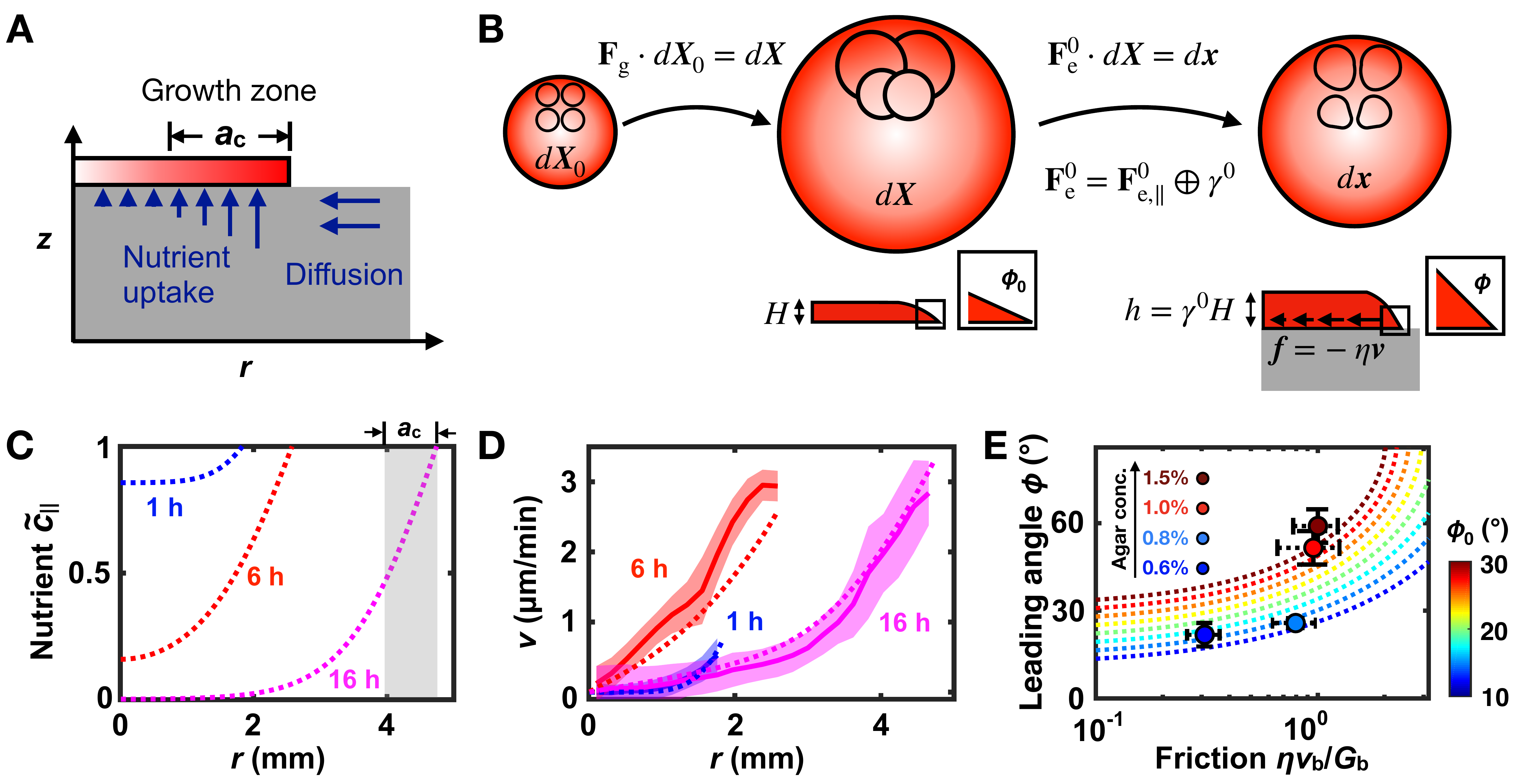}
\vspace{1em}
\caption{
{\bf A chemo-mechanical growth model captures the kinematics of biofilm expansion.}
({\bf A}) Schematic of nutrient diffusion-uptake dynamics. Nutrients diffuse through the agar substrate (gray) and are taken up by the bacterial biofilm (red), where blue arrows indicate the magnitudes of nutrient fluxes.
Bacterial growth rate is proportional to nutrient uptake, which in turn depends on the local nutrient concentration.
The established nutrient concentration profile (see {\bf C}) sets a nutrient-rich annular periphery (its width denoted by $a_c$) where cells actively grow. Lighter (darker) red color indicates slower (faster) growth. $r$ indicates the lateral distance from the center of the biofilm. 
({\bf B}) Schematic of the plane-stress elasto-growth model (color code as in {\bf A}). Starting from an initial stress-free configuration ({\it left}), local growth of the biofilm $\mathbf{F}_\mathrm{g}$ creates a virtual stress-free intermediate state ({\it middle}), which is further deformed by elastic deformation $\Feo$, to ensure its compatibility (no overlap between marked regions), into a stressed current configuration ({\it right}). The elastic deformation $\Feo$ is decomposed into an in-plane compression, denoted by $\Feop$, and a stretch $\gammao$ of the film thickness $H$ ({\it bottom right}). As the biofilm expands and moves relative to the substrate, it experiences a surface friction (black arrows) $\bs{f} = -\eta \bs{v}$, where $\eta$ is the friction coefficient and $\bs{v}$ is the expansion velocity. In the bulk, friction impedes biofilm expansion and is balanced by internal stresses; at the rim, friction increases the biofilm leading angle from $\phi_0$ to $\phi$ ({\it bottom middle} and {\it bottom right}). 
({\bf C}) Nutrient concentration $\tilde{c}_\parallel$ and ({\bf D}) radial expansion velocity $v$ versus the radial coordinate $r$ at the designated times. $\tilde{c}_\parallel$ is normalized by the concentration at the edge of the biofilm. Shaded gray area in {\bf C} indicates the active growth zone where $\tilde{c}_\parallel>0.5$. Solid curves with shaded error bands in {\bf D} represent experimental data (mean $\pm$ std) for a biofilm grown on a 0.7\% agar substrate. The radial velocity was extracted by averaging over a ring of the biofilm at radius $r$ from the center. Dotted curves represent simulation results for the parameters chosen by fitting the simulation velocity profiles to the experimental data (see Methods and Fig.~S1).
({\bf E}) Theoretical predictions for the biofilm leading angle $\phi$ (Eq. (\ref{eq_angle}), dotted curves) as a function of the dimensionless friction $\eta v_b/G_b$ and initial angle $\phi_0$ (colorbar), where $v_b$ is the expansion velocity at the biofilm's edge and $G_b$ is the biofilm shear modulus. Theory curves are computed with the circumferential compression of the biofilm edge set to $\mathbf{F}^0_\mathrm{e,\theta\theta} = 0.8$, but the results depend only minimally on this choice (Fig.~S9).
Colored circles show experimental data (mean $\pm$ std, $n$ = 3) at the designated agar concentrations. $\phi$ is measured at $t = 36$ h. Horizontal error bars are dashed because the friction coefficient $\eta$ is not directly measured, but rather is inferred by fitting (see Methods for details).
\label{Fig2}
}
\end{figure*}

To understand the observations described in the previous section, we developed a chemo-mechanical model of biofilm development that takes into account the diffusion of nutrients and their uptake by bacteria (Fig.~\ref{Fig2}A), growth of the biofilm, mechanical deformation of the biofilm and the agar substrate, and the friction between them (Fig.~\ref{Fig2}B). In this section, we focus on the early stage of  development, when the biofilm surface is still flat. We denote by superscript 0 the deformations of the flat biofilm. The modifications of the model required to describe the wrinkled morphologies are discussed in a later section. 

The kinematics of biofilm development are described by a time-varying mapping between an internal material coordinate system $\bs{X}_0$ and the laboratory frame $\bs{x}$, i.e. $\bs{x} = \bs{x}(\bs{X}_0,t)$. Following the finite-strain formalism \cite{ogden1997non}, we define the deformation gradient ${\bf F}=\partial \bs{x}/ \partial \bs{X}_0$, which captures the local change in shape and volume of a biofilm relative to its initial configuration. The overall change in shape arises from both growth and mechanical deformation. Accordingly, we follow the convention of multiplicative decomposition \cite{rodriguez1994stress,goriely2005differential,dervaux2008morphogenesis}, and decompose $\mathbf{F}=\Fe\Fg$ into a contribution $\Fg$ due to growth (which results in a post-growth intermediate configuration $\bs{X}$, where neighboring regions may overlap creating incompatibility; Fig.~\ref{Fig2}B) and a contribution $\Fe$ due to elastic deformation, characterizing the reorganization required to ensure compatibility (deformed contours in Fig.~\ref{Fig2}B). Using this theoretical framework, we next specify our model of biofilm growth and mechanics (see also Supplementary Notes I and II for details).

During development, \vc biofilms on agar stabilize at a thickness of roughly 100 $\mu$m, which is set by the penetration depth of oxygen \cite{costerton1995microbial}, and subsequently extend primarily in 2D along the substrate. Experimentation shows that bacterial cells in \vc biofilms grown on agar do not locally order in the horizontal directions \cite{yan2018bacterial}. Therefore, we model the growth part of the deformation gradient as $\Fg = \lamg {\bf I}_\parallel \oplus 1$, reflecting an isotropic increase in size in the planar direction by a factor $\lamg$, and neglecting growth in the vertical direction (the thickness $H$ of the undeformed biofilm is assumed to be constant). The thin-film geometry also permits a simplified 2D representation of a biofilm in which physical quantities are expressed as functions of an in-plane coordinate $\bs{x}_\parallel$.

In order to account for nutrient-dependent biofilm growth, we consider the kinematics of a 2D nutrient field $c_\parallel(\bs{x}_\parallel,t)$: 
\begin{equation}
\frac{\partial c_\parallel}{\partial t}=D \nabla^2_\parallel c_\parallel - Q_0 \Jep^{-1}\, \frac{ c_\parallel}{(K+c_\parallel)}.
\end{equation}
Here, $D$ is the diffusion constant, spatial derivatives are taken with respect to $\bs{x}_\parallel$, and the final term describes the uptake of nutrients by bacteria according to the Monod law~\cite{monod1949growth}, where $K$ is the concentration of nutrients at the half-maximal uptake rate, $Q_0$ is the maximum uptake rate per unit area in the intermediate grown configuration, and the $\Jep=\det(\Fep)$ factor is included to account for the change in areal density of bacteria due to elastic deformation (see Methods and description below). The growth field $\lamg(\bs{x}_\parallel,t)$ evolves in time according to the consumption of nutrients $\partial \lamg/\partial t = \kg (c_\parallel) \lamg$, where the growth rate $\kg (c_\parallel)$ is related to the Monod law described above (see Supplementary Note II for details). This reduced 2D model provides a reasonable approximation to the full 3D diffusion dynamics of nutrients in the agar (see Supplementary Note IIE and Figs.~S2 and S3). Importantly, the 2D model is sufficient to capture the spatially non-uniform growth that plays an essential role in biofilm morphodynamics. 

Mechanically, we model the biofilm as a plane-stress thin film, where it is assumed that the stress components perpendicular to the biofilm surface are negligible. The plane-stress simplification allows for the elastic deformation $\Fe^0=\Fep^0 \oplus \gamma^0$ to be decomposed into the in-plane compression $\Fep^0$ and the vertical stretch $\gamma^0$ (see Supplementary Note IIB and Fig. 2B), leading to a quasi-2D description of a biofilm with varying thickness $h(\bs{x}_\parallel,t)=\gamma^0(\bs{x}_\parallel,t) H$.
 
In our model, mechanical stresses in the biofilm arise from elastic deformation, and are specified by the constitutive relation $\bs{ \sigma}_\parallel (\Fe)$, where $\bs{ \sigma}_\parallel$ is the in-plane stress tensor. Biofilms are complex hydrogel-like materials, whose constitutive relations are well approximated by nearly incompressible neo-Hookean elasticity (see Supplementary Note IIF and Fig.~S4). Here, we modeled the biofilm as an incompressible neo-Hookean elastic material \cite{budday2014mechanical,shyer2013villification,tallinen2016growth}, but our results are largely insensitive to any plausible choice of rheological model for the biofilm (see Supplementary Note VII and Fig.~S14).

We obtain the expansion velocity $\bs{v}(\bs{x}_\parallel)$ of a growing biofilm from a differential equation for local force balance,
\begin{equation}
\nabla_\parallel\cdot(h \bs{ \sigma}_\parallel) - \eta \bs{v} = 0.
\label{eq:ForceBalance}
\end{equation}
Here, we assume that friction between the growing biofilm and the agar arises from binding and unbinding of biofilm matrix polymers with the adhesive biofilm proteins that have been secreted onto the agar surface \cite{berk2012molecular}. In particular, we model the friction as viscous friction, and we assume the friction coefficient $\eta$ to be proportional to the shear modulus $\Gs$ of the agar substrate \cite{walcott2010mechanical,sens2013rigidity,schwarz2013physics} (see Supplementary Note IIG and Fig.~S5). The partial differential equations for this model were solved numerically in the Lagrangian coordinate system with the open source computing platform FEniCS (see Methods for details). Next, we present model results for the early stages of biofilm development, prior to wrinkling.

\subsection*{Biofilm expansion has three kinematic stages}

In the model, nutrients are gradually depleted underneath the growing biofilm (Fig.~\ref{Fig2}C). Once a steadily expanding state is achieved after about 15 h, most of the growth is restricted to the narrow nutrient-rich zone $a_\mathrm{c} \approx 1\,\textrm{mm}$ near the rim of the biofilm (Figs.~\ref{Fig2}C and S3), which is consistent with experiment~\cite{yan2019mechanical}.

Our  model predicts three stages of biofilm expansion with distinct radial velocity profiles $v(r,t)$ (Figs.~\ref{Fig2}D and S6), where $r$ is the distance from the center of the biofilm. In the initial stage, the interior of the biofilm is stationary while the biofilm edge moves slowly outward. The magnitude of radial velocity and the size of the moving region gradually increase until, at the second stage, the entire biofilm undergoes uniform expansion with a radial velocity $v$ that is linearly proportional to $r$. At later times, in the third stage, the biofilm expansion in the central region slows due to the depletion of nutrients while the edge of the biofilm continues to move outward with a steady velocity (Fig.~S6). These three kinematic stages are also observed in experiments and our model predictions closely match the measured velocity profiles (Fig.~\ref{Fig2}D).

The three kinematic stages can be understood in the following way: During the first stage, which corresponds to early times ($t \ll 1/\kg^\mathrm{max}$ where $\kg^\mathrm{max}$ is the maximal growth rate of a biofilm), friction with the agar substrate prevents the growing biofilm from expanding radially in the central region. As stresses gradually build up (Fig.~\ref{Fig3}A and B), the width of the mobile annular zone at the biofilm edge increases in proportion to $(t/\eta)^{1/2}$ (see Supplementary Note III and Fig.~S7). The second stage ensues once this mobile zone spreads through the entire biofilm, so that the radial velocity becomes approximately a linear function of the distance from the center (see Supplementary Note III and Figs.~\ref{Fig2}D and S8).  The third stage follows once the nutrients in the central region are depleted, which slows down biofilm growth and reduces the radial velocity in that region (Fig.~\ref{Fig2}C and D). 

\subsection*{Higher friction increases the biofilm leading angle}
We next investigate how friction shapes the edge of an expanding biofilm. Specifically, we consider the local deformation of the wedge-shaped edge of a biofilm when sliding on a surface with velocity $v_\mathrm{b}$. The surface provides a frictional shear force of magnitude $\eta v_\mathrm{b}$ acting on  the bottom of the biofilm edge, and thus generates a simple shear parallel to the horizontal plane. This shear deformation increases the leading angle from $\phi_0$ in the rest state to $\phi$ in the deformed state (Fig.~\ref{Fig2}B).

To quantify how the biofilm leading angle increases with friction, we decomposed the elastic part of the deformation gradient $\Fe$ into the product of rotations $\mathbf{R}$ and principal stretches $\mathbf{U}$ to connect the geometry, characterized by the angles  $\phi$ and $\phi_0$, to the stress state of the biofilm edge (see Supplementary Note IV and Fig.~S9). This analysis yields the following relation
\begin{gather}
    \tan^2 \phi = \frac{\tan \phi_0 + \zeta}{1/\tan \phi_0 - \zeta} \label{eq_angle}
\end{gather}
between the leading angle $\phi$ and friction, where $\zeta = (\mathbf{F}_\mathrm{e,\theta\theta})\eta v_\mathrm{b}/\Gb$ denotes the scaled friction normalized by the biofilm shear modulus $\Gb$, and $\mathbf{F}_{\mathrm{e,\theta\theta}}$ describes the circumferential compression at the biofilm edge. In the absence of friction, i.e., when $\zeta = 0$, Eq.~(\ref{eq_angle}) reduces to $\phi = \phi_0$. In the presence of friction, our analysis predicts that the leading angle $\phi$ increases with $\zeta$ when $\zeta < 1/\tan\phi_0$, while the biofilm edge bulges out and constantly tumbles (no steady-state translation) if $\zeta > 1/\tan\phi_0$.

The experimental difficulty in measuring the friction coefficient $\eta$ and the circumferential compression  $\mathbf{F}_\mathrm{e,\theta\theta}$ precludes a direct quantitative comparison with theory; nevertheless, it is clear from the measured $\phi$ of experimental biofilms that higher friction (i.e., higher concentration agar) increases the biofilm leading angle (see Methods and Figs.~\ref{Fig2}E and S9).

\subsection*{Non-uniform biofilm growth results in anisotropic stress}

\begin{figure*}[t!]
\centering
\includegraphics[width=.8\textwidth]{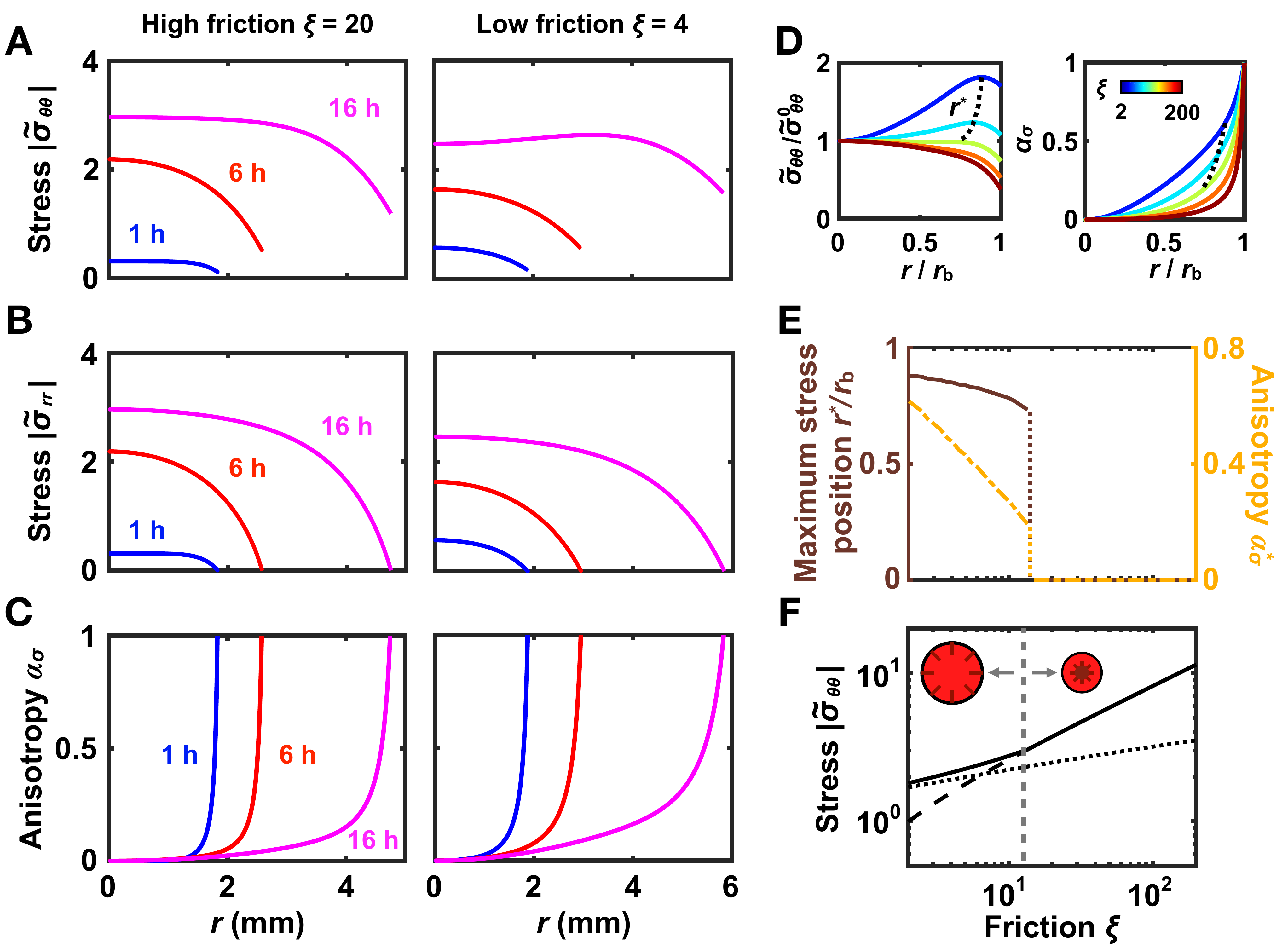}
\vspace{1em}
\caption{
{\bf Spatiotemporal evolution of the stress field in a growing biofilm.} 
({\bf A-C}) Magnitude of ({\bf A}) the circumferential stress $\tilde{\sigma}_{\theta\theta}$, ({\bf B}) the radial stress $\tilde{\sigma}_{rr}$, and ({\bf C}) the stress anisotropy $\alpha_\sigma=(\tilde{\sigma}_{\theta\theta} - \tilde{\sigma}_{rr})/(\tilde{\sigma}_{\theta\theta} + \tilde{\sigma}_{rr})$, plotted against the radial coordinate $r$ at the designated times for high friction (left, $\xi=20$) and low friction (right, $\xi=4$). All the stresses are normalized by the biofilm shear modulus $\Gb$, i.e., $\tilde\sigma = \sigma / G_b$. Identical simulation parameters were used as in Figs.~\ref{Fig2}{\bf C} and {\bf D}. 
({\bf D})~Circumferential stress $\tilde{\sigma}_{\theta\theta}$ ({\it left}, normalized by the stress at the center of the biofilm $\tilde{\sigma}_{\theta\theta}^0$) and stress anisotropy $\alpha_\sigma$ ({\it right}) versus the radial coordinate $r$ normalized by the biofilm radius $r_\mathrm{b}$ plotted for different dimensionless friction parameters, defined as $\xi = \frac{\eta (\kg^\mathrm{max}R_\mathrm{b0})}{\Gb (H/R_\mathrm{b0})}$, where $\eta$ denotes the friction coefficient, $\kg^\mathrm{max}$ denotes the maximum growth rate of the biofilm, and $R_\mathrm{b0}$ and $H$ denote, respectively, the initial biofilm radius and thickness. Color scale indicates the values of $\xi$ on a logarithmic scale. The five curves (colored from blue to cyan) correspond to $\xi = e^{0.7}, e^{1.7}, e^{2.7},e^{3.7},$ and $e^{4.7}$, respectively. $r^*$ denotes the radial position where the circumferential compressive stress reaches a maximum (black dotted curves).  
({\bf E})~The normalized radial coordinate $r^*/r_\mathrm{b}$ and the stress anisotropy $\alpha_\sigma^*$ at the position of maximum circumferential compressive stress, plotted as functions of the dimensionless friction parameter $\xi$. 
({\bf F})~The magnitude of the largest circumferential compressive stress $\tilde{\sigma}_{\theta\theta}^*$ ($r = r^*$, solid curve) compared to the circumferential stress at the edge ($r = r_\mathrm{b}$, dotted curve) and at the center ($r = 0$, dashed curve) for different dimensionless friction parameters $\xi$. When the circumferential stress at the edge of the biofilm is larger (smaller) than the circumferential stress at the center, radial patterns start forming near the edge (near the center) as indicated by the inset figures. In experiments, the transition between two different morphologies occurs at 0.7\% agar concentration (gray dashed line). Panels {\bf D-F} show simulation results at time $t$ = 30 h, which is roughly when the experimental biofilms start to form periodic wrinkles.\label{Fig3}
}
\end{figure*}

The evolution of mechanical stresses during the early stages of biofilm growth dictates the onset of mechanical instability and the consequent morphology of the wrinkles. Thus, we investigate the evolution of the magnitude of radial stress $\sigma_{rr}$, the magnitude of circumferential/hoop stress $\sigma_{\theta\theta}$, and the stress anisotropy defined as $\alpha_\sigma = (\sigma_{\theta\theta} - \sigma_{rr})/(\sigma_{\theta\theta} + \sigma_{rr})$ \cite{chen2018stress}. The isotropic stress state corresponds to $\alpha_\sigma =0$, while pure hoop stress corresponds to $\alpha_\sigma =+1$ and pure radial stress corresponds to $\alpha_\sigma = -1$. 

The spatial distributions of stresses have distinctive characteristics during each of the three kinematic stages of biofilm expansion. Initially, the inner core of the biofilm only minimally expands ($\Fep^{0} \Fg \approx {\bf I}_\parallel$), which, given the material growth of the biofilm, $\Fg = \lamg \mathbf{I}_\parallel$, must result in a compensating isotropic in-plane deformation ($\Fep^{0} \approx \lamg^{-1} {\bf I}_\parallel$), and thus an isotropic compressive stress state with $\sigma_{rr} = \sigma_{\theta\theta}$ (Figs.~\ref{Fig3}A-C). Moreover, stresses are approximately uniform in magnitude throughout the immobile core region of the biofilm, but decline in the outer mobile region. Note that the value of radial stress $\sigma_{rr}$ necessarily decreases to zero at the edge of the biofilm, while the hoop stress $\sigma_{\theta\theta}$ can be nonzero. Therefore, the stress anisotropy is initially localized to the outer mobile region.

As the biofilm continues to grow, internal stresses increase exponentially in time, and eventually overcome friction, enabling the entire biofilm to expand uniformly (Figs.~\ref{Fig2}D and \ref{Fig3}A-C). During this second stage, mechanical stresses continue to increase exponentially and acquire a characteristic parabolic profile (see Supplementary Note III and Fig.~\ref{Fig3}). During the third stage when nutrients become depleted, stresses increase more slowly near the center of the biofilm due to the reduced rate of biofilm growth, while the magnitude of hoop stress near the edge still increases exponentially due to continuous biofilm growth in this nutrient-rich region (Fig.~\ref{Fig2}C). As a result, when friction is low, the location of the maximum hoop stress shifts away from the center of the biofilm during the later stages (Fig.~\ref{Fig3}A). Note that the stress anisotropy is always positive (Fig.~\ref{Fig3}C), meaning that the compressive hoop stress is always larger in magnitude than the radial stress.  

We found that the region of the biofilm under anisotropic stress becomes larger during the third stage of development (Fig.~\ref{Fig3}C). Thus, we hypothesized that the non-uniform growth pattern due to depletion of nutrients plays an important role in generating anisotropic stresses. To quantify the extent of stress anisotropy for the entire biofilm, we computed the normalized range of anisotropy $\Delta r_\alpha/r_\mathrm{b}$, defined as the radial range of the area where $\alpha_\sigma > 0.1$ relative to the biofilm radius $r_\mathrm{b}$, as a function of time. According to our model, the increase of $\Delta r_\alpha / r_\mathrm{b}$ is accompanied by a narrowing of the nutrient-rich zone (Fig.~S10). To further explore this connection, we also computed the normalized anisotropy range for a uniformly growing biofilm, which we found to be close to zero (Fig.~S10). We conclude that the faster growth at the biofilm edge promotes predominantly circumferential stress (see Supplementary Note V), which explains the appearance of radial wrinkles in the peripheral regions in experiments (Fig.~\ref{Fig1}A).

\subsection*{Friction favors isotropic stress and shifts the position of maximal circumferential stress}
How does friction $\xi$ affect the distribution of mechanical stresses in a growing biofilm? To address this question, we compared the distribution of circumferential stress $\sigma_{\theta\theta}$ and stress anisotropy $\alpha_\sigma$ for a series of simulations with different friction coefficients
(see Fig.~\ref{Fig3}D).
Notably, at a typical time when biofilms start to form patterns in experiments, our simulations show that the radial position $r^*$, corresponding to the maximal circumferential stress, varies with the magnitude of friction: $r^*$ is near the biofilm edge when friction is small, while $r^*$ is near the biofilm center when friction is large (Figs.~\ref{Fig3}D and E).
Moreover, the stress anisotropy $\alpha_\sigma^*$ at $r^*$ decreases towards zero (isotropic stress state) with increasing friction (Fig.~\ref{Fig3}D and E).

Intuitively, these differences in stress distribution result from the counteracting effects of friction and non-uniform growth. Friction impedes biofilm expansion ($\mathbf{F}_\parallel \to \mathbf{I}_\parallel$ when $\eta \to \infty$), retards the relaxation of growth-induced isotropic compression ($\Feop \to \lamg^{-1}\mathbf{I}_\parallel$ when $\eta \to \infty$), and thus favors isotropic stress in the biofilm center. By contrast, non-uniform growth favors peripheral circumferential stress due to the mismatch between the biofilm perimeter that increases only linearly in time, and the exponential material growth of the biofilm at the edge. The fact that when friction is small the circumferential stress close to the biofilm rim is larger than that at the center (Fig.~\ref{Fig3}D and F) explains why, in experiments, the wrinkle pattern emerges from the outer region (Fig.~\ref{Fig1}A and B). In contrast, in experiments with high concentration agar, the wrinkle pattern first appears in the center of the biofilm because the large friction results in strong isotropic compression in that region.

\subsection*{The in-plane stress field determines the morphology of biofilm wrinkle patterns}
\label{sec:wrinkling}

\begin{figure*}[t!]
\centering
\includegraphics[width=.8\textwidth]{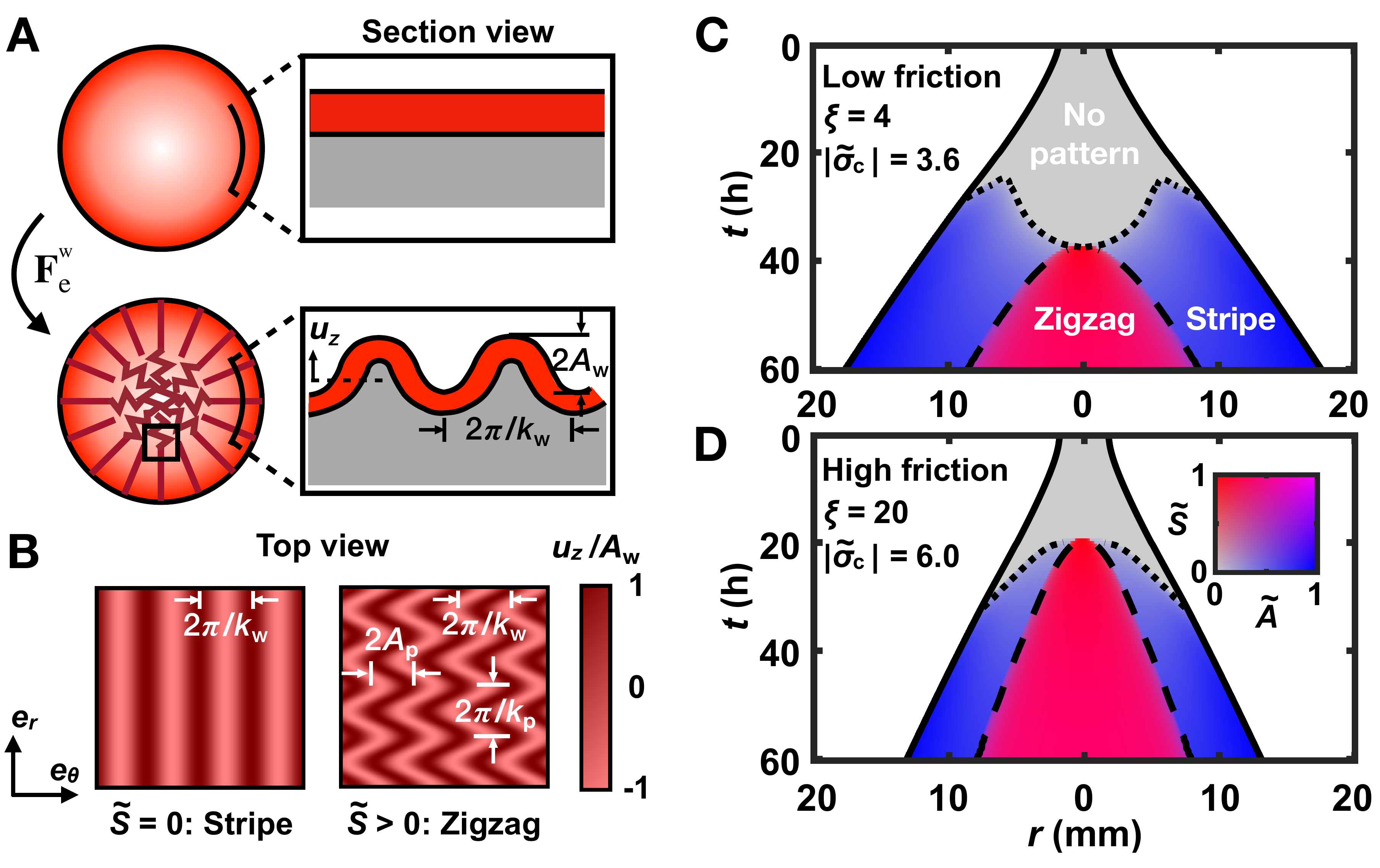}
\vspace{1em}
\caption{
{\bf Morphology and spatiotemporal dynamics of biofilm wrinkle patterns.}
({\bf A},{\bf B}) Schematic of the wrinkling model. The wrinkling deformation tensor, denoted by $\Few$, maps a pre-stressed, flat biofilm ({\it top}) to a wrinkled biofilm ({\it bottom}). Color code as in Fig. 2{\bf B}. The wrinkling pattern is characterized by two dimensionless scalar fields: the normalized amplitude $\tilde{A}$ and the shape factor $\tilde{S}$. $\tilde{A}$ is defined as the product of the wave number (denoted by $k_\mathrm{w}$) and the amplitude (denoted by $A_\mathrm{w}$) of periodic wrinkles. $\tilde{S}= k_\mathrm{p} A_\mathrm{p}/\sqrt{6}$ where $k_\mathrm{p}$ and $A_\mathrm{p}$ denote, respectively, the wave number and the amplitude of in-plane wiggles. The shape parameter $\tilde{S}$ was normalized such that its values are restricted to the interval $[0,1]$. {\it Right}: Panels in {\bf A} show close-up cross-sections of a flat ($\tilde{A} = 0$) and a wrinkled ($\tilde{A}>0$) biofilm. $u_z$ denotes the out-of-plane displacement. {\bf B} shows top view schematics of the herringbone ansatz, $u_z = A_\mathrm{w}\cos[k_\mathrm{w}(x_\theta - A_\mathrm{p}\cos k_\mathrm{p} x_r)]$, for a straight striped pattern ($\tilde{S}=0$) and a zigzag pattern ($\tilde{S}>0$). Here, $x_\theta = r\theta$ and $x_r = r$ denote, respectively, the linear coordinates along the circumferential and radial directions.
({\bf C}, {\bf D}) Kymograph representation of the evolution of patterns of the modeled biofilm for designated parameters ($\tilde{\sigma}_c$ denotes the critical stress, normalized by the biofilm modulus $G_\mathrm{b}$, for the onset of the wrinkling instability). The top and bottom kymographs can be interpreted as biofilms grown on different agar concentrations (see Fig. 1{\bf B}). $\tilde{\sigma}_c$ in {\bf C} is chosen such that the wrinkling instability occurs at a time similar to that in experiment. $\tilde{\sigma}_c$ in {\bf D} is then inferred according to the dependence of critical stress on $\Gs/\Gb\propto\xi$. See Methods for details. Solid, dotted, and dashed curves, respectively, denote the boundaries of the entire biofilm, regions with a radial stripe pattern ($\tilde{A} > 0, \tilde{S} = 0$), and regions with a zigzag pattern ($\tilde{A} > 0, \tilde{S} > 0$). Gray denotes the region without any pattern. For the patterned regions, the color of each spatiotemporal bin indicates the local amplitude $\tilde{A}(r,t)$ and shape $\tilde{S}(r,t)$ of the pattern. The color code is shown in the inset of {\bf D}. \label{Fig4}
}
\end{figure*}

Lastly, we address how the stress profiles discussed above dictate the morphology of biofilm wrinkles. As the biofilm grows, the magnitude of compressive stresses increases (Fig.~\ref{Fig3}). Once compressive stresses reach a critical value $\sigma_c$, the flat state becomes unstable to the formation of wrinkles
\cite{cai2011periodic,audoly2008buckling, yan2019mechanical}. The critical compressive stress $\sigma_c$ increases with $\Gs(\propto\xi)$, and scales as $\sigma_\mathrm{c}\sim \Gs^{2/3}\Gb^{1/3}$ in the asymptotic limit where the shear modulus of the biofilm $\Gb$ is much larger than that of the agar substrate $\Gs$ (see Methods and \cite{yan2019mechanical} for details). It was previously shown that, for highly anisotropic stresses, wrinkles are oriented orthogonal to the direction of maximum compressive stress, whereas for isotropic stresses, wrinkles form zigzag herringbone-like patterns~\cite{cai2011periodic,audoly2008buckling} (Fig.~S11). The stress profiles in Fig.~\ref{Fig3} are thus consistent with the experimental observations in Fig.~\ref{Fig1}A that radial wrinkles form in the outer regions, where the stress is predominantly circumferential,  whereas zigzag wrinkles form in the core region where the stress is largely isotropic.

In order to more quantitatively understand the spatiotemporal evolution of biofilm wrinkle patterns, we developed a 2D coarse-grained model that employs two scalar order parameter fields $\tilde{A}$ and $\tilde{S}$ to describe, respectively, the amplitude and the shape of the wrinkle patterns. Specifically, $\tilde{A} = 0~(\tilde{A} > 0)$ corresponds to the flat (wrinkled) state, and $\tilde{S}=0$ ($\tilde{S}>0$) corresponds to striped (zigzag) wrinkles. The total elastic deformation of a biofilm $\Fe = \Few \Feo$ is decomposed as the superposition of the wrinkling deformation $\Few$, which also deforms the agar, and the planar compression $\Feo$ (Fig.~\ref{Fig4}A). We follow previous work \cite{cai2011periodic,audoly2008buckling} to describe wrinkles of different morphologies, and we use the herringbone ansatz to approximate the out-of-plane displacement in coarse-grained patches of the biofilm (see Supplementary Note VI and Fig. 4B). The primary sinusoidal wrinkling with amplitude $A_\mathrm{w}$ (in the vertical $z$ direction) and wavelength $\lambda_\mathrm{w} = 2\pi / k_\mathrm{w}$ occurs along the direction that corresponds to the maximum compressive stress (circumferential direction ${\bf e}_\theta$ in our case). The secondary sinusoidal wiggles with amplitude $A_\mathrm{p}$ (in the horizontal direction) and wavelength $\lambda_\mathrm{p} = 2\pi / k_\mathrm{p}$ appear in the orthogonal direction (Fig.~\ref{Fig4}B). In terms of these quantities, the relevant dimensionless order parameters are $\tilde{A} = k_\mathrm{w}A_\mathrm{w}$ and $\tilde{S} = k_\mathrm{p} A_\mathrm{p}/\sqrt{6}$.

The formation of wrinkles relaxes the elastic compressional energy of the biofilm, but at the expense of the bending energy of the biofilm and the elastic deformation energy of the agar. By taking into account these energy contributions and using the above ansatz for the shape of wrinkles, we derived the following total free-energy density per unit area (see Supplementary Note VI)
\begin{eqnarray}
\frac{\Psi_\parallel^\mathrm{total} }{\Gb H} &\approx& - \frac{1}{4}\Big((|\tilde{\sigma}_{\theta\theta}^{0}| - \tilde{\sigma}_\mathrm{c})(1+3 b\tilde{S}^2) + 3  (|\tilde{\sigma}_{rr}^{0}| - \tilde{\sigma}_\mathrm{c})\tilde{S}^2\Big)\tilde{A}^2 \nonumber \\
&& + \frac{1}{8}\Big(1 + (6 b + 3) \tilde{S}^2 \Big)\tilde{A}^4,
\label{eq:FreeEnergy}
\end{eqnarray}
which is valid for stresses near the critical stress $\tilde{\sigma}_\mathrm{c}=(3\Gs/\Gb)^{2/3}$ of the wrinkling instability. Here, stresses $\tilde{\sigma}\equiv \sigma/\Gb$ are normalized by the biofilm shear modulus $\Gb$. $\tilde{\sigma}_{\theta\theta}^{0}<0$ and $\tilde{\sigma}_{rr}^0<0$, respectively, denote the circumferential and radial pre-stress due to the planar compression $\Feo$. We set $b \approx 2/3$ to ensure that the relaxed stresses due to wrinkling remain isotropic when the imposed pre-stress is isotropic (see Supplementary Note VI).

Here, we assume that the dynamics of wrinkling is determined by the slower dynamics of the stress field. Under this approximation, the predicted wrinkled morphology corresponds to the minimum of the free energy in Eq.~(\ref{eq:FreeEnergy}). The wrinkling instability occurs via two successive continuous phase transitions controlled by the magnitude and by the anisotropy of the pre-stress (see Supplementary Note VI and Fig.~S11). The primary bifurcation from the planar state ($\tilde{A}=0$) to the wrinkled state ($\tilde{A}>0$) occurs if the magnitude of the maximum compressive stress $|\tilde{\sigma}_{\theta\theta}^{0}|$ exceeds the critical value $\tilde{\sigma}_\mathrm{c}$, while a secondary bifurcation from striped wrinkles ($\tilde{S} = 0$) to zigzag wrinkles ($\tilde{S} > 0$) occurs only when the stress anisotropy is sufficiently small, $\alpha_\sigma^0 < ( |\tilde{\sigma}_{\theta\theta}^{0}|  - \tilde{\sigma}_\mathrm{c}) /(3|\tilde{\sigma}_{\theta\theta}^{0}| + \tilde{\sigma}_\mathrm{c})$. Note that for isotropic compressive pre-stresses ($\tilde{\sigma}_{\theta\theta}^{0}=\tilde{\sigma}_{rr}^{0}$) the free-energy density is minimized by $\tilde{S}=1$.

\subsection*{The evolution of the stress field determines the biofilm wrinkling morphodynamics}
Wrinkling relaxes the mechanical stresses in the biofilm by releasing the in-plane compressive strain through out-of-plane deformation. 
Once wrinkling occurs, this relaxation mechanism prevents the magnitude of compressive stresses from increasing beyond the critical stress (see Methods, Supplementary Note VI and Fig.~S11).

We incorporated the above mean-field description of the wrinkling instability and consequent stress relaxation into our chemo-mechanical model (see Methods and Supplementary Note~VI for details). Consistent with the experimental observations in Fig.~\ref{Fig1}B, we find that for biofilms grown on low concentration agar (low shear modulus of the substrate and small friction since we assume $\eta \propto \Gs$), radial wrinkles initiate near the outer edge, then propagate inward and once they reach the center, zigzag wrinkles form in the core region (Fig.~\ref{Fig4}C). On the other hand, for biofilms grown on high concentration agar (high shear modulus of the substrate and large friction) radial wrinkles initiate in the center and expand outward, while zigzag wrinkles simultaneously appear in the core region (Figs.~\ref{Fig1}B and \ref{Fig4}D).
According to our model, compared to the case where the wrinkling instability is prevented, the expansion of the wrinkled biofilm is slowed, the stress anisotropy is reduced, and the magnitude of compressive stress is reduced as well (Figs.~S11 and S12). Thus, our model suggests that wrinkling due to a growth-induced mechanical instability feeds back and further influences biofilm expansion and pattern formation by modifying the distribution of internal stress.

\section*{Discussion}
Our experimental and modeling results highlight the connections between nutrient supply, bacterial growth, biofilm and substrate mechanics, and friction in shaping the morphology of developing bacterial biofilms on soft substrates. The depletion of nutrients beneath the center of the biofilm leads to localized growth primarily near the biofilm edge, consistent with previous experiments \cite{yan2019mechanical,srinivasan2019multiphase, liu2015metabolic}. This uneven growth profile, in turn, produces anisotropic compressive stresses, which are predominantly circumferential at the periphery of the biofilm, but are largely isotropic in the central region. The consequence of such a stress profile is the formation of radial wrinkles in the outer region of the biofilm and a zigzag herringbone-like pattern in the central region. Moreover, the location of the maximum circumferential stress -- where wrinkles first appear once the magnitude of the stress reaches a critical value -- varies with the magnitude of friction, from near the outer edge when friction is small to near the center when friction is large.
As a result, for biofilms grown on soft agar substrates with low friction, wrinkles first appear in the peripheral region and propagate inward. In contrast, for biofilms grown on stiff agar substrates with high friction, wrinkles first appear in the central region and propagate outward.

What are the biological implications of forming 3D biofilm structures? One possibility is that the wrinkled thin film structure provides a larger surface area-to-volume ratio compared to a flat film, thereby enhancing access to nutrients and conferring growth advantages to the bacterial population \cite{kempes2014morphological,stewart2008physiological}. Furthermore, under adverse nutrient conditions, biofilms disassemble through a process called dispersal, and dispersing cells primarily depart from the biofilm's outer surface \cite{singh2017vibrio}. Therefore, the large surface area of wrinkled biofilms may facilitate dispersal when submerged. The convoluted 3D structure of biofilms also reduces the average distances between cells compared to a flat film of the same area, which might enhance communication between bacterial cells, e.g. via quorum-sensing signaling \cite{miller2001quorum,hammer2003quorum}. Finally, the 3D biofilm structure positions biofilm cells at different heights,  potentially generating a ``bet hedging'' strategy under particular conditions. For example, the rough surfaces of wrinkled biofilms exposed to external flows will alter the flow field, forming large (small) shear stress zone near the peaks (valleys) of wrinkles. Consequently, the cells near the peaks will exhibit a larger probability to detach while the cells near the valleys will tend to stay attached to the surfaces. \cite{shen2015role}. 

Our results provide insight into the spatiotemporal development of \vc biofilm morphology, but it remains to be explored  whether and how the increasing mechanical stress and/or the formation of the 3D biofilm structure affect the proliferation rate of bacteria, alter biofilm matrix production, or promote survival success of cells in particular biofilm regions. Additional experimental studies will be required to complete our understanding of how growth, mechanical stresses, and morphological transitions are coupled to drive biofilm development. Furthermore, our study focused only on the initial stages of wrinkling, during which the amplitudes of wrinkles are small and biofilms remain in contact with the agar substrate. We previously demonstrated that at later stages of development, biofilms can locally delaminate from the agar substrate, which significantly influences the subsequent development of morphological patterns~\cite{yan2019mechanical}.

In this work, we modeled the rheology of growing \vc biofilms as that of a hyperelastic material. However, our previous measurements show that \vc biofilms actually behave as more complex viscoelastic media that yield upon large shear deformation~\cite{yan2018bacterial}. Indeed, our elastic material model leads to stresses that are larger than the measured yield stress. We thus suspect that yielding constantly occurs during \vc biofilm growth. More generally, reorganization and yielding of growing biological materials are commonly observed during morphogenesis, for example in plants~\cite{dumais2006anisotropic}, fruit flies \cite{he2014apical,guirao2017biomechanics} and brain tissues \cite{foubet2018mechanical}.
Thus, the effects of viscoelasticity~\cite{matoz2019wrinkle} and elastoplasticity (Figs.~S13 and S14) of biofilms on their morphological development will be an important topic for future studies.

The concepts we presented here to analyze the development of \vc biofilms should also be applicable to biofilms of many other bacterial species that form similar morphological patterns \cite{wilking2013liquid, dervaux2014growth, haussler2013biofilms}. However, there are also examples of biofilms with distinct morphologies, such as the distorted concentric rings observed in wild-type {\it Pseudomonas aeruginosa} PA14 biofilms~\cite{madsen2015facultative} and  biofilms formed by {\it Escherichia coli} K-12 strain W3110~\cite{serra2013microanatomy}. Our model suggests that if biofilm growth and/or matrix production is faster in the central region than the outer region, one expects a region in which the radial compressive stress surpasses the circumferential stress. In this case, our model predicts that wrinkles will form as concentric (possibly distorted) rings (see Fig.~S11). Such a pattern of matrix production is indeed reported in the two biofilm formers mentioned above. For example, in wild-type {\it P. aeruginosa} biofilms, cells at the biofilm center display upregulated matrix production due to oxygen limitation, whereas cells located in the oxygen rich periphery downregulate matrix production~\cite{madsen2015facultative}. In biofilms formed by {\it E. coli} K-12 strain W3110, cells generate matrix components (amyloid curli fibers) only upon entry into stationary phase when nutrients are depleted, which typically occurs first at the biofilm center~\cite{serra2013microanatomy}.
Thus, we expect that similar physical mechanisms to those underlying the dynamics of expansion and pattern formation of \vc biofilms may be widely applicable to other bacterial biofilms, including those with distinct morphologies.

\section*{Materials and Methods}

\subsection*{Growing and imaging experimental biofilms} 
\subsubsection*{Bacterial strain and biofilm growth}
The \vc strain used in this study is a derivative of the \vc O1 biovar El Tor strain C6706str2 \cite{thelin1996toxin} that harbors a missense mutation in the {\it vpvC} gene (VpvC W240R), which elevates the levels of c-di-GMP and confers a rugose biofilm phenotype~\cite{beyhan2007smooth}. 
Standard lysogeny broth (LB) medium solidified with different percentages of agar was used as the solid support on which biofilms were grown. Initially, \vc was streaked onto LB plates containing 1.5\% agar and grown at 37\degreeC\  overnight. Individual colonies were selected and inoculated into 3~mL of LB liquid medium containing $\sim$10 glass beads (MP Biomedicals Roll and Grow Plating Beads, 4~mm diameter) and these cultures were then grown at 37\degreeC\  with shaking to mid-exponential phase (about 5 h). Subsequently, the cultures of bacteria were mixed by vortex to break change{clumps} up into individual cells, the OD$_{600}$ was measured with a cell density meter (Amersham Biosciences Ultrospec 10), and then the cultures were diluted back to an OD$_{600}$ value of 0.5. 1 $\mu$L of these preparations were spotted onto pre-warmed agar plates made with different concentrations of agar.  Subsequently, the plates were incubated at 37\degreeC. During the first 10 h after inoculation, bacterial colonies formed the initial biofilms without extending beyond the inoculated circle (radius $R_0\approx 2$~mm). These biofilms were used as the initial/reference configurations for modeling, with $t=0$ in our simulations corresponding to the time when a biofilm starts expanding radially.
Four biofilms were grown per agar plate for the surface topography measurements, while for the time-lapse imaging, one biofilm was grown on each agar plate.

\subsubsection*{Time-lapse transmission imaging}
The imaging system has been described previously \cite{yan2019mechanical}. Briefly, an agar plate containing the inoculum was placed on an LED illumination pad (Huion L4S Light Box) and imaged with a Nikon D3300 SLR camera equipped with a Sigma 105 mm F2.8 Macro Lens. The entire setup was placed in a 37\degreeC\ environmental room and was covered to exclude light. The camera was controlled with DigiCamControl software. Imaging started 5~h after inoculation when the camera was capable of focusing on the growing biofilms. Image snapshots were taken automatically every 15~min for 3~days.

\subsubsection*{Image processing} 
The protocol and Matlab codes used to analyze the morphological features of biofilms have been reported previously \cite{yan2019mechanical}. In brief, an intensity-based thresholding method was used to binarize the pre-processed transmission images (using a built-in thresholding function in Matlab) and to separate the biofilm region from the background. For each biofilm, the transmission image taken 12~h after inoculation was used to define the biofilm center for the entire time course. The biofilm radius $r_\mathrm{b}$ was computed by averaging the distance between each point on the circumference and the center. 

Regions of the biofilm were binned into rings of width $0.2~\textrm{mm}$ for further analysis. First, we took the Fourier transform of the image intensity in the circumferential direction for each separate ring at distance $r$ from the center of the biofilm. Radial stripes appear in the resulting power spectrum as a sharp maximum at non-zero spatial frequency $f(r,t)$. The radial coordinate, at which the peak power instead appears at zero spatial frequency, was defined as the boundary of the region with a radial stripe pattern. Next, the radial intensity distribution $I(r)$ was obtained by averaging the intensity values over the circumferential direction for each ring. The intensity $I(r)$ for the disordered core is distinctly different (darker) from that of the outer region of the biofilm. Thus, we set a threshold intensity value for each biofilm to identify the central region with a disordered zigzag pattern, enabling us to measure this central region's radius as a function of time.

The velocity field of an expanding biofilm was measured by particle image velocimetry (PIV) performed with the open source tool PIVlab~\cite{thielicke2014pivlab}. The 2D displacement field between two successive frames (separated by 30 min) was computed via a Fourier transform correlation with three passes. The sizes of  interrogation windows for the three passes were chosen to be 128 pixels, 64 pixels, and 32 pixels, respectively. By averaging the radial components of the coarse-grained velocity vectors over the circumferential direction for each ring, we obtained the radial velocity field in Fig.~2D. Error bands correspond to standard deviations of the means. 

\subsubsection*{3D confocal profiling and leading angle measurement}
The surface profiles of biofilms grown for different times were analyzed with a Leica DCM 3D Micro-optical System. A 10$\times$ objective was used to image a roughly 3 mm $\times$ 3 mm region of the biofilm, with a step size of 2 $\mu$m in the $z$ direction. Subsequent processing and analyses were performed using Leica Map software. First, the three-point flattening procedure was performed on the agar surface to level the image. 3D views of biofilms were then rendered with a built-in function in the software. 

To measure the leading angle of an expanding biofilm, line profiles perpendicular to the biofilm periphery and spanning the region from the agar surface to the top surface of the biofilm were generated at five different locations. For each line profile, two points were manually selected on the biofilm edge, which were used to obtain a sloped line. The leading angle was then extracted with a built-in function in the software from the slope of this line.  The measurements of the initial angle were performed 12 h after inoculation. The steady-state leading angles were measured every 4~h from 24~h to 48~h after inoculation.

\subsection*{Modeling biofilms}
\subsubsection*{Continuum modeling}
To model the combined role of growth and mechanics in the morphological transition of \vc biofilms, we adopted the formulation of elastic growth \cite{rodriguez1994stress, goriely2005differential}, where the total geometric stretches $\Ft = \partial \bs{x}/\partial \bs{X}_0$, defined as the deformation gradient from the initial configuration $\bs{X}_0$ to the current configuration $\bs{x}$, are decomposed into stretches $\Fg$  due to growth and stretches $\Fe$ due to elastic deformation. Below, we describe a 2D chemo-mechanical model of  biofilm development that includes diffusion of nutrients and their uptake by bacteria, bacterial growth/extracellular matrix production, and mechanical deformation. The subscript $\parallel$ is used to denote the in-plane components of 3D vectors/tensors and reduced 2D variables, while a tilde $\sim$ above a variable is used to denote a dimensionless variable.

{\it Growth:}  Local growth of the \vc biofilm is treated as horizontal isotropic growth, i.e., $\Fg = \begin{pmatrix}\lambda_\mathrm{g}(t)\mathbf{I}_\parallel & 0 \\ 0 & 1\end{pmatrix}$, where $\mathbf{I}_\parallel$ denotes the 2D identity matrix. $\lambda_\mathrm{g}(t)$ describes the stretch in the horizontal direction due to growth, while there is no growth in the $z$ direction because biofilms maintain approximately constant thickness. The stretch due to growth evolves as $\partial \lamg/ \partial t = \kg \lamg$, where $\kg$ denotes the local growth rate. To capture the nutrient-dependent spatially non-uniform growth, $\kg(\tilde c_\parallel)$ is assumed to be a function of the normalized 2D nutrient field $\tilde{c}_\parallel (\bs{x}_\parallel,t)$ (normalized by the concentration of nutrients $c_0$ at the edge of the biofilm, i.e., $\tilde{c}_\parallel (\bs{x}_\parallel,t)\equiv c_\parallel (\bs{x}_\parallel,t)/c_0$), which reflects the nutrient availability in the agar medium. The equation that describes diffusion and uptake of nutrients is $\partial_t \tilde{c}_\parallel = D \bs{\nabla}_\parallel \tilde{c}_\parallel - Q(\Fe)\phi(\tilde{c}_\parallel) = Q_0(a_c^2\bs{\nabla}_\parallel \tilde{c}_\parallel - \Jep^{-1}\phi(\tilde{c}_\parallel))$, where $D$ is the nutrient diffusion coefficient, $Q_0$ characterizes the maximum nutrient uptake rate by bacteria in the undeformed grown configuration, and $\Jep = \det(\Fep)$ is introduced to account for the increased areal density of bacterial cells upon elastic deformation of the biofilm.  The characteristic width of the nutrient-rich annulus near the biofilm edge is given by $a_c = (D/Q_0)^{1/2}$. We assume that the uptake of nutrients depends on the local availability of nutrients via the Monod law, i.e., $\phi(\tilde{c}_\parallel) = \tilde{c}_\parallel / (\tilde{c}_\parallel + \tilde{K})$, where $\tilde{K} = 0.5$ is the concentration of nutrients at the half-maximal uptake rate \cite{monod1949growth} (Note that our model results are insensitive to the specific choice of $\tilde{K}$: for a different $\tilde{K}$, a similar $\tilde{c}_\parallel$ profile can be obtained by adjusting the uptake rate $Q_0$; see below for the fitting procedure). Finally, the growth rate is specified as $\kg(\tilde{c}_\parallel) = k_0\phi(\tilde{c}_\parallel) + \kr$, where $k_0$ is the maximum rate of nutrient-dependent growth and a small, constant nutrient-independent growth rate $\kr$ is added to account for the residual biofilm growth due to the vertical diffusion of nutrients (see Supplementary Note II and Figs.~S2 and S3).

{\it Mechanics:} The growth of the biofilm drives its expansion. As the biofilm moves relative to the agar, the friction $\bs{f}$ between the biofilm and the agar impedes biofilm expansion and induces internal mechanical stresses $\bs{\sigma}$. Friction is modeled as a viscous drag, i.e., $\bs{f}= -\eta \bs{v}_\parallel$, which is proportional to the expansion velocity of the biofilm $\bs{v}_\parallel = (\partial \bs{x}_\parallel /\partial t)_{\bs{X}_0}$, and the drag coefficient $\eta$ is assumed to be proportional to the agar shear modulus $\Gs$ \cite{walcott2010mechanical,zemel2010optimal,marcq2011rigidity,sens2013rigidity,schwarz2013physics} (see Supplementary Note IIG and Fig. S5). In order to relate stress $\bs{\sigma}$ to elastic deformation $\Fe$, we leverage the fact that the thickness of the biofilm ($\sim 100~\mu$m) is always about 10 to 100 times smaller than its radius ($\gtrapprox 5$ mm), and we treat the biofilm as a {\it plane stress} thin film made from nearly incompressible hyperelastic material. Thus, the thin-film deformation $\Fe=\Few \Feo$ is decomposed into the product of a wrinkling deformation $\Few$ and a planar deformation $\Feo = \begin{pmatrix}\Feop & 0 \\ 0 & \gammao \end{pmatrix}$, where $\Feop$ denotes the in-plane compression and $\gammao$ denotes the resulting vertical stretch. Lateral force balance yields $\eta q \bs{v}_\parallel = H\bs{\nabla}_\parallel\cdot(\gammao\bs{\sigma}_\parallel)$, where $H$ denotes biofilm thickness in the undeformed configuration, and $q = q(\Few)$ is a factor that accounts for the increase of the contact area between the biofilm and the agar due to the wrinkling profile.

Prior to wrinkling, the biofilm is flat and thus $\Few =\mathbf{I}$ and $q = 1$. The deformation $\Feo$ can be obtained from $\Feop = \partial \bs{x}_\parallel/\partial \bs{X}_{0,\parallel}$, and from $\gammao = 1/\text{det}(\Feop)$ due to incompressibility. The in-plane stresses are calculated to be $\bs{\sigma}_\parallel^0 = \Gb[\Feop(\Feop)^T - (\gammao)^2 \mathbf{I}_\parallel]$, where $\Gb$ denotes the biofilm shear modulus (see Supplementary Notes I,II for details). A wrinkling instability occurs once compressive stresses reach the critical value. To describe the wrinkling deformation $\Few$ we use two coarse-grained scalar fields, the amplitude $\tilde{A}$ and the shape $\tilde{S}$ (see Fig.~\ref{Fig4} and Supplementary Note VI). The fields $\tilde{A}$ and $\tilde{S}$ are computed from a Landau-Ginzburg-type free energy density Eq.~(\ref{eq:FreeEnergy}) and the factor $q$ is approximated as $q(\tilde{A}) = 1+\frac{1}{4}\tilde{A}^2[1+3(b+1)\tilde{S}^2]$, where $b\approx 2/3$ (see Supplementary Note VI for details). The stress relaxation from the pre-stress $\bs{\sigma}_\parallel^0$ to the true stress $\bs{\sigma}_\parallel$ due to wrinkling is described by (see Supplementary Note VI)
\begin{equation}
\begin{split}
{\sigma}_{\theta\theta} &\approx {\sigma}_{\theta\theta}^0 + \tilde{A}^2 \Big[ 1+(3 b  + 3/2) \tilde{S}^2 \Big]G_b,\\
\tilde{\sigma}_{rr} &\approx  {\sigma}_{rr}^0 + \tilde{A}^2\Big[1/2+(3b/2  + 3) \tilde{S}^2 \Big]G_b.
\end{split}
\label{Eq:StressRelaxation}
\end{equation} 
Note that for isotropic compressive pre-stresses (${\sigma}_{\theta\theta}^{0}={\sigma}_{rr}^{0}$), the relaxed stresses remain isotropic ($\tilde S=1$, $b=2/3$).

{\it Dimensionless governing equations:}
We define dimensionless variables $\tilde{\bs{\sigma}} = \bs{\sigma}/\Gb$, $\tilde{\bs{x}} = \bs{x}_\parallel / R_\mathrm{b0}$, and $\tau = t/\tau_0$,  where the shear modulus of the biofilm $\Gb$ was chosen as the scale for stresses, the initial biofilm radius $R_\mathrm{b0}$ as the characteristic length scale, and the inverse of the growth rate at the edge of the biofilm $\tau_0 = (\kg^{\max})^{-1} = [\kg(r = r_\mathrm{b})]^{-1}$ as the characteristic time scale associated with biofilm expansion. Upon non-dimensionalizing the equations describing biofilm growth and mechanics discussed above, we obtain the following equations
\begin{subequations}
\label{eq_ndim_all}
\begin{align}
    \begin{gathered}
        \text{Nutrient diffusion\quad}\\
        \text{\quad \quad \quad and uptake:\quad}
    \end{gathered}
    &~\partial_\tau \tilde{c}_\parallel = \tilde{Q}_0\Big[\tilde{a}_c^{2}\tilde{\bs{\nabla}}_\parallel^2\tilde{c}_\parallel - J_{e,\parallel}^{-1}\phi(\tilde{c}_\parallel)\Big], \label{eq_ndim_nc}\\
    \begin{gathered}
    \text{Nutrient limited\quad }\\
    \qquad\qquad\text{growth:\quad}
    \end{gathered}
    &~\partial_\tau\lamg =  \Big[(1-\tilde{k}_\mathrm{r})\frac{\phi(\tilde{c}_\parallel)}{\phi (1)} + \tilde{k}_\mathrm{r} \Big] \lamg, \label{eq_ndim_growth}\\
    \text{Force balance:\quad}&~\tilde{\bs{\nabla}}_\parallel\cdot(\gammao \tilde{\bs{\sigma}}_\parallel)  = \xi q(\tilde{A})\tilde{\bs{v}}_\parallel, \label{eq_ndim_force_balance}\\
    \begin{gathered}
           \text{Constitutive\quad }\\
           \text{relation:\quad}
    \end{gathered}
           &~ \tilde{\bs{\sigma}}^0_\parallel  = \Feop(\Feop)^T- (\gammao)^2 \mathbf{I}_\parallel, \label{eq_ndim_constitutive}
\end{align}
\end{subequations}
where $\tilde{Q}_0 = Q_0\tau_0$, $\tilde{k}_\mathrm{r} = k_\mathrm{r}\tau_0$, and the dimensionless friction $\xi = \frac{\eta (R_\mathrm{b0}/\tau_0)}{\Gb (H/R_\mathrm{b0})}$ is identified to be a control parameter of the model. Before wrinkling occurs, $\tilde{\bs{\sigma}}_\parallel = \tilde{\bs{\sigma}}^0_\parallel$. After wrinkling occurs, Eq. (\ref{eq_ndim_constitutive}) describes the pre-stress $\tilde{\bs{\sigma}}^0_\parallel$, and the actual stress $\tilde{\bs{\sigma}}_\parallel$ is computed from Eq. (5).
Taken together, the set of dimensionless governing equations is able to describe both the planar expansion of the biofilm ($\tilde{A} = 0$) and the 3D biofilm wrinkling morphology ($\tilde{A} > 0$). The parameters in our model are either estimated directly from experiment or are obtained by fitting to experimental data (see Fig.~S1 and Table~S2).

\subsubsection*{Numerical simulations}
The numerical solutions of  Eq.~(\ref{eq_ndim_all}) were obtained by performing finite element simulations. Rather than solving Eqs. (\ref{eq_ndim_nc}) and (\ref{eq_ndim_force_balance}) in the Eulerian frame, these equations were rewritten and solved in the Lagrangian frame of reference (see Supplementary Note IID for details).
We further assumed axisymmetric solutions and expressed the governing equations in polar coordinates to numerically solve for six scalar fields $\tilde{r}$, $\lamg$, $\tilde{c}_{\parallel}$, $\tilde{\gamma}$, $\tilde{A}$, and $\tilde{S}$ as functions of the dimensionless initial radial coordinate $\tilde{R}_0$. The validity of the axisymmetry assumption was verified by comparing to simulations on 2D circular domains with no assumption of symmetry. We used a fixed 1D domain of $\tilde{R}_0\in [0,1]$ that was discretized and generated by Gmsh \cite{geuzaine2009gmsh}. The geometric stretch near the edge $\tilde{R}_0 = 1$ is larger than that near the center $\tilde{R}_0 = 0$ due to the non-uniform growth. Therefore, we used a finer discretization of the domain near $\tilde{R}_0 = 1$ to ensure high precision numerical solutions.

The initial conditions are $\tilde{r} (\tilde{R}_0, \tau = 0)=\tilde{R}_0$, $\tilde{c}_\parallel (\tilde{R}_0, \tau = 0)\equiv 1$, $\lamg (\tilde{R}_0, \tau = 0)\equiv 1$, $\gammao(\tilde{R}_0, \tau = 0) \equiv 1$, $\tilde{A}(\tilde{R}_0, \tau = 0) =\tilde{S}(\tilde{R}_0, \tau = 0) \equiv 0$ , and boundary conditions $\tilde{c}_\parallel (\tilde{R}_0=1, \tau) = 1$, $\tilde{\sigma}_{rr} (\tilde{R}_0=1, \tau) = 0$. Partial differential equations were then converted to their equivalent weak forms and computationally discretized by first-order (2 noded) linear elements \cite{langtangen2019introduction}, and implemented in the open-source computing platform FEniCS~\cite{alnaes2015fenics}. The time increment was set to be $\Delta \tau = 0.01$. At each time step, we used the standard Crank-Nicolson method to perform the numerical integration \cite{crank1947practical}. To ensure numerical convergence, we checked explicitly whether the wrinkling instability occurred  by evaluating the difference between the circumferential stress and the critical stress $\delta_\sigma = |\tilde{\sigma}_{\theta\theta}^0| - \tilde{\sigma}_\mathrm{c}$. We required that $\delta_\sigma^{(n)} > 0$ at $\tau = n\Delta \tau$ for the wrinkling to occur at $\tau = (n+1)\Delta \tau$.

\subsubsection*{Choice of parameters}
{\it Critical stress for wrinkling:} Our previous study revealed that a trilayer model quantitatively captures the biofilm wrinkle wavelength \cite{yan2019mechanical}. The trilayer theory also predicts how the critical stress varies with the stiffness contrast between the biofilm and the substrate $\Gs/\Gb \propto \xi$ \cite{lejeune2016understanding} (see also \cite{yan2019mechanical} for the calculated values of critical stress and $\Gs/\Gb$ for different agar concentrations). However, our chemo-mechanical model (Eq. (\ref{eq_ndim_all})) reaches the theoretical critical stress earlier than the time when wrinkling occurs in the experiments because we model biofilms as elastic materials and do not consider viscoelasticity and plasticity (see Discussion and Supplementary Note VII). In practice, we rescale the critical stress $\tilde{\sigma}_c$ in Fig.~\ref{Fig4}C and D such that wrinkling instability in the simulations occurs at a time similar to that in experiment.

{\it Fitting parameters from the velocity profiles:} The dimensionless friction parameter $\xi$ and the dimensionless maximum rate of nutrient uptake $\tilde{Q}_0$ were determined by fitting the radial velocity profiles of the modeled biofilm to those extracted from experiments at different times. The similarity between the radial velocity profiles was assessed in terms of the normalized mean squared distance (MSD). In experiments, we measured the radial velocity profiles for a biofilm grown on 0.7\% agar concentration at 40 different time points $t_j$ separated by 30 min from $t=0$~h to $t = 20$~h (before the wrinkling instability occurs) as described above in the \emph{Image Processing} section. At each time $t_j$, the experimental data were represented as $(\hat{r}_{i,j}, \hat{v}_{i,j})$ ($i = 1,\dots, N_j$; $\hat{v}_{i,j}$ averaged over the circumferential direction). The number of data points $N_j$ at each time point is equal to the ratio of the biofilm radius to the width of radial bins. For a particular set of parameters $(\xi, \tilde{Q}_0)$, we first numerically computed the velocity profiles $v_r(r,t_j)$ of the modeled biofilm. For each time point $t_j$ we computed the normalized squared distance (SD) $\Delta\tilde{s}^2_{i,j}$ between the experimental data points $(\hat{r}_{i,j}, \hat{v}_{i,j})$ and the simulated profile $v_r(r,t_j)$  as $\Delta\tilde{s}^2_{i,j} = \min\limits_{r}\big\{(\frac{\hat{r}_{i,j} - r}{L_0})^2 + (\frac{\hat{v}_{i,j} - v_r(r,t_j)}{V_0})^2\big\}$ where we used a characteristic length scale $L_0 = 5$ mm and a characteristic velocity $V_0 = 3~\mu$m/min. The normalized SD between the radius and edge velocity for the experimental biofilm and those of the modeled biofilm was used as one additional data point $\Delta\tilde{s}^2_{N_j + 1,j}$ associated with time $t_j$.
 Finally, the normalized MSD was calculated as $\frac{1}{(N_{j+1})N_t}\sum_{j=1}^{N_t}\sum_{i = 1}^{N_j+1}\Delta\tilde{s}^2_{i,j}$. We searched the parameter space to find the optimal parameter values $\xi^*$ and $\tilde{Q_0}^*$ that minimize the normalized MSD (Fig.~S1). For simulations with different friction, we varied the parameter $\xi$ keeping all the other parameters fixed.

\subsubsection*{Analysis of the biofilm leading angle} To compare the biofilm leading angles in experiments with theoretical predictions, we inferred the value of $\eta/\Gb$ for biofilms grown on 0.7\%  agar by fitting the velocity profiles as described above, i.e., $(\eta/G_b)^* = \frac{\xi^* (H/R_\mathrm{b0})}{R_\mathrm{b0} / \tau_0}$. Next, we inferred the normalized friction $(\eta/\Gb)v_\mathrm{b}$ for biofilms grown on different agar concentrations (agar shear modulus denoted by $\Gs$) by making the assumption that $(\eta/\Gb)\propto \Gs/\Gb$, i.e., $(\eta/\Gb)v_\mathrm{b} = (\eta/\Gb)^*\times\frac{(\Gs/\Gb)}{(\Gs/\Gb)|_\text{agar conc. = 0.7\%}} \times v_\mathrm{b}$. The uncertainty of these values (Fig.~\ref{Fig2}E horizontal error bars) was estimated by taking into account the measurement errors of $\Gs$, $\Gb$, and $v_\mathrm{b}$. The value of the circumferential compression $\Fe_{,\theta\theta}$ in Eq.~(\ref{eq_angle}) remains undetermined. Nevertheless, we can estimate $\Fe_{,\theta\theta} \in (0.7,0.9)$ from the wrinkling instability analysis~\cite{yan2019mechanical}. The specific choice of $\Fe_{,\theta\theta}$ in this range only minimally affects the results (Fig.~S9).

\subsection*{Data and software availability}

\subsubsection*{Code availability} Matlab codes for the image processing have been described in a previous publication~\cite{yan2019mechanical}. The simulation codes used to model the biofilm are available on GitHub (\hyperlink{https://github.com/f-chenyi/biofilm-mechanics-theory}{https://github.com/f-chenyi/biofilm-mechanics-theory}). 

\subsubsection*{Data availability} The data are available upon request.


\section*{Acknowledgements}
This work was supported by the Howard Hughes Medical Institute (B.L.B., etc.), National Science Foundation Grants MCB-1713731 (B.L.B.), MCB-1853602 (B.L.B., H.A.S., and N.S.W.), NIH Grant 1R21AI144223 (B.L.B., H.A.S., and N.S.W.), NIH Grant
2R37GM065859 (B.L.B.), NIH Grant GM082938 (N.S.W.), the NSF through the Princeton University Materials Research Science and
Engineering Center DMR-1420541 (B.L.B., H.A.S., A.K.), and the Max Planck Society-Alexander von Humboldt Foundation (B.L.B.). J.Y. holds a Career Award at the Scientific Interface from the Burroughs Wellcome Fund. R.A. acknowledges support from the Human Frontiers of Science Program (LT-000475/2018-C). We thank Dr. Maria Holland for helpful discussions.

\bibliographystyle{ieeetr}
\bibliography{Biofilm_Ref2.bib}

\end{document}


\title{Supplementary Information}
\maketitle

\section{\label{sec_general_formulation}General Formulation of Growing Elastic tissues}
A general formulation that takes into account the combined effect of mechanics and growth, was originally proposed by Rodriguez \textit{et al.} \cite{rodriguez1994stress}, and has since then been widely used for modeling  different biological systems, such as in \cite{goriely2005differential,dervaux2008morphogenesis,de2016constitutive,genet2015heterogeneous}. Briefly, the model considers three configurations (see Fig. 2 in main text): a stress-free initial configuration, a stress-free virtual configuration after growth, and a deformed current configuration. In a fixed orthonormal Cartesian basis $\lbrace \boldsymbol{e}_1, \boldsymbol{e}_2, \boldsymbol{e}_3\rbrace$, the coordinates of material points in the three configurations are denoted by $\boldsymbol{X}_0$,  $\boldsymbol{X}$, and $\boldsymbol{x}$, respectively. A complete kinematic description of deformed shapes for growing tissues is given by the mapping $\boldsymbol{x}=\boldsymbol{x}(\boldsymbol{X}_0,t)$ from the initial configuration ($\boldsymbol{X}_0$) to the current deformed configuration ($\boldsymbol{x}$). Equivalently, one can define the total deformation gradient $\Ft = \partial \boldsymbol{x} /\partial \boldsymbol{X}_0 $ to describe local changes in shape and density. The fundamental idea of Rodriguez \textit{et al.}~\cite{rodriguez1994stress} is that the total deformation $\mathbf{F}$ can be decomposed into a deformation $\Fg$ due to growth to the virtual stress-free reference state, and an elastic deformation $\Fe$ that ensures compatibility of the growing tissue. Formally, the multiplicative decomposition reads
\begin{gather}
    \Ft = \Fe \cdot\Fg = \frac{\partial\bs{x}}{\partial\bs{X}}\cdot \frac{\partial\bs{X}}{\partial\bs{X}_0},
\end{gather} where $\boldsymbol{A} \cdot \boldsymbol{B}$  denotes the matrix multiplication $(\boldsymbol{A} \cdot \boldsymbol{B})_{ij} = A_{ik}B_{kj}$ and summation over repeated indices is implied. Besides the assumption above, the model of growing elastic tissues generally consists of the following three ingredients:
\begin{itemize}
\item{\textbf{A governing equation for growth}: The rate of deformation $d\Fg/dt$ due to growth can in general be a function of many physical fields. For example, the growth rate $d\Fg/dt$ may depend on external fields, such as a nutrient concentration field. It may also depend on internal stresses, if the organisms can sense and respond to their mechanical environment (see for example \cite{uyttewaal2012mechanical,matsumoto1999cyclic}).
} 

\item{\textbf{A stress-strain constitutive relation}: A stress-strain constitutive relation relates tissue deformation to stresses. Here, we assume that the internal Cauchy stress $\bs{\sigma}$ arises solely from the elastic deformation, i.e. $\bs{\sigma}=\bs{\sigma}(\Fe)$. 
For a hyperelastic material, this relation can be determined from the elastic free energy density $\Psi = \Psi(\Fe)$. 
Specifically, for isotropic materials the energy density can be expressed in terms of the two invariants $\Psi = \mathcal{E}(\Ic,\Je)$, where $\Ic = \text{tr}(\Fe^T\Fe)$ is the trace (first invariant) of the right Cauchy-Green deformation tensor, and $\Je = \det(\Fe)$ denotes the volumetric change due to elastic deformation \cite{ogden1997non}. Different stress measures can be defined based on whether they report the force in an undeformed or a deformed configuration. For example, the first Piola-Kirchhoff (PK) stress $\mathbf{P}$ measures the force per area in the undeformed configuration, which is evaluated as
\begin{gather}
\mathbf{P}  = \frac{\partial \Psi}{\partial \Fe} = 2\frac{\partial \mathcal{E}}{\partial \Ic} \Fe + \frac{\partial \mathcal{E}}{\partial \Je} \big(\mathbf{adj}(\Fe)\big)^T,
\label{Eq:PKStress}
\end{gather}
where $\mathbf{adj}$ denotes the adjugate of a matrix~\cite{magnus1999matrix}. The PK stress $P_{iJ}$ is asymmetric, because it has one index $i$ attached to the current configuration, and one index $J$ attached to the undeformed virtual configuration.
In contrast, the Cauchy stress $\sigma_{ij}$ measures the force per area in the current \textit{deformed} configuration, and it is symmetric due to conservation of angular momentum~\cite{landau1989theory}. The Cauchy and PK stresses are related by 
\begin{equation}
\boldsymbol{\sigma} = \Je^{-1}\mathbf{P}\Fe^T = 2\Je^{-1}\left(\frac{\partial \mathcal{E}}{{\partial \Ic}}\right) \, \mathbf{B} + \left(\frac{\partial \mathcal{E}}{\partial \Je}\right)\, \mathbf{I},
\label{Eq:CauchyStress}
\end{equation}
where $\mathbf{B} = \Fe \Fe^T$ denotes the left Cauchy-Green deformation tensor \cite{ogden1997non}.}

\item{\textbf{A force-balance equation}: The balance of acceleration and forces is given by 
\begin{equation}
\rho_m\dot {\bs {v} }-\bs {\nabla }\cdot \bs{\sigma }-\rho_m\bs{b} =0,
\label{Eq:ForceBalance}
\end{equation}
where $\rho_m (\bs {x} ,t)$ denotes the mass density, $\dot{\bs{v}}(\bs {x} ,t)$ the acceleration ($\dot{\boldsymbol{v}}$ denotes total time derivative of velocity $\boldsymbol{v}$), $\bs {\nabla}$ the derivative with respect to the coordinates $\bs {x}$, $\bs {\sigma}(\bs {x} ,t)$ is the Cauchy stress, and $\rho_m\bs {b} $ the body force density (per volume). The acceleration term can often be neglected in small systems and it is negligible for the biofilms considered here (see Sec.~\ref{sec_plane_stress}}).
\end{itemize}

In the next section, we describe a chemo-mechanical model of {\it V. cholerae} biofilm development and present the corresponding three key relations discussed above. Unless otherwise specified, we use the notation where the lower case letters, the upper case letters, and the upper case letters with the subscript 0 denote quantities in the current, the virtual, and the initial configurations, respectively. In addition, we use the subscript $\parallel$ for reduced 2D quantities. Greek letter indices are used to denote the components of in-plane 2D vectors/tensors and Latin letters are used to denote the components of 3D vectors/tensors.
Summation over repeated indices is to be assumed unless otherwise specified. Table~\ref{tb_symbol} lists all symbols and describes their meaning.

\begin{table}[b!]
\captionsetup{labelfont=bf,singlelinecheck=off}
    \caption{Summary of the used symbols and their descriptions.
} \label{tb_symbol}
    \centering
    \renewcommand{\arraystretch}{1.25}
    \begin{tabularx}{16.5cm}{YZ|YZ}
         \hline\hline
          \multicolumn{4}{l}{\textbf{General notations (Secs. \ref{sec_general_formulation} -- \ref{sec_plasticity})}}\\
         \hline\hline
         \multicolumn{1}{Y|}{\textbf{Symbols}} & \multicolumn{1}{C|}{\textbf{Descriptions}} & \multicolumn{1}{Y|}{\textbf{Symbols}} & \multicolumn{1}{C}{\textbf{Descriptions}} \\ \hline 
         $\parallel$ & {\small Subscript: reduced 2D variables or the in-plane components of a 3D vector/tensor} & $\sim$ & {\small Normalized variales or dimensionless parameters}\\
         \hline
         $\mathbf{Grad}_0,\mathbf{Grad}$, $\bs{\nabla}_\parallel, \nablap$ & {\small Spatial gradient taken in the initial, the virtual, the current, and the deformed flat configuration} & $\mathbf{adj}, \text{sym}$ & {\small The adjugate or the symmetric part of a matrix}\\
         \hline\hline

         \multicolumn{4}{l}{\textbf{General formulation of growing elastic biofilms (Secs. \ref{sec_general_formulation},~\ref{sec_plane_stress} and \ref{sec_summary})}}\\
         \hline\hline
         \multicolumn{1}{Y|}{\textbf{Symbols}} & \multicolumn{1}{C|}{\textbf{Descriptions}} & \multicolumn{1}{Y|}{\textbf{Symbols}} & \multicolumn{1}{C}{\textbf{Descriptions}} \\
         \hline 
         $\bs{X}_0,\bs{X},\bs{x}$ & {\small Coordinates of the initial, virtual, and current configuration} & $\Ft,\Fe,\Fg$ & {\small Total, elastic, and growth deformation gradients}\\
         \hline
         $\mathbf{S}_0,\mathbf{P},\bs{\sigma}$ & {\small Stress measured in the initial, virtual, and current configuration} & $\mathbf{C}, I_C$ & {\small The right Cauchy-Green deformation tensor and its first invariant}\\
         \hline
         $\mathbf{B}$ & {\small The left Cauchy-Green deformation tensor} & $J, \Je$ & {\small The determinant of $\Ft$ and $\Fe$}\\
         \hline
         $\lamg$ & {\small Geometric stretch due to growth} & $\kg,k_\mathrm{r}$ & {\small Total and residual growth rate}\\
         \hline
          $\gamma$ & {\small Vertical stretch of biofilm} & $\Gb,\Gs$ & {\small Biofilm and agar shear modulus}\\
          \hline
          $\eta, \bs{f}, \xi$ & {\small Drag coefficient, dimensional and dimensionless friction} & $r_\mathrm{b}$ & {\small Biofilm radius}\\

         \hline\hline
         \multicolumn{4}{l}{\textbf{Nutrient diffusion-uptake dynamics (Secs. \ref{sec_nutrient} and \ref{sec_discussion_nd})}}\\
         \hline\hline
         \multicolumn{1}{Y|}{\textbf{Symbols}} & \multicolumn{1}{C|}{\textbf{Descriptions}} & \multicolumn{1}{Y|}{\textbf{Symbols}} & \multicolumn{1}{C}{\textbf{Descriptions}} \\ \hline 
         $c, c_\parallel$ & {\small Nutrient concentration in a 2D or a 3D model} & $D$ & {\small Nutrient diffusion constant}\\
         \hline
         $Q_0, J_0$ & {\small Maximum nutrient uptake rate in a 2D or a 3D model} & $a_c$ & {\small Width of the nutrient-rich annulus}\\
         \hline 
         $K$ & {\small Half-rate concentration of nutrient uptake} & $H_\mathrm{s},R_{s}$ & {\small Radius and thicknesss of the agar substrate}\\
         \hline\hline
         \multicolumn{4}{l}{\textbf{Model discussion: Lagrangian specification (Sec. \ref{sec_discussion_formulation})}}\\
         \hline\hline
         \multicolumn{1}{Y|}{\textbf{Symbols}} & \multicolumn{1}{C|}{\textbf{Descriptions}} & \multicolumn{1}{Y|}{\textbf{Symbols}} & \multicolumn{1}{C}{\textbf{Descriptions}} \\ \hline 
         $\rho, \rho_0$ & {\small Material density in the deformed and the undeformed configuration} & $\bs{\chi}$ & {\small The reference vector denoting the initial coordinate of a material point}\\
         \hline\hline
         & & & {\bf Continued on next page}\\
         \hline\hline
    \end{tabularx}
\end{table}

\begin{table}[]
    \centering
    \renewcommand{\arraystretch}{1.25}
    \begin{tabularx}{16.5cm}{YZ|YZ}
         \hline\hline
         \multicolumn{4}{l}{\textbf{Model discussion: Constitutive model (Sec. \ref{sec_discussion_constitutive})}}\\
         \hline\hline
         \multicolumn{1}{Y|}{\textbf{Symbols}} & \multicolumn{1}{C|}{\textbf{Descriptions}} & \multicolumn{1}{Y|}{\textbf{Symbols}} & \multicolumn{1}{C}{\textbf{Descriptions}} \\ \hline 
         $\Pi$ & {\small Osmotic pressure} & $\phi_0$ & {\small Equilibrium dextran concentration}\\
         \hline 
         $v_\mathrm{w}, \mu_\mathrm{w}$ & {\small Volume and chemical potential of water molecule} & $\bs{X}_\mathrm{d}, \Fd$ & {\small Coordinate and deformation gradient associated with the dry state}\\
         \hline 
         $\mathbf{F}_0, \lambda_0$ & {\small Equilibrium swelling} & $\Jd$ & {\small Volumetric change of biofilm matrix compared to its dry state}\\
         \hline 
         $\chi$ & {\small  Flory interaction parameter} & $\Psi^J$ & {\small Free energy increase associated with volumetric change}\\
         \hline\hline
         \multicolumn{4}{l}{\textbf{Model discussion: Friction (Sec. \ref{sec_discussion_friction})}}\\
         \hline\hline
         \multicolumn{1}{Y|}{\textbf{Symbols}} & \multicolumn{1}{C|}{\textbf{Descriptions}} & \multicolumn{1}{Y|}{\textbf{Symbols}} & \multicolumn{1}{C}{\textbf{Descriptions}} \\ \hline 
         $K_\mathrm{p}, K_\mathrm{s}, K^\prime$ & {\small Spring constant of the polymer spring, the substrate, and them connected in series} & $k_\mathrm{on},k_\mathrm{off}$ & {\small The binding/unbinding rate between biofilm polymers and biofilm proteins}\\
         \hline 
         $\tau_\mathrm{on},\tau_\mathrm{off}$ & {\small Time scales associated with the binding/unbinding processes} & $\rho_\mathrm{p}$ & {\small Average density of bound polymers}\\
         \hline 
         $E_\mathrm{adh}, \epsilon$ & {\small Dimensional and dimensionless adhesion energy parameter} & $\xi_T^\prime$ & {\small Typical stretch of the polymer springs activated by thermal energy}\\
         \hline\hline
         
            \multicolumn{4}{l}{\textbf{Coarse-grained wrinkling model (Sec. \ref{sec_wrinkle})}}\\
         \hline\hline
         \multicolumn{1}{Y|}{\textbf{Symbols}} & \multicolumn{1}{C|}{\textbf{Descriptions}} & \multicolumn{1}{Y|}{\textbf{Symbols}} & \multicolumn{1}{C}{\textbf{Descriptions}} \\ \hline 
         $\Fe^0, \Few$ & {\small Horizontal compression and wrinkling deformation} & $\bs{u}, w$ & {\small In-plane and out-of-plane displacement}\\
         \hline 
         $\bs{x}^\prime$ & {\small Coordinate in the deformed flat configuration} & Superscript 0 & {\small Quantities without wrinkling}\\
         \hline 
         $\Pi, \Psi$ & {\small Elastic energy} & Superscript c & {\small Quantities in the coarse-grained description}\\
         \hline 
         $A, \beta_1\beta_2$ & {\small Amplitude and shape of wrinkles} & $k_\mathrm{w}, \lambda_\mathrm{w}$ & {\small Wrinkle wavenumber and wavelength}\\
         \hline 
         $a_{11}^0, a_{22}^0$ & {\small Principal stretches of $\Fe^0$} & $\sigma_c$ & {\small Critical stress of the wrinkling instability}\\
         \hline 
         $\tilde{A}, \tilde{S}$ & {\small Dimensionless order parameter of wrinkles} & $q$ & {\small Area factor accounting for the area increase in the wrinkled state}\\
         \hline\hline
         
           \multicolumn{4}{l}{\textbf{Model discussion: Elasto-plasticity (Sec. \ref{sec_plasticity})}}\\
         \hline\hline
         \multicolumn{1}{Y|}{\textbf{Symbols}} & \multicolumn{1}{C|}{\textbf{Descriptions}} & \multicolumn{1}{Y|}{\textbf{Symbols}} & \multicolumn{1}{C}{\textbf{Descriptions}} \\ \hline 
         $\mathbf{E}, \mathbf{s}$ & {\small Deviatoric deformation and stress} & $\sigma_\mathrm{Y}$ & {\small Yielding stress}\\
         \hline 
         $\epsilon_\mathrm{E}, \sige$ & {\small Scalar measures of the deviatoric deformation and stress} & $\tilde{G}^\prime$ & {\small Normalized plastic modulus}\\
         \hline\hline
    \end{tabularx}
\end{table}
\clearpage

\section{A Chemo-Mechanical Model for Bacterial Biofilms}
\label{sec:ChemoMechanicalModel}
To investigate how mechanical stresses build up in {\it V. cholerae} biofilms when growing on agar, we model the biofilm as a thin elastic film that consumes nutrients, proliferates, and expands. The effective surface friction between the expanding biofilm and the agar generates compressive stresses inside the biofilm. Although viscoelastic relaxation and plastic deformations are quite likely to occur during biofilm development \cite{yan2018bacterial}, we neglect all sources of internal dissipation in our model (see Sec.~\ref{sec_plasticity} for further discussion). This section is organized as follows: in Sec.~\ref{sec_nutrient} we begin by modeling the nutrient-dependent biofilm growth; in Sec.~\ref{sec_plane_stress} we propose a material constitutive model for the  biofilm and derive the force-balance equation;   in Secs.~\ref{sec_summary} we summarize the governing equations and relevant model parameters, which are converted to dimensionless form; in Secs. \ref{sec_discussion_formulation} -- \ref{sec_discussion_friction} we discuss different aspects of the model, including the general formulation, the nutrient diffusion-uptake dynamics, the constitutive model, and the friction between biofilm and agar.

\subsection{\label{sec_nutrient}Nutrient-limited biofilm growth}
In experiments we previously found that there is an annulus of width $\sim$1 mm at the (outer) edge of the biofilm, in which the fraction of dead cells is significantly lower compared to the interior part of the biofilm~\cite{yan2019mechanical}. This indicates that cell growth occurs primarily at the edge of the biofilm, presumably due to nutrient limitation elsewhere. To account for the non-uniform growth pattern observed in experiments, we model nutrient-dependent growth as follows. 

{\bf Nutrient uptake-diffusion dynamics: } The diffusion of nutrients in the agar substrate, denoted by concentration $c(\bs{x}_\parallel,z,t)$, 
is described by Fick's law $\partial c/\partial t = D\bs{\nabla}^2 c$, where $D$ is the nutrient diffusion constant and $\bs{ x}_\parallel\equiv(x,y)$. As shown in Fig.~2 in the main text, biofilms are located at the top surface $z = 0$, and consume nutrients from the agar. This sets the boundary condition $-D\partial_z c = \Jep^{-1}J_0 \phi(c)$ at $z = 0$ and $ |\bs{x}_\parallel|<r_\text{b}$, where $J_0$ is the maximum nutrient uptake rate per unit area of cells in the undeformed configuration, $\Jep^{-1}$ accounts for the increase of biofilm thickness due to compression, and $r_\text{b}$ denotes the radius of the biofilm in the current configuration. The rate of nutrient uptake is assumed to depend on local nutrient availability via the Monod law $\phi(c) = c/(K + c)$ where $K$ is the concentration of nutrients at the half-maximal uptake rate \cite{monod1949growth}. Zero flux boundary conditions are imposed elsewhere on the boundary of the agar substrate, i.e. ${\bs n} \cdot {\bs \nabla} c=0$, where ${\bs n}$ is the normal vector to the agar boundary.

Simulating the full diffusion dynamics of nutrients requires tracking the concentration field in the three-dimensional (3D) space of the agar medium in the Eulerian formulation, while the mechanical stresses in the biofilm are more readily computed in a two-dimensional (2D) plane in the Lagrangian formulation (see Sec. \ref{sec_plane_stress} and \ref{sec_discussion_nd}). However, to  capture the nutrient-limited non-uniform growth of the biofilm, it is sufficient to consider only the lateral diffusion of nutrients. Therefore, we introduced a reduced 2D model to simplify the full 3D dynamics of nutrients and to make the computations more tractable. The reduced 2D model is given by
\begin{gather}
    \frac{\partial c_\parallel}{\partial t} = D\bs{\nabla}_\parallel^2c_\parallel - \Jep^{-1}Q_0\phi(c_\parallel), \label{eq_nc_2d}
\end{gather}
where $c_\parallel(\bs{x}_\parallel,t)$ denotes the reduced 2D nutrient concentration and $Q_0$ denotes the maximum uptake rate. The factor $\Jep^{-1}$ is introduced to account for the change of areal density of the biofilm upon deformation. The concentration of nutrients at the edge of the biofilm ($|\bs{x}_\parallel| = r_b(t)$) is set to a fixed value $c_{0,\parallel}$.
As the biofilm grows and expands, nutrients get depleted underneath the biofilm (see Fig. 2C in the main text). The concentration profile of nutrients has a maximum at the edge of the biofilm and decreases toward the interior of the biofilm with a penetration depth $a_c \sim \sqrt{Dc_{0,\parallel}/Q_0}$.
To verify that the reduced 2D model provides a reasonably accurate approximation of the full 3D dynamics of nutrients, we compared the concentration profiles of nutrients for both models in Sec.~\ref{sec_discussion_nd}.

{\bf Nutrient-dependent growth: }  During development \vc biofilms expand primarily in the horizontal plane \cite{stewart2008physiological, yan2019mechanical}, while the biofilm thickness is approximately constant, being set by the oxygen penetration depth~\cite{costerton1995microbial}.  Thus, we consider only horizontal (in-plane) growth in our model. Growth is assumed to be {\it isotropic} because our experiments reveal that the orientation of bacteria inside the biofilm is disordered and so the growth, division, and matrix production by the bacteria should result in isotropic expansion. The deformation due to growth is thus $\Fg =\begin{pmatrix} \lamg \mathbf{I}_\parallel & \bs{0} \\ \bs{0} & 1 \\ \end{pmatrix} $, where $\lamg(\bs{x}_\parallel,t)$ denotes the local stretch due to growth, and $\mathbf{I}_\parallel$ is the 2D identity matrix. Nutrient dependent growth is described by $\partial\lamg/\partial t=\kg (c_\parallel) \lamg$, where the growth rate is ${k}_\mathrm{g}({c}_\parallel) = (k_g^\textrm{max}-{k}_\mathrm{r})\phi({c}_\parallel) + k_\mathrm{r}$, where $k_g^\textrm{max}$ is the maximum growth rate and we added a small nutrient-independent growth rate $k_r$ to account for the fact that in 3D simulations the concentration of nutrients does not drop to zero beneath the center of the biofilm (see Sec. \ref{sec_discussion_nd} for details). This nutrient-dependent growth combined with the diffusion-uptake dynamics of nutrients, captures our experimental finding that the active growth zone is restricted to a finite annulus near the biofilm edge (see Fig. 2 in the main text and Fig.~\ref{Fig:nutrientDynamics2Dvs3D}).

\subsection{\label{sec_plane_stress}Thin plate mechanics of biofilms}
{\bf Plane stress assumption: } The \vc biofilms in our experiments can be considered as thin-film structures ($\sim 100\text{ }\mu$m in height and $5-15$ mm in diameter). In continuum mechanics, a plane stress assumption is often used for the analysis of thin films, such that the stress components perpendicular to the biofilm surface are negligible throughout the whole biofilm. Thus, we use the plane stress model to describe the mechanics of our biofilms. In this section, we consider only a flat biofilm configuration, before wrinkling occurs. The out-of-plane deformation induced by the wrinkling instability is discussed in Sec.~\ref{sec_wrinkle}.

Consider the full 3D configuration $\bs{x}(\bs{X_0},t)$ of a thin, flat, elastic film. The elastic deformation tensor $\Fe$ describes deformation from the virtual stress-free grown configuration $\bs{X}$ to the final deformed configuration $\bs{x}$ (see Fig.~2 in the main text), which can be written explicitly as 
\begin{gather}
\Fe = \begin{pmatrix}
\Fe_{,\parallel} & \partial_Z \bs{x}_\parallel \vspace{0.5em}\\
(\mathbf{Grad}_\parallel\,z)^T & \partial_Z z
\end{pmatrix},
\end{gather} 
where $\Fe_{,\parallel} = \mathbf{Grad}_\parallel\,\bs{x}_\parallel$ is the in-plane deformation tensor, and $\mathbf{Grad}_\parallel$ denotes gradient with respect to the spatial coordinates $\bs{X}_\parallel$ in the intermediate virtual stress-free configuration. For typical biofilms in our experiments, we estimate that $\mathbf{Grad}_\parallel\,z \sim \Delta h_\mathrm{b} / r_\mathrm{b} \lesssim 100\mu\text{m} / 5\text{mm} = 0.02$, where $\Delta h_\mathrm{b}$ denotes the height difference between the biofilm center and the biofilm edge. Thus, we assume that $\mathbf{Grad}_\parallel\,z$ is negligible compared to the in-plane deformation $\Fe_{,\parallel}$ and to the elastic change of thickness $\partial_Z z$. The plane stress assumption also implies that $\partial_Z \bs{x}_\parallel$ must be small (see for example, F\"{o}ppl-von Karman plate theory or Kirchhoff-Love plate theory \cite{landau1989theory,reddy2006theory,audoly2010elasticity}). 
For simplicity, we set $\mathbf{Grad}_\parallel\,z= \partial_Z \bs{x}_\parallel = 0$ and define the vertical stretch $\gamma = \partial_Z z \equiv \gamma(\bs{X}_\parallel)$, which describes the relative change of thickness. This yields the plane stress elastic deformation tensor 
\begin{gather}
    \Fe \approx 
    \begin{pmatrix}
     \Fe_{,\parallel} & \mathbf{0}\\[0.5em]
     \mathbf{0}^T & \gamma \\
    \end{pmatrix}
\end{gather}
for a thin, flat biofilm. 
Finally, the relation between $\gamma$ and $\Fe_{,\parallel}$ can be derived from the plane stress condition $\sigma_{zz} = 0$ (see Eq.~(\ref{Eq:CauchyStress}) in Sec.~\ref{sec_general_formulation} for the expression for $\bs{\sigma}$), which yields:
\begin{gather}
    2\Je^{-1}\left(\frac{\partial \mathcal{E}}{\partial \Ic}\right)\gamma^2+\left(\frac{\partial \mathcal{E}}{\partial \Je} \right) = 0, \label{eq_gamma}
\end{gather}
where $\Ic = \text{tr}(\Fe^T\Fe)=  \gamma^2 + I_{C,\parallel}$  is the trace (first invariant) of the right Cauchy-Green deformation tensor and $\Je = \det(\Fe) = \gamma \Jep$ denotes the volumetric change due to elastic deformation. Here we have introduced the invariants $I_{C,\parallel}= \text{tr}(\Fe_{,\parallel}^T\Fe_{,\parallel})$ and $\Jep= \det(\Fe_{,\parallel})$ that are related to the in-plane part of the elastic deformation $\Fe_{,\parallel}$.

{\bf Constitutive relation: } Next, we specify the constitutive model that relates the stress to the deformation of a biofilm. 
Here, we use a common neo-Hookean material model \cite{budday2014mechanical,shyer2013villification,tallinen2016growth} to approximate biofilm elasticity; this model will be motivated and discussed in more detail in Sec.~\ref{sec_discussion_constitutive}. The strain energy density for the neo-Hookean model is given by
\begin{eqnarray}
\mathcal{E}(\Ic,\Je) &=& \frac{\Gb}{2}\Big[\Ic-3-2\ln \Je\Big] + \frac{\lambdab}{2} \big(\ln 
\Je\big)^2, \label{Eq:Energy_neoHookean}
\end{eqnarray}
where $\Gb$ and $\lambdab$ are the Lam\'{e} elastic constants. Note that $\Psi$ in Eq.~(\ref{Eq:Energy_neoHookean}) above is the energy density per unit volume in the virtual stress-free configuration $\bs{X}$. For an incompressible neo-Hookean material, the energy density reads $\mathcal{E}(\Ic,\Je) = (\Gb/2)(\Ic-3) - p (\Je - 1)$, where a pressure $p$ is introduced as the Lagrange multiplier to ensure $\Je = 1$.

For a given in-plane deformation $\Fep$, the vertical stretch $\gamma$ is obtained by inserting Eq.~(\ref{Eq:Energy_neoHookean}) into Eq.~(\ref{eq_gamma}) and solving for $\gamma$. The in-plane components of the Cauchy stress $\bs{\sigma}_\parallel$ are obtained from Eq.~(\ref{Eq:CauchyStress}). Results for the compressible and incompressible case are:
\begin{itemize}
    \item \emph{Compressible case}: $\gamma$ is obtained by solving the nonlinear equation $\Gb(\gamma^2 -1) + \lambdab\ln(\gamma \Jep) = 0$. The in-plane components of the Cauchy stress are
    \begin{equation}
    \bs{\sigma}_\parallel = (\gamma \Jep)^{-1}\big[\Gb\Fe_{,\parallel}\Fe_{,\parallel}^T + \lambdab\ln (\gamma \Jep)\mathbf{I}_\parallel-\Gb\mathbf{I}_\parallel\big].
    \end{equation}
\item \emph{Incompressible case}: Constraint $\Je\equiv 1 = \gamma \Jep$ specifies the relation $\gamma=1/\Jep = 1/\det(\Fe_{,\parallel})$. The pressure $p$ can be calculated as $p = \Gb \gamma^2$, and therefore the in-plane components of the Cauchy stress are
\begin{equation}
\bs{\sigma}_\parallel = \Gb(\Fep\Fep^T- \gamma^2 \mathbf{I}_\parallel).
\end{equation}
\end{itemize}

{\bf Force-balance equation: } A growing biofilm moves relative to the substrate and experiences an effective friction. We model this friction as a viscous drag \cite{walcott2010mechanical,sens2013rigidity}, i.e., we assume that the friction $\bs{f}$ (force per unit area in the current deformed configuration) takes the form $\bs{f} = - \eta \bs{v}_\parallel$, where $\bs{v}_\parallel$ is the expansion velocity of the biofilm, and $\eta$ the drag coefficient. The justification for the assumption of viscous friction is discussed in Sec.~\ref{sec_discussion_friction} where we present a microscopic model for the origin of this friction. 

The general force-balance relation is presented in Eq.~(\ref{Eq:ForceBalance}) in Sec.~\ref{sec_general_formulation}. Here we first discuss the relative importance of inertial effects. For biofilms the mass density is estimated to be $\rho_\mathrm{m} \sim 10^{3}~\text{kg/m}^3$ since they contain mostly water. The typical velocity and time scales associated with biofilm growth and expansion are $U_0\sim 3~\mu$m/min and $\tau_0\sim~1$~h. The shear modulus of the biofilm is $\Gb\sim 10^3$~Pa, and the typical radius of biofilm is $r_\mathrm{b}\sim 10$~mm. The magnitude of inertial forces $\rho_\mathrm{m}\dot{\bs{v}}$ can be estimated as $\rho_\mathrm{m} U_0/\tau_0$, while the magnitude of forces related to internal stress $\bs{\nabla}\cdot \bs{\sigma}$ is estimated as $\Gb/r_\mathrm{b}$. Comparing these estimates we find $\frac{\rho_\mathrm{m} U_0/\tau_0}{\Gb / r_\mathrm{b}}\sim 10^{-15}$ and thus the inertial forces are negligible compared to the forces resulting from internal stresses. A similar argument can be made concerning the gravitational body force, which can be neglected as well. Thus the 3D force-balance equation for the biofilm simplifies to $\bs{\nabla}\cdot\bs{\sigma} = \bs{0}$ with the traction-free boundary condition ($\bs{\sigma}\cdot\bs{e}_z = 0$) at the top of the biofilm and with the frictional traction ($\bs{\sigma}\cdot(-\bs{e}_z)=\bs{f}$) at the biofilm-agar interface. Note that $\bs{\nabla}$ are derivatives with respect to coordinates in the deformed configuration. By integrating the force-balance equation in the $z$ direction over the deformed biofilm thickness $h=\gamma H$, where $H$ is thickness of the undeformed biofilm, we obtain the horizontal force-balance relation
\begin{equation}
\bs{\nabla}_\parallel\cdot(\gamma H \bs{\sigma}_\parallel) + \bs{f} = \bs{0}.
\label{Eq:2DForceBalanceCurrent}
\end{equation}

\subsection{\label{sec_summary} Summary}
{\bf Governing equations, nondimensionalization, and control parameters: } We define dimensionless variables $\tilde{\bs{\sigma}} = \bs{\sigma}/\Gb$, $\tilde{\bs{x}} = \bs{x}_\parallel / R_\mathrm{b0}$, $\tau = t/\tau_0$, and $\tilde{c}_\parallel = c_\parallel/c_{0,\parallel}$,  where the shear modulus of the biofilm $\Gb$ was chosen as the scale for stresses, the initial biofilm radius $R_\mathrm{b0}$ as the characteristic length scale,  the inverse of the growth rate at the edge of the biofilm $\tau_0 = \left(k_\text{g}^\text{max}\right)^{-1}=[\kg(r = r_\mathrm{b})]^{-1}$ as the characteristic time scale associated with biofilm expansion, and the concentration $c_{0,\parallel}$ of nutrients at the edge of the biofilm as the relevant concentration scale. Upon nondimensionalizing the equations describing biofilm growth and mechanics discussed above, we obtain the following equations
\begin{align}
   \text{Nutrient diffusion-uptake:}&~\partial_\tau \tilde{c}_\parallel = \tilde{Q}_0\big[\tilde{a}_c^{2}\tilde{\bs{\nabla}}_\parallel^2\tilde{c}_\parallel - \Jep^{-1}\phi(\tilde{c}_\parallel)\big], \label{eq_ndim_nc}\\
    \text{Nutrient limited growth:}&~\partial_\tau\lamg =  \tilde{k}_\mathrm{g} (\tilde{c}_\parallel) \lamg, \label{eq_ndim_growth}\\
    \text{Force balance:}&~\tilde{\bs{\nabla}}_\parallel\cdot(\gamma \tilde{\bs{\sigma}}_\parallel)  = \xi \tilde{\bs{v}}_\parallel, \label{eq_ndim_force_balance}\\
    \text{Constitutive relation:}&~ \tilde{\bs{\sigma}}_\parallel  = \begin{cases}
    \Fep\Fep^T- \gamma^2 \mathbf{I}_\parallel\quad\text{(incompressible)}\\
    (\gamma \Jep)^{-1}\Big[\Fe_{,\parallel}\Fe_{,\parallel}^T + \frac{\lambdab}{\Gb}\big(\ln (\gamma \Jep)-1\big)\mathbf{I}_\parallel\Big]\quad\text{(compressible)},
    \end{cases}\label{eq_ndim_constitutive}
\end{align}
where $\tilde{Q}_0 = Q_0 / (c_{0,\parallel}/\tau_0)$ characterizes the rate of nutrient uptake, $\tilde{a}_c =  R_\mathrm{b0}^{-1}(Dc_{0,\parallel}/Q_0)^{1/2}$ denotes the relative width of the nutrient-rich zone, and $\phi(\tilde{c}_\parallel) = \tilde{c}_\parallel / (\tilde{K} + \tilde{c}_\parallel)$. The nutrient-dependent growth rate takes the form $\tilde{k}_\mathrm{g}(\tilde{c}_\parallel) = \phi(1)^{-1}(1-\tilde{k}_\mathrm{r})\phi(\tilde{c}_\parallel) + \tilde{k}_\mathrm{r}$. We define a dimensionless friction parameter $\xi = \frac{\eta (R_\mathrm{b0}/\tau_0)}{\Gb(H/R_\mathrm{b0})}$, which serves as a control parameter of the model and  quantifies the relative ratio of the typical frictional forces to the characteristic forces resulting from internal stresses in the biofilm. In Sec.~\ref{sec_discussion_friction}, we show that $\xi$ increases monotonically with the substrate modulus $\Gs$. Therefore, increasing the value of $\xi$ in the model corresponds to increasing the agar concentration and thus the substrate stiffness in experiments. We assume axial symmetry and thus all the fields only depend on the distance $r$ from the center of the biofilm. Thus, the dimensionless governing equations are solved with the following initial conditions and boundary conditions:
\begin{itemize}
    \item {Initial conditions: $\tilde{{r}} (\tilde{{R}}_{0}, \tau=0) = \tilde{{R}}_{0}$, $\tilde{c}_\parallel(\tilde{{R}}_{0}, \tau=0) \equiv 1$, $\lamg(\tilde{{R}}_{0}, \tau=0)\equiv 1$ and $\gamma(\tilde{{R}}_{0}, \tau=0) \equiv 1$.}
    \item {Boundary conditions: $\tilde{{\sigma}}_{\parallel,rr}(\tilde{R}_0,\tau) = 0$ and $\tilde{c}_\parallel (\tilde{R}_0,\tau) = 1$ at the edge of the biofilm.}
\end{itemize}
Further details about the numerical simulation of the above equations can be found in the Methods section of the main text.

\begin{table}[b]
\captionsetup{labelfont=bf,singlelinecheck=off,justification = justified}
    \caption{Summary of the measured/estimated quantities for biofilms in experiments and of the numerical values of parameters used in the chemo-mechanical model of biofilm growth.
} \label{tb_parameter}
    \centering
    \renewcommand{\arraystretch}{1.25}
    \begin{tabularx}{11.5cm}{YZY}
         \hline\hline
         \multicolumn{3}{l}{\textbf{Measured/estimated quantities for biofilms in experiments}}\\
         \hline\hline
         \multicolumn{1}{Y|}{\textbf{Symbols}} & \multicolumn{1}{C|}{\textbf{Descriptions}} & \multicolumn{1}{Y}{\textbf{Values}} \\
         \hline
         $R_\mathrm{b0}$ & {\small Initial biofilm radius} & $2$ mm \\
         \hline
         $r_\mathrm{b}$ & {\small Typical biofilm radius} & $5 - 15$ mm\\
         \hline
        $H$ & {\small Undeformed biofilm thickness} & $50~\mu$m \\
         \hline
         $h_\mathrm{b}$ & {\small Typical biofilm thickness} & $100~\mu$m\\
        \hline
        $a_c$ & {\small Width of nutrient-rich annulus} & $1$ mm \\
         \hline
        $\Gb$ & {\small Biofilm shear modulus} & $\approx 1$ kPa \\
         \hline
         $\nu_\mathrm{b}$ & {\small Biofilm Poisson's ratio} & $0.37\sim 0.46$\\
         \hline
         $\tau_0$ & {\small Biofilm development time scale} & $10$ h \\
         \hline
         $v_\mathrm{b}$ & {\small Typical expansion velocity} & $3~\mu$m/min \\
          \hline
         $R_\mathrm{s}$ & {\small Agar substrate radius} & 45 mm \\
          \hline
         $H_\mathrm{s}$ & {\small Agar substrate thickness} & $6$ mm \\
         \hline\hline
         \multicolumn{3}{l}{\textbf{Numerical values for parameters used in the model}}\\
         \hline\hline
         $\tilde{a}_c$ & {\small Normalized width of the nutrient-rich zone} & 0.5 \\
         \hline
         $\tilde{k}_\mathrm{r}$ & {\small Normalized residual growth rate} & 0.15 \\
         \hline
         $\tilde{Q}_0$ & {\small Normalized 2D nutrient uptake} & 2.0 \\
         \hline
         $\xi$ & {\small Dimensionless friction parameter} & 2 $\sim$ 200 \\
         \hline\hline
    \end{tabularx}
\label{tab:parameters}
\end{table}

\begin{figure}[t]
\centering
\includegraphics[width=.8\textwidth]{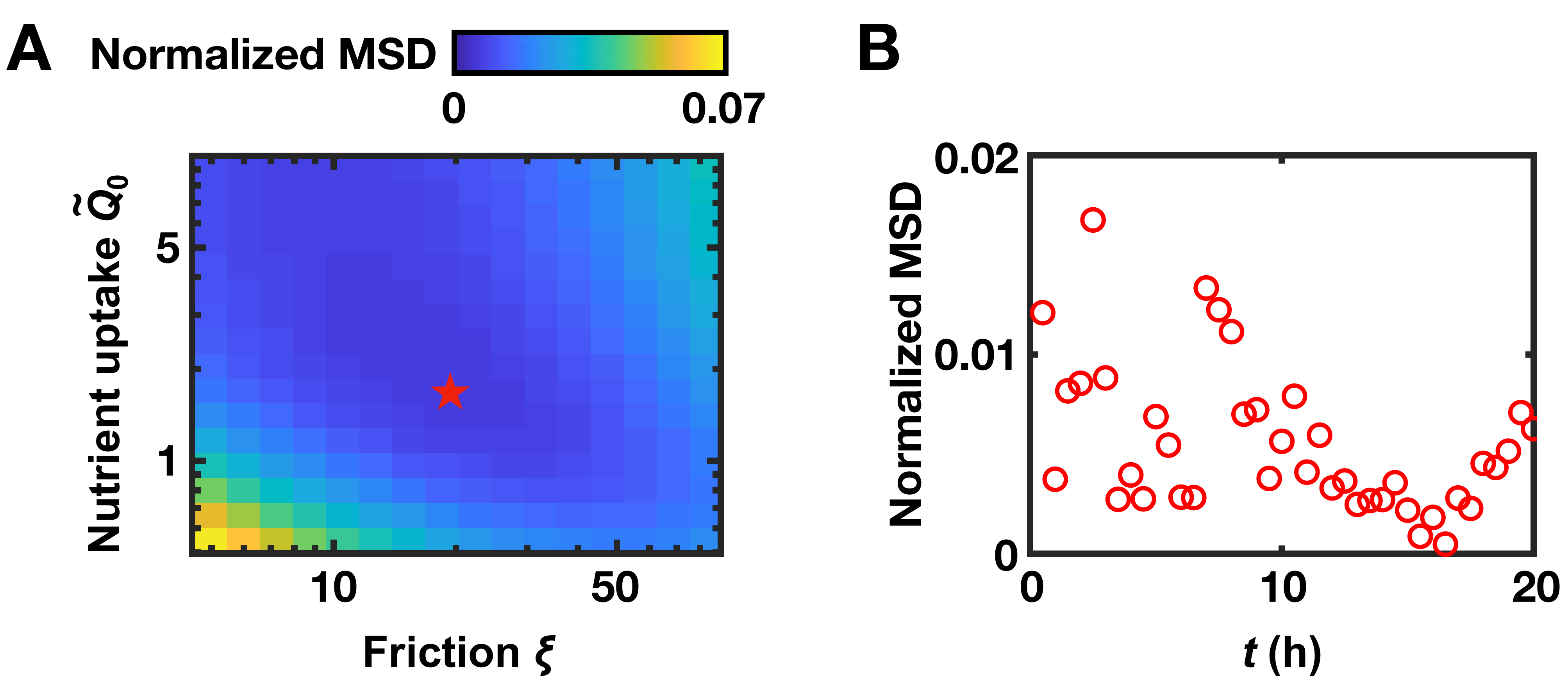}
\captionsetup{justification=justified,labelfont=bf,singlelinecheck =  false}
\caption{
{\bf Fitting of biofilm model parameters.} ({\bf A}) Time averaged normalized MSD values (see Methods for details) between the time series of velocity profiles for the modeled biofilm and the experimental biofilm grown on 0.7 \% agar upon variation of the dimensionless friction $\xi$ and the maximal nutrient uptake $\tilde{Q}_0$. The red star denotes the optimal parameter values, $\xi = 20$ and $\tilde{Q}_0=2.0$, which minimize the normalized MSD. ({\bf B}) Normalized MSD values plotted versus time for $\xi = 20$ and $\tilde{Q}_0=2.0$.}
\label{Fig:fittingParameters}
\end{figure}

{\bf Parameters of the model (see Table~\ref{tab:parameters}): } The initial radii of biofilms in experiments are measured to be $R_\mathrm{b0} \approx 2.0$ mm. The width of the active growth zone can be estimated from experiments to be $a_c \approx 1.0$ mm (see experiments in \cite{yan2019mechanical}). Thus, we fix $\tilde{a}_c  = 0.5$ in our simulations. The growth rate at the edge of the biofilm $\kg (r=r_\mathrm{b})=k_\text{g}^\text{max}$, which sets the time scale $\tau_0$, can be estimated from the steady-state expansion velocity of the biofilm grown on a soft substrate (0.4\% agar), where friction is small. When biofilms in experiments enter the third kinematic stage, the radii of biofilms increase linearly in time with the radial velocity $v_\mathrm{b}\approx 2a_c k_\text{g}^\text{max}=3~\mu$m/min. From this we estimate the characteristic time scale as $\tau_0 = 1/k_\text{g}^\text{max} \approx 2a_c/v_\mathrm{b} \approx 10 ~\mathrm{h}$. The thickness of a biofilm in the undeformed configuration $H \approx 50~\mu$m can be estimated from experiments with a soft substrate (For two-day-old biofilms grown on 0.4\% agar, the thickness is measured to be $h = 55 \pm 4~\mu$m \cite{yan2019mechanical}). The concentration of nutrients at the half-maximal uptake rate is set to $\tilde K=0.5$.
Note that our model results are insensitive to the specific choice of $\tilde{K}$: for a different $\tilde{K}$, a similar $\tilde{c}_\parallel$ profile can be obtained by adjusting the free parameter $\tilde{Q}_0$.
The two remaining parameters, the friction parameter $\xi$ and the nutrient uptake rate $\tilde{Q}_0$, are difficult to probe experimentally. Thus, we treat them as fitting parameters and their values were determined by fitting the radial velocity profiles $v_r(r)$ from the model to the experimental data for a biofilm grown on 0.7\% agar (see Fig.~2D in the main text, Fig.~\ref{Fig:fittingParameters}, and the Methods section in the main text). From the optimal fitting values, we obtain $\tilde{Q}_0 = 2$, and $\xi = 20$. When numerically simulating the growth of biofilms on agar substrates with different concentrations, we varied the value of the friction parameter $\xi$, with the values of other parameters fixed. Assuming that the friction coefficient $\eta$ increases with the substrate shear modulus $\Gs$ as $\eta \propto \Gs$, we find that the dimensionless friction parameter scales as $\xi \propto \Gs/\Gb$. Then we compute the friction parameter as $\xi = 20\frac{\Gs/\Gb}{\Gs/\Gb(\mathrm{agar~conc.=0.7\%})}$ from the measured values of the shear modulus of the substrate $\Gs$ and of the biofilm $\Gb$~\cite{yan2019mechanical}.

\subsection{\label{sec_discussion_formulation}Model discussion: Eulerian versus Lagrangian description}

In Sec.~\ref{sec_summary},
the governing equations (\ref{eq_ndim_nc}) -- (\ref{eq_ndim_constitutive}) for the chemo-mechanical model mix the \textit{Eulerian} and the \textit{Lagrangian} frames of reference. The nutrient diffusion-uptake dynamics in Eq.~(\ref{eq_ndim_nc}) and the force-balance condition in Eq.~(\ref{eq_ndim_force_balance}) are stated in the \textit{current} coordinates $\tilde {\bs{x}}_\parallel$ of the deformed biofilm, i.e. in the \textit{Eulerian} frame of reference. On the other hand the nutrient limited growth in Eq.~(\ref{eq_ndim_growth}) and the constitutive relation in Eq.~(\ref{eq_ndim_constitutive}) are stated in the \textit{initial} coordinates $\tilde {\bs{X}}_{0,\parallel}$ of biofilm, i.e. in the \textit{Lagrangian} frame of reference. In order to numerically solve the governing equations we have to rewrite them all either in the Lagrangian or the Eulerian frame of reference. For numerical simulations we used the Lagrangian frame of reference, as described below.

{\indent \bf Lagrangian formulation of the chemo-mechanical model:} Since the nutrient limited growth in Eq.~(\ref{eq_ndim_growth}) and the constitutive relation in Eq.~(\ref{eq_ndim_constitutive}) are already stated in the Lagrangian frame of reference, what remains is to rewrite the nutrient diffusion-uptake dynamics in Eq.~(\ref{eq_ndim_nc}) and the force-balance condition in Eq.~(\ref{eq_ndim_force_balance}) in terms of the initial coordinates $\tilde {\bs{X}}_{0,\parallel}$ of biofilm.

We start with the 3D force-balance equation. In the absence of body forces, the force-balance condition  for an arbitrary element with volume $\omega$ of an elastic material in the current configuration yields: $\int_\omega (\text{d}^3\bs{x})\bs{\nabla}\cdot\bs{\sigma} = \oint_{\partial \omega} \boldsymbol{\sigma}\cdot \bs{n}\,\text{d}a = 0$, where $\partial \omega$ is the boundary surface of that element and $\bs{n}$ is the unit vector normal to the boundary. The coordinate transformation must guarantee that the force acting on a particular area element is the same regardless of the choice of coordinate system. The corresponding stress measure $\mathbf{S}_0$ in the initial configuration must thus satisfy $\mathbf{S}_0\cdot \bs{N}_0\,\text{d}A_0=\boldsymbol{\sigma}\cdot \bs{n}\,\text{d}a$ for any area element, where $\bs{N}_0$ is the unit vector normal to the area element $\text{d}A_0$ in the initial configuration. Consequently, the force-balance condition for the same volume element in the initial configuration reads, $\oint_{\partial \omega} \boldsymbol{\sigma}\cdot \bs{n}\text{d}a= \oint_{\partial \Omega_0} \mathbf{S}_0\cdot \bs{N}_0\text{d}A_0 = \int_{\Omega_0} (\text{d}^3\bs{X}_0)\mathbf{Grad}_0~\mathbf{S}_0  = 0$, or equivalently, $\mathbf{Grad}_0~\mathbf{S}_0  = 0$. Here $\mathbf{Grad}_0$ represents derivatives with respect to the initial coordinates ${\bs{X}}_{0}$. The Nanson's relation, $(\text{d}a)\bs{n} = (J\text{d}A_0)\Ft^{-T}\cdot\bs{N}_0$,~\cite{de2011computational} can be used to relate the stress measure $\mathbf{S}_0$ to the Cauchy stress $\bs{\sigma}$ as $\mathbf{S}_0=\bs{\sigma} \Ft^{-T} J$, where $\Ft=\partial {\bs x}/\partial {\bs{X}_0}$ is the total deformation gradient and $J=\det{\Ft}$ is the Jacobian due to total deformation. The stress measure can also be expressed in terms of the PK stress $\mathbf{P}$ as $\mathbf{S}_0=\mathbf{P} \Fg^{-T} J_g$, where $\Fg=\partial {\bs X}/\partial {\bs{X}_0}$ is the deformation due to growth and $J_g=\det{\Fg}$ is the Jacobian due to growth.

Similar to our derivation of the 2D force balance in Eq.~(\ref{Eq:2DForceBalanceCurrent}) in the current coordinates $\bs{x}$, we integrate the 3D equation $\mathbf{Grad}_0~\mathbf{S}_0=\bs{0}$ in the $z$ direction over the biofilm thickness $H$ in the reference initial coordinates to obtain the dimensionless 2D force-balance equation 
\begin{gather}
\widetilde{\mathbf{Grad}}_{0,\parallel}~\tilde{\mathbf{S}}_{0,\parallel} = \widetilde{\mathbf{Grad}}_{0,\parallel}(\lamg\tilde{\mathbf{P}}_\parallel) = J_\parallel \xi \tilde{\bs{v}}_\parallel,\label{eq_ndim_fb_initial}
\end{gather}
where $\widetilde{\mathbf{Grad}}_{0,\parallel}$ refers to the derivative with respect to nondimenzionalized initial coordinates $\tilde{\bs{X}}_{0,\parallel}$, $\lambda_g$ is the local stretch due to growth, and the Jacobian factor $J_\parallel=\det(\bs{F}_\parallel)$ on the right side results from the transformation of the area elements. In the Lagrangian description, the radial velocity of the biofilm is  $\bs{v}_\parallel = (\partial_t \bs{x}_\parallel)_ {\bs{X}_{0,\parallel}}$, where the subscript means that the time derivative is taken at fixed ${\bs{X}_{0,\parallel}}$, and the elastic part of the deformation gradient $\Fe_{,\parallel} = \lamg^{-1}\Ft_{\parallel}$ is used to calculate the PK stress $\mathbf{P}_\parallel(\Fe_{,\parallel})$ (see Eq.~(\ref{Eq:PKStress})).

The same method as above is applied to transform the diffusion-uptake equation~(\ref{eq_nc_2d}) for the nutrient field $c_\parallel$. The integral form of the equation for any given area element $s$ in the current configuration is $\int_s \partial_t c_\parallel\,\text{d}a = \oint_{\partial s}D\bs{\nabla}_\parallel c_\parallel~\cdot~ \bs{n}_\parallel\,\text{d}l - \int_s \Jep^{-1} Q_0\phi(c_\parallel)\,da$, where we used the divergence theorem to transform the diffusion term $\int_{s}D\bs{\nabla}^2_\parallel c_\parallel \cdot \,\text{d}s=\oint_{\partial s}D\bs{\nabla}_\parallel c_\parallel~\cdot~\bs{n}_\parallel\,\text{d}l$. Here, $\bs{n}_\parallel$ is the unit normal to the boundary element $\text{d}l$. To transform the first and the third terms in the above equation, we use the Jacobian $J_\parallel$ to transform the area elements as $\text{d}a = J_\parallel\text{d}A_0$. In order to transform the time derivative of the concentration field we note that $c_\parallel(\bs{X}_{0,\parallel},t)=c_\parallel(\bs{x}_\parallel(\bs{X}_{0,\parallel},t),t)$. The time derivative of the concentration field is thus transformed as 
$(\partial_t c_\parallel)_{\bs{X}_{0,\parallel}} = (\partial_t c_\parallel)_{\bs{x}_\parallel} + (\bs{\nabla}_\parallel c_\parallel)_t \cdot \bs{v}_\parallel = (\partial_t c_\parallel)_{\bs{x}_\parallel} + (\mathbf{Grad}_{0,\parallel}~ c_\parallel)_t \cdot \Ft_{\parallel}^{-1}\cdot \bs{v}_\parallel$. In the last step we used the chain rule $\partial c_\parallel/\partial \bs{x}_\parallel = (\partial c_\parallel/\partial \bs{X}_{0,\parallel}) \cdot (\partial \bs{X}_{0,\parallel}/\partial \bs{x}_\parallel)$.

To transform the term with the diffusion constant, we first again employ the Nanson's relation $(\text{d}l)\bs{n}_\parallel=(J_\parallel\text{d}L_0)\Ft_\parallel^{-T}\cdot \mathbf{N}_{0,\parallel}$ and the chain rule to derive $\bs{\nabla}_\parallel c_\parallel \cdot \bs{n}_\parallel\text{d}l=J_\parallel\text{d}L_0(\mathbf{Grad}_{0,\parallel} c_\parallel)\cdot \big(\Ft_\parallel^{-1}\Ft_\parallel^{-T}\big)\cdot \bs{N}_{0,\parallel}.$ By combining all of the results above, we derive the transformed equation for the diffusion and uptake of nutrients in terms of the initial coordinates as
\begin{equation}
(\partial_t c_\parallel)_{\bs{X}_{0,\parallel}} - (\mathbf{Grad}_{0,\parallel}~ c_\parallel)_t \cdot \Ft_{\parallel}^{-1}\cdot \bs{v}_\parallel = J_\parallel^{-1}\mathbf{Grad}_{0,\parallel}\bs{\cdot}\big[J_\parallel D(\mathbf{Grad}_{0,\parallel} c_\parallel) \cdot (\Ft_\parallel^{-1}\Ft_\parallel^{-T})\big] - \Jep^{-1}Q_0\phi(c_\parallel).    
\end{equation}
The dimensionless equation is then easily obtained from the equation above.

{\bf Eulerian formulation of the chemo-mechanical model:} Finally, we give the equivalent formulation of the chemo-mechanical model using the \textit{Eulerian} formulation, where the variables of interest are the velocity  profile of the biofilm $\bs{v}_\parallel\left(\bs{x}_\parallel ,t\right)$, the areal mass density of the biofilm $\rho_\parallel\left(\bs{x}_\parallel ,t\right)$, and the concentration of nutrients $c_\parallel\left(\bs{x}_\parallel ,t\right)$. Below we discuss how to rewrite the nutrient limited growth in Eq.~(\ref{eq_ndim_growth}) and the constitutive relation in Eq.~(\ref{eq_ndim_constitutive}) in the Eulerian frame of reference.

First, we show that the nutrient limited growth equation in the Lagrangian formulation $\partial_t \lamg = \kg \lamg$ is equivalent to the mass conservation equation 
\begin{equation}
\partial_t \rho_\parallel + \nabla_\parallel \cdot(\rho_\parallel \bs{v}_\parallel) = 2 \kg \rho_\parallel \label{Eq:MassConservation}
\end{equation}
in the Eulerian formulation. We will do so by rewriting the mass conservation equation above in the Lagrangian formulation. The factor of 2 is related to the assumption that the growth of the biofilm is two dimensional, while the biofilm thickness remains unchanged. We assume that the areal mass density of the undeformed biofilm is constant $\rho_{0,\parallel}$. Elastic deformation of the biofilm then changes the  density to $\rho_{\parallel} = \rho_{0,\parallel} / \Jep$, which follows from the fact that the mass of a small element in the current coordinates $\rho_\parallel \,d^2{\bs {x}_\parallel}$ must be equal to the mass of the element in the virtual coordinates $\rho_{0,\parallel} \,d^2{\bs {X}_\parallel}$. Here we used the Jacobian $\Jep =\det(\Fep)$.

The time derivative of the areal mass density of the biofilm can be transformed from the current to the initial coordinates as $(\partial_t \rho_\parallel)_{\bs{X}_{0,\parallel}} = (\partial_t \rho_\parallel)_{\bs{x}_\parallel} + (\bs{\nabla}_\parallel \rho_\parallel)_t \cdot \bs{v}_\parallel$, which is identical to the transformation for the concentration of nutrients described in the previous section. By combining this equation with the mass conservation equation (\ref{Eq:MassConservation}) we obtain 
$(\partial_t \rho_\parallel)_{\bs{X}_{0,\parallel}} + \rho_\parallel (\nabla_\parallel\cdot\bs{v}_\parallel) = 2\kg \rho_\parallel$.
By inserting the relation $\rho_{\parallel} = \rho_{0,\parallel} / \Jep$ in the above equation we obtain 
\begin{equation}
    (\partial_t \Jep^{-1})_{\bs{X}_{0,\parallel}} + \Jep^{-1}(\bs{\nabla}_{\parallel}\cdot \bs{v}_{\parallel}) = 2\kg \Jep^{-1}.
    \label{Eq:Jep}
\end{equation}
The divergence term in the equation above can  be transformed to the Lagrangian formulation as
\begin{gather}
\bs{\nabla}_{\parallel}\cdot \bs{v}_{\parallel} = \frac{\partial {v}_{\alpha}}{\partial x_{\alpha}}  = \frac{\partial {v}_\alpha} {\partial X_{0,\beta}}\ \frac{\partial X_{0,\beta}}{\partial x_\alpha}  = (\dot{\Ft}_\parallel)_{\alpha \beta} (\Ft_{\parallel}^{-1})_{\beta \alpha} = \tr(\dot{\Ft}_\parallel \cdot \Ft_{\parallel}^{-1}) =
\dot{\Ft}_{\parallel} : \Ft_{\parallel}^{-T},
\label{Eq:DivV}
\end{gather}
where we used the definition of the deformation gradient $\Ft_{\parallel}=\partial {\bs {x}}_\parallel/\partial {\bs X}_{0,\parallel}$ and we defined the tensor contraction $\bs{A}:\bs{B}=\tr(\bs{A} \cdot \bs{B}^T)=A_{\alpha \beta} B_{\alpha \beta}$. Note that the Greek indices in above equations denote the components of in-plane 2D vectors and tensors. By combining Eqs.~(\ref{Eq:Jep}) and (\ref{Eq:DivV}) we obtain
\begin{gather}
-\Jep^{-1}\ (\partial_t \Jep)_{\bs{X}_{0,\parallel}}  + \dot{\Ft}_{\parallel}: \Ft_{\parallel}^{-T} = 2\kg. \label{Eq:Jep2}
\end{gather}
The time derivative term of the Jacobian $\Jep$ can be expressed using Jacobi's formula \cite{magnus1999matrix} as 
\begin{equation}
\partial_t \Jep = \text{tr}(\mathbf{adj} (\Fe_{,\parallel}) \cdot \dot{\bs{F}}_{e,\parallel}) = \tr(\Jep \Fe_{,\parallel}^{-1} \cdot \dot{\bs{F}}_{e,\parallel}) = \Jep (\dot{\bs{F}}_{e,\parallel} : \Fe_{,\parallel}^{-T}).  \label{Eq:Jep3}
\end{equation}
Next, we leverage the fact that the total in-plane deformation can be decomposed as $\Ft_{\parallel} = (\Fe\cdot\Fg)_{\parallel} = \lamg\Fe_{,\parallel}$. Thus we obtain  $\dot{\Ft}_{\parallel} = (\partial_t \lamg)\Fe_{,\parallel} + \lamg \dot{\Fe}_{,\parallel}$ and $\Ft_{\parallel}^{-T} = \lamg^{-1}\Fe_{,\parallel}^{-T}$. By combining these equations with Eqs. (\ref{Eq:Jep2}) and (\ref{Eq:Jep3}) we obtain
\begin{gather}
 \lamg^{-1} (\partial_t \lamg)  (\Fe_{,\parallel} : \Fe_{,\parallel}^{-T})= 2\kg.
\end{gather}
Note that $\Fe_{,\parallel}:\Fe_{,\parallel}^{-T}=\tr(\Fe_{,\parallel}\cdot \Fe_{,\parallel}^{-1}) = \tr(\mathbf{I}_{\parallel}) = 2$. Therefore, the equation above is equal to the nutrient limited growth equation $\partial_t \lamg = \kg \lamg$, which is thus indeed equivalent to the mass conservation law in Eq.~(\ref{Eq:MassConservation}).

To complete the Eulerian formulation of the chemo-mechanical model we also need to express the constitutive relation $\bs{\sigma}_\parallel(\Fep)$ in the current coordinates $\bs{x}_\parallel$.
We start by defining a map $\bs{\chi}(\bs{x}_\parallel,t)=\bs{X}_{0,\parallel}$ from the current coordinates $\bs{x}_\parallel$ to the initial coordinates  $\bs{X}_{0,\parallel}$. By definition the material derivative of this map is zero, i.e.
\begin{gather}
\partial_t \bs{\chi} + \bs{v}_\parallel\cdot\nabla_\parallel\bs{\chi} = \mathbf{0}.
\label{Eq:ChiMap}
\end{gather}
The gradient of the map $\bs{\chi}$ is related to the total deformation gradient as $\bs{\nabla}_\parallel\bs{\chi} = \partial \bs{X}_{0,\parallel}/\partial \bs{x}_{\parallel}=\Ft_{\parallel}^{-1}$. The elastic part of the deformation gradient can then be written as $\Fep = \Ft_{\parallel} \cdot \Fg_{\parallel}^{-1} = \lambda_g^{-1} \Ft_{\parallel}$. Note that the determinant of the elastic deformation gradient is $\Jep = J_\parallel / J_{g,\parallel} = \lambda_g^{-2} J_\parallel$.
The final step is to express the growth factor $\lamg$ in the Eulerian formulation. We recall that the areal mass density of the biofilm is $\rho_\parallel = \rho_{0,\parallel}/\Jep$ and thus we obtain the relation for the growth factor $\lambda_g^{-2} = \Jep / J_\parallel = (\rho_{0,\parallel}/\rho_{\parallel}) \det(\bs{\nabla}_\parallel\bs{\chi})$. By combining these relations we can express the elastic part of the deformation gradient in the Eulerian formulation
\begin{gather}
\Fe_{,\parallel} =  \lamg^{-1}\Ft_{\parallel} = \sqrt{\rho_\parallel^{-1} \rho_{0,\parallel}\,\text{det}\big(\bs{\nabla}_\parallel\bs{\chi}\big)}\,  (\bs{\nabla}_\parallel\bs{\chi})^{-1}. \label{eq_eul_fep}
\end{gather}
Using the above expression we can then calculate the Cauchy stress $\bs{\sigma}_\parallel (\Fe_{,\parallel})$ from the constitutive relation in Eq.~(\ref{eq_ndim_constitutive}).

In the Eulerian formulation of the chemo-mechanical model the unknowns are $\bs{\chi}$, $\bs{v}_{\parallel}$, $\rho_\parallel$, and $c_\parallel$. They can be obtained by solving Eqs.~(\ref{eq_ndim_nc}),  (\ref{eq_ndim_force_balance}), (\ref{Eq:MassConservation}), and (\ref{Eq:ChiMap}) together with the 
constitutive relation in Eq.~(\ref{eq_ndim_constitutive}) and the elastic part of the deformation gradient in Eq.~(\ref{eq_eul_fep}).
Solving these equations is complicated, because of the moving boundary conditions at the biofilm edge, and therefore requires special numerical treatment, such as the phase field approach~\cite{yu2005iterative,camley2013periodic,shao2010computational,kockelkoren2003computational}.

\subsection{\label{sec_discussion_nd}Model discussion: Comparison of 3D and 2D nutrient dynamics}

To compare the full 3D dynamics of nutrients   with the reduced 2D model, we simulate the dynamics of nutrients in the 3D substrate, in the presence of a growing biofilm on top of the substrate. For simplicity, we set the friction coefficient to zero, i.e. $\eta=0$.
We consider dimensionless variables $\tilde{\bs{x}} = \bs{x}/R_\mathrm{b0}$, $\tau = t/\tau_0$, and $\tilde{c}= c/c_0$, where $c_0$ is the initial concentrations in the 3D substrate, and the length scale $R_\mathrm{b0} = 2$ mm and the time scale $\tau_0 = 1/\kg(c=c_0) = 10$~h are the same as in the 2D model.
The nondimensionalized equation for the diffusion of nutrients is $\partial_\tau\tilde{c} = \tilde{D}\tilde{\nabla}^2 \tilde{c}$, where $\tilde{D} = D\tau_0/R_\mathrm{b0}^2$ is fixed to be the same value as in the 2D model. A biofilm of radius $\tilde{r}_\mathrm{b}(\tau)$ is located at the top surface $z=0$ and it consumes nutrients from the substrate, which sets the boundary condition $-\partial_{\tilde{z}} \tilde{c} = \tilde{J_0} \phi(\tilde{c})$ at $\tilde{z}=0$ and $|\tilde {x}_\parallel|<\tilde{r}_\mathrm{b}(\tau)$, where we used the Monod law $\phi(\tilde c)=\tilde c/(\tilde K + \tilde c)$ as described previously. Zero flux boundary conditions are imposed elsewhere on the boundary of the substrate, i.e. ${\bs n} \cdot {\bs \nabla} c=0$, where ${\bs n}$ is the vector normal to the substrate boundary.

To simplify the treatment of biofilm expansion we consider a 2D incompressible biofilm with an initial radius $\tilde{r}_\mathrm{b}(\tau=0)=1$. Due to the consumption of nutrients the total area of the biofilm $\tilde A(\tau)=\pi (\tilde{r}_\mathrm{b}(\tau))^2$ increases at a rate $\partial_\tau {\tilde A} = \int_0^{\tilde r_\mathrm{b}} \tilde k_a(\tilde r) \,2 \pi \tilde r \,d\tilde r$, where $\tilde{k}_\mathrm{a}(r)$ denotes the 2D areal growth rate and is related to the linear growth rate $\tilde{k}_\mathrm{g}$ by $\tilde{k}_\mathrm{a} = 2\tilde{k}_\mathrm{g}$. The growth rate $\tilde{k}_\mathrm{a}(r)$ was chosen to be proportional to the local nutrient uptake rate, specifically $\tilde{k}_\mathrm{a}(\tilde r)=  2\phi\big( \tilde{c} (\tilde r,\tilde z=0)\big)$. The radius of the biofilm thus expands with velocity
\begin{gather}
\partial_\tau {\tilde{r}}_\mathrm{b} = \tilde{r}_\mathrm{b}^{-1} \int_0^{\tilde{r}_\mathrm{b}} k_a(\tilde r) \, \tilde r \,d\tilde r. \label{Eq:biofilmRadius}
\end{gather}

 The 3D geometry of the substrate was chosen to be a cylinder with radius $\tilde{R}_\mathrm{s}$ and thickness $\tilde{H}_\mathrm{s}$, and the initial concentration of nutrients was set to $\tilde{c}(\tilde r, \tilde z, \tau=0)\equiv 1$. Note that the 3D model of nutrient dynamics depends on two parameters $\tilde{K}$ and $\tilde{J}_0$ related to the uptake of nutrients; the corresponding parameters in the reduced 2D model are $\tilde{K}$ and $\tilde{Q}$.

In simulations we compared the following three scenarios, where we used the same  Eq.~(\ref{Eq:biofilmRadius}) for the expansion velocity of biofilm:
\begin{enumerate}[label=(\Alph*)]
 {\item the reduced 2D model of nutrient dynamics in Eq.~(\ref{eq_ndim_nc}); }
{ \item the full 3D model of nutrient dynamics with a thin substrate $\tilde{R}_\mathrm{s} = 22.5,$ and $\tilde{H}_\mathrm{s} = 0.05$ (thin substrate); and }
{ \item the full 3D model of nutrients with a thick substrate $\tilde{R}_\mathrm{s} = 22.5,$ and $\tilde{H}_\mathrm{s} = 3$ (these dimensions are comparable to the experimental agar substrates, $R_\mathrm{s} \approx 45$ mm and $H_\mathrm{s} \approx 6$ mm).}
\end{enumerate}

\begin{figure}[t]
\centering
\includegraphics[width=\textwidth]{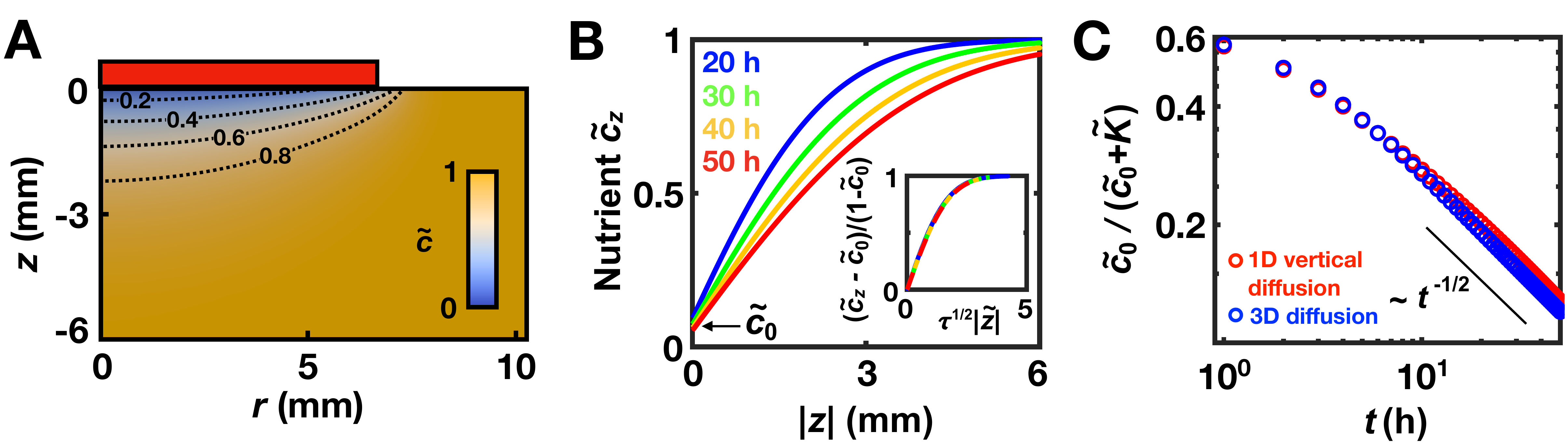}
\captionsetup{justification=justified,labelfont=bf,singlelinecheck =  false}
\caption{
{\bf Dynamics of nutrient diffusion in the agar substrate and uptake by biofilm cells.} ({\bf A}) Snapshot of the nutrient concentration field in the substrate, denoted by $\tilde{c}(r,z)$, computed from the full 3D nutrient diffusion model (see Secs. \ref{sec_nutrient} and \ref{sec_discussion_nd} for details) at time $t = 20$ h. The characteristic length scale $R_\mathrm{b0} = 2$ mm and the time scale $\tau_0 = 10$~h were used to convert nondimensional coordinates and time. Color scale indicates the dimensionless nutrient concentration $\tilde{c}$. Dashed curves are the contours for $\tilde{c} = 0.2,~0.4,~0.6,$ and $0.8$. Red bar on top of the substrate indicates the extent of the biofilm at the corresponding time, within which a non-zero flux boundary condition is applied. ({\bf B}) The nutrient field at the center of the biofilm/substrate shown in {\bf A}, defined as $\tilde{c}_z = \tilde{c}(r=0,z,t)$, versus $z$ coordinate at time $t =$ 20 h (blue), 30 h (green), 40 h (yellow), and 50 h (red). The nutrient concentration at $r = 0$ and $z= 0$, denoted by $\tilde{c}_0$, shows a slow decay over time. Inset: the evolution of the nutrient field $\tilde{c}_z$ can be collapsed onto a master curve $\tilde{c}_z = \tilde{c}_0(t) + (1-\tilde{c}_0)f(\tau^{-1/2}\tilde{z})$, where $\tau$ and $\tilde{z}$ denote the dimensionless time and $z$-coordinate, respectively. ({\bf C}) The nutrient-dependent Monod factor at $r=0$ and $z = 0$, expressed as $\tilde{c}_0/(\tilde{c}_0 + \tilde{K})$, versus time $t$, for the full 3D diffusion model (blue circles) and a 1D vertical diffusion model (red circles).  The black line indicates a slope of -1/2 on a log-log scale. This value of the slope indicates that $\tilde{c}_0$ scales as $t^{-1/2}$ in the long time limit, and the slow decay arises from the vertical diffusion dynamics. Simulation model parameters are: $\tilde{K} = 0.4$ and $\tilde{J} = 2.2$. 
}
\label{Fig:nutrientDynamics3D}
\end{figure}

\begin{figure}[t]
\centering
\includegraphics[width=.6\textwidth]{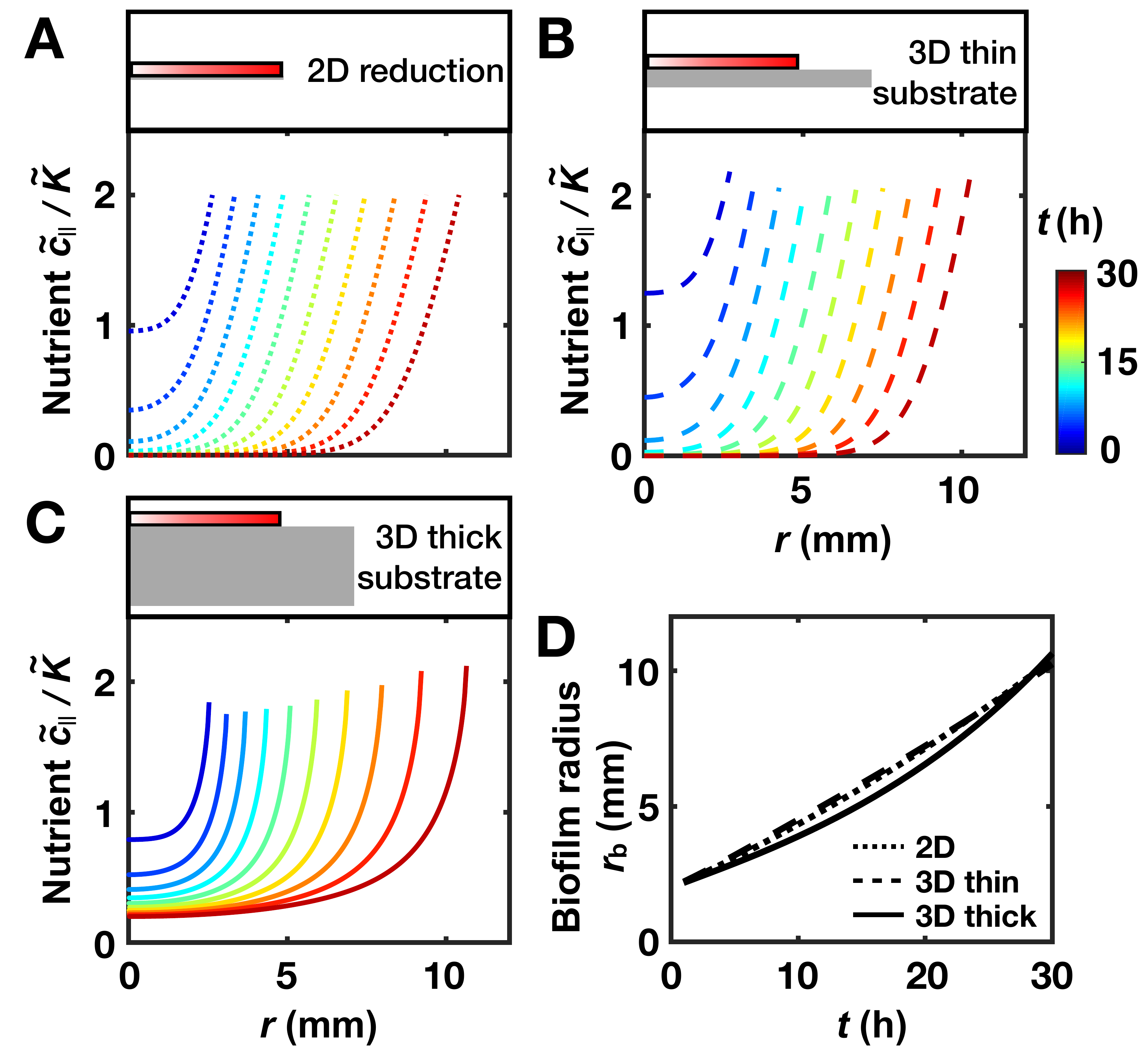}
\captionsetup{justification=justified,labelfont=bf,singlelinecheck=false}
\caption{
{\bf Validation of the reduced 2D model for nutrient diffusion-uptake dynamics.} ({\bf A-C}) Model schematic ({\it top}) and the spatiotemporal evolution of the nutrient field ({\it bottom}) are shown for ({\bf A}) the reduced 2D diffusion-uptake model (dotted curves), ({\bf B}) the full 3D diffusion model with a thin substrate (substrate thickness ${H}_\mathrm{s}$ = 0.1 mm, dashed curves) and ({\bf C}) the full 3D diffusion model with a thick substrate (experimental substrate thickness ${H}_\mathrm{s}$ = 6 mm, solid curves). For each model, the nutrient concentration at the biofilm-substrate interface, defined as $\tilde{c}_\parallel = \tilde{c}(r,z=0)$, versus radial coordinate $r$ is plotted within the biofilm regime ($r<r_\mathrm{b}$, where $r_\mathrm{b}$ denotes the biofilm radius) for different times. $\tilde{c}_\parallel$ is normalized by the half-rate constant $\tilde{K}$ to visualize the spatial differences in growth/uptake rates. Color scale indicates time. The characteristic length scale $R_\mathrm{b0} = 2$ mm and the time scale $\tau_0 = 10$~h were used to convert nondimensional coordinates and time. See text for details about the implementation of biofilm expansion dynamics. ({\bf D}) Biofilm radius $r_\mathrm{b}$ plotted against time $t$ for the three models. Line styles are the same as in {\bf A-C}. Simulation parameters were chosen such that the three models result in similar biofilm expansion dynamics as follows: ({\bf A}) $\tilde{K} = 0.5$, $\tilde{a}_c^{-2} = 3.0$; ({\bf B}) $\tilde{K} = 0.4$, $\tilde{J} = 0.26$; and ({\bf C}) $\tilde{K} = 0.4$ and $\tilde{J} = 2.2$ as in Fig.~\ref{Fig:nutrientDynamics3D}.
}
\label{Fig:nutrientDynamics2Dvs3D}
\end{figure}
Fig.~\ref{Fig:nutrientDynamics2Dvs3D} shows the profile of nutrients $\tilde{c}(\tilde r, \tilde z=0, \tau)$ and the radius of the biofilm $\tilde{r}_\mathrm{b}(\tau)$ for the three scenarios described above.
 Note that if the substrate is sufficiently thin, i.e. $\tilde{H}_\mathrm{s} \ll 1$, the diffusion of nutrients across the substrate thickness is very fast and therefore the concentration of nutrients $\tilde{c}$ becomes nearly uniform across the thickness and thus independent of coordinate $z$. In this case the reduced 2D model of nutrient dynamics approximates the averaged nutrient concentration $\tilde{c}_\parallel = \int_{-H_\text{s}}^0\tilde{c}\,dz$ with the maximum uptake rate $\tilde Q_0 = \tilde J_0/\tilde H_\text{s}$ (see for example, \cite{srinivasan2019multiphase}). Fig.~\ref{Fig:nutrientDynamics2Dvs3D} shows that the results for models A and B are in fact nearly identical except at very early times $\tau \ll \tilde H_\text{s}^2$ related to the characteristic time for diffusion of nutrients across the substrate thickness.

For the thick substrate (model C) we find a nutrient rich annulus at the periphery of the biofilm, similar to the other two models. However, for model C the concentration of nutrients directly beneath the center of the biofilm decreases very slowly (see Fig.~\ref{Fig:nutrientDynamics3D}A), due to the vertical diffusion of nutrients. To quantify this effect, we consider a 1D diffusion problem in the $z$ direction, $\partial_\tau \tilde{c} = D\partial_{\tilde{z}}^2\tilde{c}$ with the same uptake of nutrients $-\partial_{\tilde{z}}\tilde{c} =\tilde{J}_0\tilde{c}/(\tilde{c} + \tilde{K})$ at $\tilde{z} =0$. The asymptotic solution for this 1D problem is $\tilde{c}(\tau,\tilde z) = \tilde{c}_0(t) + (1-\tilde{c}_0)f(t^{-1/2}z)$, where $\tilde{c}_0(t)\sim t^{-1/2}$ denotes the concentration at $\tilde{z} =0$, and the function $f(x)$ satisfies the differential equation $f''(x) + (1/2)xf'(x) = 0$ with boundary conditions $f(0) = 0$ and $f(\infty) = 1$. 
This scaling is consistent with the results of simulations of the 1D model described above and for the 3D simulations of model C (see Fig.~S2). Thus the concentration of nutrients underneath the core of biofilm evolves in time as $\tilde{c}(\tilde r=0, \tilde z=0, \tau) \sim \tau^{-1/2}$.
To account for this slowly vanishing  concentration, in the reduced 2D model we introduced a small residual growth rate in Eq.~(\ref{eq_ndim_growth}) that is independent of the reduced nutrient concentration $\tilde{c}_\parallel$.

\subsection{\label{sec_discussion_constitutive}Model discussion: Constitutive model}
In our chemo-mechanical model, we employ a phenomenological hyperelastic material model to describe biofilm mechanics. It is not known whether growing \vc biofilms are {\it effectively} compressible or not, since they are physically swelled, and can, in principle, undergo volumetric change by shifting water to the agar substrate. Our goals here are (i) to derive the constitutive relation from a model of the microscopic interactions, and discuss the use of the simple neo-Hookean model, and (ii) to estimate the effective compressibility of biofilms grown on agar. Previous works~\cite{yan2018bacterial, berk2012molecular} showed that the mechanical behavior as well as the matrix composition of \vc biofilms resemble those of hydrogels. Thus, we use a hydrogel-like constitutive model to approximate biofilm elasticity. We caution that the microscopic model is still highly simplified, and cannot capture all the complex rheological behaviors observed in the experiments~\cite{yan2018bacterial}. 

{\bf A swollen hydrogel model: }We model biofilms grown on agar as cross-linked polymer networks in contact with a reservoir of solvent (water) molecules. Similar models have been proposed before to study polymeric gels~\cite{hong2009inhomogeneous,hong2008theory}. Since the agar substrate contains a much larger total amount of water than the biofilms, we assume that the presence of agar provides a constant chemical potential of water molecules $\mu_\text{w}$. Experimentally, $\mu_\text{w}$ can be determined from the equilibrium concentration $\phi_0$ of a dextran droplet atop the agar~\cite{yan2018bacterial} as $\mu_\text{w}=-\kT \phi_0^3/3$, where the chemical potential for a theta solvent
~\cite{rubinstein2003polymer} was used, because in aqueous solution dextran assumes a conformation close to that of an ideal coil ~\cite{guner1999unperturbed,antoniou2012solution}.

\begin{figure}[t]
\centering
\includegraphics[width=.7\textwidth]{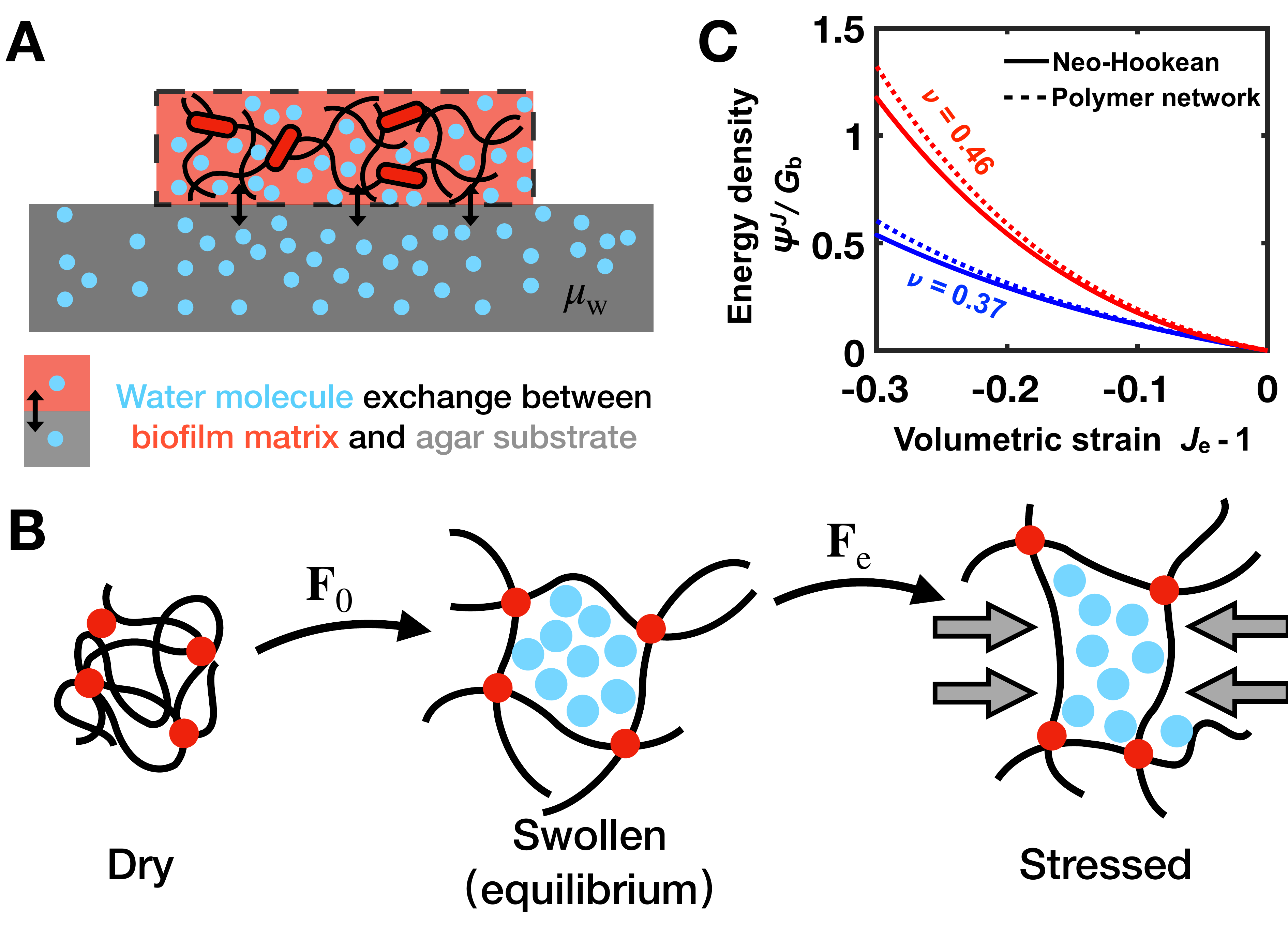}
\captionsetup{justification=justified,labelfont=bf,singlelinecheck=false}
\caption{
{\bf Compressibility of a biofilm in contact with an agar substrate that provides a water reservoir.} ({\bf A}) Schematic of the cross-linked polymer network model for the biofilm matrix. When biofilms are grown on agar substrates, they can exchange water with the agar (denoted by black double-headed arrows). We assume the total water content of the agar is much higher than that of the biofilm. Thus, we treat the agar as a reservoir of water molecules with a constant chemical potential $\mu_\mathrm{w}$. ({\bf B}) Schematic representation of the multiplicative decomposition of the deformation of a cross-linked polymer network. The polymer network is cross-linked in the dry state ({\it left}). In the presence of a water reservoir, it swells to a minimum free energy equilibrium state ({\it middle}). The equilibrium swelling is denoted by $\mathbf{F}_0$.  External mechanical forces (represented by the gray arrows) will further deform the network (denoted by $\mathbf{F}_\mathrm{e}$, {\it right}), which could accumulate or lose water accordingly. The total deformation is $\mathbf{F}_\mathrm{d}=\mathbf{F}_{e}\mathbf{F}_0$. ({\bf C}) The free-energy density increase $\Psi^J$ due to the volumetric change of the biofilm, defined as all the terms containing $J_\mathrm{e} = \mathrm{det}(\mathbf{F}_\mathrm{e})$ in the total free-energy density, normalized by the biofilm shear modulus $\Gb$, plotted versus the volumetric strain $J_\mathrm{e} - 1$ for the neo-Hookean constitutive model (solid curves, Eq.~(\ref{Eq:Energy_neoHookean})) and the cross-linked polymer network constitutive model (dashed curves, Eq.~(\ref{Eq:Energy_Hydrogel})). Poisson's ratio $\nu = 0.37$ (blue) and $\nu = 0.46$ (red) correspond to biofilms grown on 0.6\% and 1.5\% agar, respectively.
}
\label{Fig:hydrogel}
\end{figure}
Next, we consider the free energy of the system when a biofilm is in contact with agar. We take the stress-free dry cross-linked network of biofilm matrix as the reference state (coordinate denoted by $\bs{X}_\mathrm{d}$, not to be confused with the initial coordinates in the chemo-mechanical model). The dry network can swell and deform due to the presence of osmotic pressure or mechanical loads (Fig.~\ref{Fig:hydrogel}). The deformation gradient tensor can be defined as $F_{\mathrm{d},ij} = \partial{x_{i}}/\partial{X_{\mathrm{d},j}}$ where we use our standard notation $\bs{x}$ to specify the current deformed coordinates. It is commonly assumed that the volume of the swollen polymer network equals the volume of the dry network plus the volume of the absorbed solvent, i.e. both the polymers and the water molecules are incompressible and the volumetric change of the matrix network is solely due to absorbing water from (or losing water to) the agar. We adopt the same assumption and write $\Jd =\text{det}(\Fd) = 1+v_\mathrm{w}C$, where $C$ denotes the number of water molecules per unit \textit{reference volume}. The volume fraction of polymers in a swollen network is thus $\phi =\Jd^{-1}$. \footnote{Here, we assume the polymer network is cross-linked in its dry state. But this assumption is not essential for the results derived below. In fact, we can assume $\phi = \phi_\mathrm{d}\Jd^{-1}$ where $\phi_\mathrm{d}$ denotes the polymer volume fraction in the state where they are cross-linked. One can perform the same calculations outlined here and obtain the same results.}
  
Following the equilibrium swelling theory of the Flory and Rehner model \cite{flory1943statistical,flory1943statisticalswell} and previous works on swollen hydrogels \cite{hong2009inhomogeneous}, we obtain the total free energy density (per unit reference volume)
\begin{align}
    W &= W_\text{elastic}(\Fd) + W_\text{poly-sol}(\Jd) + \Delta W_\text{agar} \notag\\
    & = \frac{1}{2}N\kT \big[\text{tr}(\Fd^T\Fd) - 3 - 2\ln \Jd\big]+  \frac{\kT}{v_\mathrm{w}}\Big[ (\Jd-1)\ln(1-\Jd^{-1}) + \chi (1-\Jd^{-1})\Big] - \frac{\mu_\text{w}}{v_\mathrm{w}}(\Jd-1),\label{eq_strain_energy_hydrogel}
\end{align}
where $N$ denotes the number of polymer strands per unit reference volume, and the long chain limit is assumed \cite{rubinstein2003polymer}.
The first term arises from stretching of the polymer network, and has an entropic origin \cite{flory1943statistical}. The second term is the standard mixing free energy of polymer solutions, in which the Flory interaction parameter $\chi$ characterizes the two-body interaction between the polymer and the solvent. Note that the usual expression $f(\phi) \approx v_\mathrm{w}^{-1}\kT[ (1-\phi)\ln(1-\phi) + \chi\phi(1-\phi)]$ measures the mixing free energy per \textit{current} volume, so $W_\text{poly-sol}(\Jd) = \Jd f(\phi)= \Jd f(\Jd^{-1})$. Finally, the third term accounts for the free energy cost of absorbing water from the agar.

The results above allow us to calculate the Cauchy stress $\bs{\sigma}$ by $\boldsymbol{\sigma} = \Jd^{-1}(\delta W/\delta \Fd)\cdot \Fd^T$. Using the relation $\delta \Jd  = \Jd\Fd^{-T}:\delta\Fd$, we derive 
\begin{gather}
\boldsymbol{\sigma} = \Jd^{-1}N\kT(\Fd\Fd^T - \mathbf{I}) + \frac{\kT}{v_\mathrm{w}}\Big[\ln(1-\Jd^{-1}) + \Jd^{-1} + \chi \Jd^{-2}\Big]\mathbf{I} - \frac{\mu_\text{w}}{v_\mathrm{w}}\mathbf{I}. \label{cauchy}
\end{gather}

{\bf Equilibrium condition: } Consider the mechanical and chemical equilibrium condition of the swollen biofilm on agar. We assume the swelling is isotropic, and denote by $\mathbf{F}_0 = \lambda_0 \mathbf{I}$ the equilibrium deformation-gradient tensor, and by $J_0 = \lambda_0^3 $ its determinant. The equilibrium condition $\bs{\sigma} = \bs{0}$ yields
\begin{gather}
\lambda_0^{-1}N\kT(1-\lambda_0^{-2}) + \frac{\kT}{v_\mathrm{w}} \Big[\ln(1-\lambda_0^{-3}) + \lambda_0^{-3} + \lambda_0^{-6}\chi \Big] - \frac{\mu_\text{w}}{v_\mathrm{w}} = 0, \label{eqlib}
\end{gather}
from which the initial swelling can be determined.
We assume large equilibrium swelling \cite{sutherland2001biofilm} so that we can apply a series expansion in $\lambda_0^{-3} \ll 1$ to the second term. The resulting equation becomes $\Gb(1-\lambda_0^{-2}) +(\kT/v_\mathrm{w})\big[(\chi - 1/2) \lambda_0^{-6} + (\phi_0^3 - \lambda_0^{-9})/3 + \mathcal{O}(\lambda_0^{-12})\big] =  0$, where $\Gb = \lambda_0^{-1}N\kT$ and we shall see in the following paragraph that $\Gb$ is exactly the Lam\'{e}'s second parameter (shear modulus) of the biofilm material. 

For \emph{V. cholerae} biofilms, we argue that $\chi \approx 1/2$. In the absence of protein cross-linkers (i.e. the $\Delta ABC$ strain \cite{yan2017extracellular}), a submerged biofilm behaves like a polymer solution in contact with pure solvent. One can readily calculate the osmotic pressure in this case to be $\Pi(\phi) \approx -(\chi - 1/2)\phi^2 + \phi^3/3 + \mathcal{O}(\phi^4)$. If $\chi < 1/2$ (good solvent), then there is a tendency for the biofilm to swell indefinitely under water; if $\chi > 1/2$, there exists a stable gel fraction $\phi_\text{eq} = 3(\chi - 1/2)$ for which $\Pi = 0$. The experiments \cite{yan2017extracellular} suggest that $\phi_\text{eq} \ll 1$ and thus $\chi$ is close to (slightly greater than) 1/2, and for finite osmotic pressure, it is reasonable to neglect the order $\phi^2$ term.

With this approximation, we finally estimate, to the lowest order, the equilibrium swelling ratio to be 
\begin{gather}
J_0^{-1} = \lambda_0^{-3} \approx \Big(\phi_0^3 + \frac{\Gb}{\kT / 3v_\mathrm{w}}\Big)^{1/3} \label{initswell}
\end{gather}

{\bf Lam\'{e} parameters: }To probe the mechanical properties of the swollen polymer network, we can introduce additional deformations $\Fe$ on top of $\mathbf{F}_0$ and calculate the resulting Cauchy stress. The total deformation is denoted by $\Fd = \Fe\mathbf{F}_0$. For small deformations, $\Fe\Fe^T \approx \mathbf{I} + 2\bs{\epsilon}$, where $\bs{\epsilon}$ is the symmetric strain tensor defined in linear elasticity theories \cite{landau1989theory}. In the small deformation limit, the Cauchy stress can be specified by two Lam\'{e} parameters $\lambdab, \Gb$, according to $\bs{\sigma} = 2\Gb\bs{\epsilon} + \lambdab\text{tr}(\bs{\epsilon})\mathbf{I} $. Thus, by comparing the Cauchy stress calculated from the polymer network model to the linear elasticity result, one can determine the equivalent Lam\'{e} parameters for the swollen biofilm matrix, and consequently the Poisson's ratio $\nub$.

First, consider a shear deformation $\Fe$ with $\Je = 1$. An example of this is the shear test of biofilm material on a rheometer\footnote{For rheological measurement, the $\mu_\mathrm{w}$ term does not exist in Eq.~(\ref{cauchy}), and instead a Lagrange multiplier should be introduced to ensure that $\Je = 1$ is satisfied.}.~\cite{yan2018bacterial}  Substituting $\Fd = \lambda_0 \Fe$ and $\Jd = J_0$ into Eq. (\ref{cauchy}), one immediately finds $\Gb = \lambda_0^{-1} N\kT$ from the anisotropic part of the Cauchy stress. 

The experimentally measured Poisson's ratio of \vc biofilms scraped off a substrate is near $\nub=0.5$ \cite{yan2018bacterial,yan2019mechanical}, indicating that the biofilm is a nearly incompressible material. However, when the biofilms are grown on agar, the water molecules retained by the biofilm matrix might be redistributed into the agar upon mechanical loading and/or elastic deformation. In the present paper, since we are modeling biofilm growth and development, we are interested in the possible compressibility of a biofilm in contact with agar.  

We therefore consider a small isotropic compression/extension, i.e. $\Fe = (1 + \epsilon/3) \mathbf{I}$. From linear elasticity theory, the Cauchy stress is given by $\boldsymbol{\sigma} = (\lambdab + 2\Gb/3)\epsilon \mathbf{I}$. On the other hand, if we set $\mathbf{F} = \lambda_0 (1 + \epsilon/3) \mathbf{I}$ in Eq.~(\ref{cauchy}), and expand in series of $\epsilon$, we derive $\boldsymbol{\sigma} = \big[\Gb(\lambda_0^{-2} - 1/3) + \lambda_0^{-9} \kT/v_\mathrm{w}\big]\epsilon\mathbf{I}$. Assuming $\lambda_0 \gg 1$, we obtain $\lambdab \approx -\Gb + \lambda_0^{-9}\kT/v_\mathrm{w}$. Replacing $\lambda_0^{-3}$ with experimentally measurable quantities using (\ref{initswell}), we finally obtain
\begin{gather}
\frac{\lambdab}{\Gb}= \frac{2 \nub}{(1-2\nub)} = 2 + \frac{\kT/v_\mathrm{w}}{\Gb}\phi_0^3. \label{eq_poisson}
\end{gather}

{\bf Estimation of $\nub$ and comparison to the neo-Hookean model: }
Note that the right hand side of Eq. (\ref{eq_poisson}) is always larger than 2, which corresponds to a Poisson's ratio larger than $1/3$. Our modeling of the biofilm matrix as a swollen polymer network thus leads to the conclusion that biofilms growing on agar behave as almost incompressible hyperelastic materials. Using the experimentally measured values of biofilm shear modulus $\Gb$ and the equilibrium dextran concentration $\phi_0$ \cite{yan2018bacterial}, along with $\kT = 4 \times 10^{-21}$J, and $v_\mathrm{w} = 3\times10^{-29}\text{m}^3$, we can calculate $\nub\approx 0.37$ for low agar concentration (0.6\% agar, $\phi_0$ = 0.02, $\Gb$ = 1.1kPa) and  $\nub\approx 0.46$ for high agar concentration (1.5\% agar, $\phi_0$ = 0.045, $\Gb$ = 1.4kPa).

In our chemo-mechanical model, we proposed to use a neo-Hookean constitutive model $\Psi (\Fe) = \frac{\Gb}{2}(\text{tr}(\Fe^T\Fe)-3-2\ln \Je) + \frac{\lambdab}{2}(\ln \Je)^2$ for biofilm material, whereas here we presented a model motivated by the thermodynamics of swelling. To end the discussion on the constitutive relation, we shall compare the resulting strain energy density of the two models. Note that Eq.~(\ref{eq_strain_energy_hydrogel}) gives the strain energy density per unit volume in the dry polymer network configuration. Thus, to compare with Eq.~(\ref{Eq:Energy_neoHookean}), we rewrite Eq.~(\ref{eq_strain_energy_hydrogel}) to give the strain energy density per unit volume in the equilibrium swollen configuration, i.e. $\Psi(\Fe) = J_0^{-1}W(\lambda_0\Fe) = \frac{1}{2}\lambda_0^{-1}Nk_\text{B}T\text{tr}(\Fe^T\Fe) + \Psi^{J}(\Je) + \text{constant}$. Here, we have explicitly separated $\Psi(\Fe)$ into three parts: the first term is identical to the $\Gb\text{tr}(\Fe^T\Fe)/2$ term in $\Psi$ of the neo-Hookean model; the third term is a constant independent of $\Fe$, and does not contribute to the stress strain relation; the second term is only a function of $\Je$, and reads
\begin{align}
 \Psi^{J}(\Je) &=-J_0^{-1}Nk_\text{B}T\ln(\Je) + \frac{k_\text{B}T}{v_\mathrm{w}}\Big[(\Je - J_0^{-1})\ln(1-J_0^{-1}\Je^{-1}) - J_0^{-2}\Je^{-1}/2\Big] - \frac{\mu_\text{w}}{v_\mathrm{w}}(\Je - J_0^{-1})\notag\\
 &= \Gb\Big\{-\lambda_0^{-2}\ln(\Je) + \frac{k_\text{B}T/v_\mathrm{w}}{\Gb}\Big[(\Je - J_0^{-1})\ln(1-J_0^{-1}\Je^{-1}) - J_0^{-2}\Je^{-1}/2 + \phi_0^3(\Je-J_0^{-1})/3\Big]\Big\}. \label{Eq:Energy_Hydrogel}
\end{align}
Similarly, we obtain, for the neo-Hookean model, $\Psi^J=\Gb\{-\ln \Je+0.5(\lambdab/\Gb)(\ln \Je)^2\}$ where $\lambdab/\Gb$ can be related to the polymer network model via Eq.~(\ref{eq_poisson}). We plot $\Psi^J(\Je) - \Psi^J(1)$ for the two models in Fig.~\ref{Fig:hydrogel}. We find that the two models relate isotropic compression to energy density in a similar way.

\subsection{\label{sec_discussion_friction}Model discussion: A microscopic model for surface friction}
\begin{figure}[t]
\centering
\includegraphics[width=.6\textwidth]{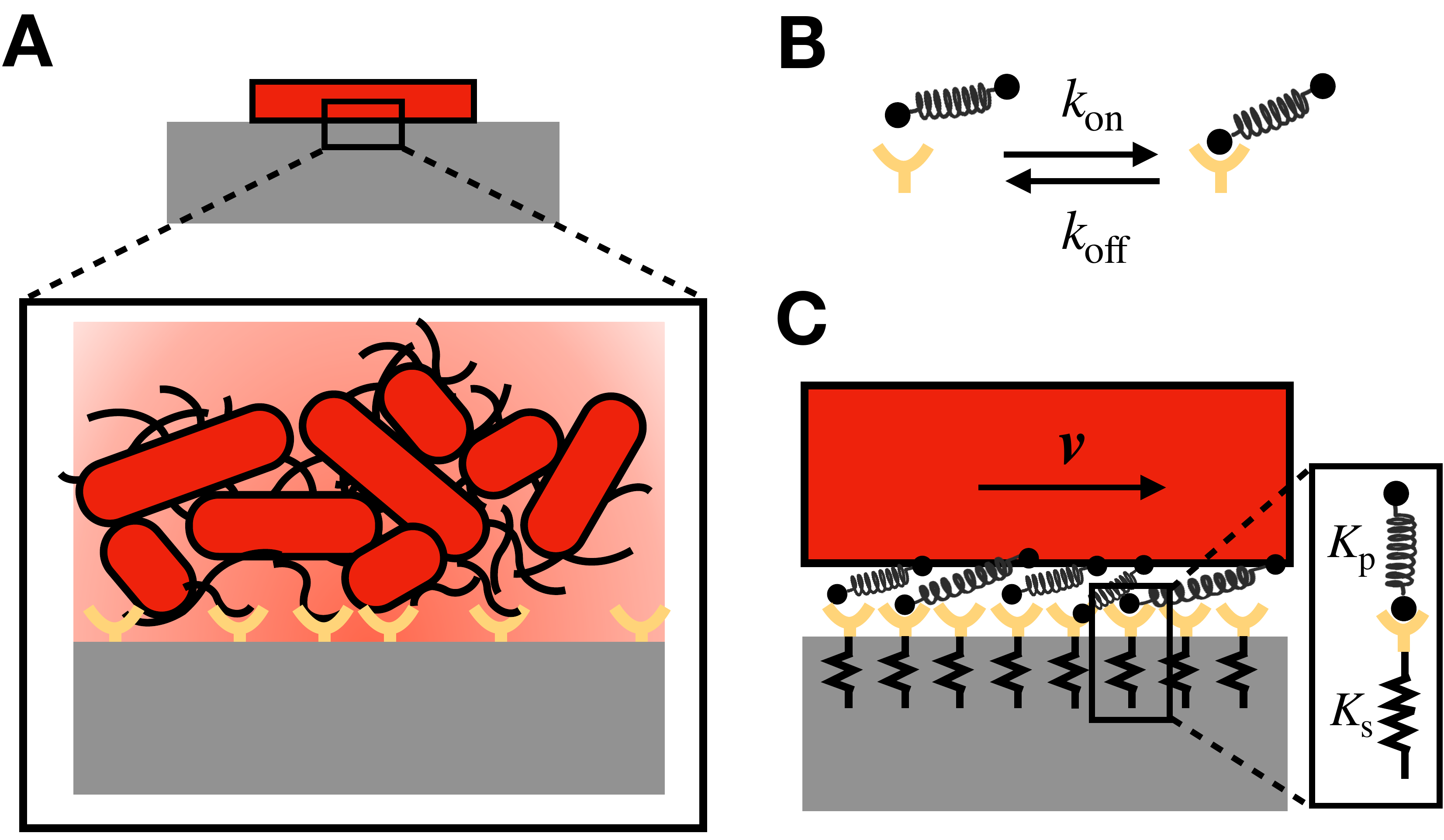}
\captionsetup{justification=justified,labelfont=bf,singlelinecheck =false}
\caption{
{\bf Microscopic model for surface friction between a biofilm and the agar.} ({\bf A}) Close-up schematic of the mechanical interactions between a biofilm and the substrate surface. Bacterial cells deposit biofilm proteins (yellow Y-forks) on the substrate surface, to which extracellular polysaccharides (EPS, black curvy lines) can adhere \cite{fong2007rbmbcdef,berk2012molecular,yan2018bacterial}. ({\bf B}) A two-state toy model for the binding/unbinding kinetics of biofilm polymers and proteins. Polymers, represented by springs (black), can be either unbound ({\it left}) or bound to surface-adhered biofilm proteins ({\it right}). Binding and unbinding of the polymer springs occur at rates $k_\mathrm{on}$ and $k_\mathrm{off}$, respectively. ({\bf C})~Schematic of the stick-slip process underlying surface friction between a biofilm and the agar substrate. As the biofilm expands relative to the substrate, polymer springs that are bound to the substrate surface (via biofilm-derived proteins) become stretched and they deform the substrate. This activity gives rise to the frictional forces that resist biofilm expansion. Inset: a simplified representation of a polymer bound to the elastic substrate showing a polymer spring (stiffness $K_\mathrm{p}$) connected in series with a substrate spring (stiffness $K_\mathrm{s}$).
}
\label{Fig:friction}
\end{figure}

{\bf Stick-slip mechanism of surface friction: } Adhesion and friction are closely related in many natural processes \cite{tian2006adhesion,kweon2011friction}. We assume that the friction between biofilm and agar arises from the binding and unbinding of biofilm matrix polymers and/or cell surface polymers with the adhesive biofilm proteins that bacteria deposit on the substrate surface (Fig.~\ref{Fig:friction}). Specifically, we model polymers as “springs” with spring constant $K_\mathrm{p}$. For polymer blobs of radius $R_\mathrm{p}$, the spring constant is $K_\mathrm{p} \sim \kT/R_\mathrm{p} ^2$ \cite{rubinstein2003polymer}. For simplicity, we only consider the horizontal stretch of these springs. The rates of binding ($k_\mathrm{on}$) and unbinding ($k_\mathrm{off}$) at equilibrium are related by $k_\mathrm{off} ^0 = k_\mathrm{on}  \exp(-E_\text{adh}/\kT)$
where $E_\text{adh}$ denotes the energy difference of the bound and unbound state for a single unstretched polymer spring. 
When the biofilm moves relative to the agar substrate at a velocity $v$, the adhered polymer springs become stretched. 
The total displacement of the anchor point of a single spring on the biofilm is $\Delta x = v t$ at time $t$ after binding. The resulting elastic force on this spring is $f_\text{el} = K^\prime \Delta x$ where $K^\prime = K_\mathrm{p} K_\mathrm{s} /(K_\mathrm{p}  + K_\mathrm{s}) $ accounts for the elasticity of both the substrate and the polymer (Fig.~\ref{Fig:friction}). The effective spring constant of the elastic substrate is given by the Boussinesq solution  $K_\mathrm{s}  = \pi E_\mathrm{s}  a/((1+\nu_\mathrm{s} )(2-\nu_\mathrm{s}))$ \cite{landau1989theory},
in which $E_\mathrm{s} $ and $\nu_\mathrm{s} $ denote, respectively, the Young's modulus and the Poisson's ratio of the substrate, and $a$ denotes the radius of the interaction area. The bound polymers ultimately detach from the substrate and this stick-slip process gives rise to the kinetic friction when the biofilm moves relative to agar.

Consider an area element $\Delta A$ in which $N_\mathrm{b}$ polymer springs are attached to the substrate. The kinetic friction per unit area of biofilm can be calculated as $ f = (\Delta A)^{-1}\sum_{i = 1}^{N_\mathrm{b}}K^\prime v t_i = \rho_\text{p} K^{\prime}v \tau_\mathrm{off}  $ where $ \rho_\text{p}=N_\mathrm{b}/\Delta A$ is the average density of polymers bound to the substrate and $\tau_\mathrm{off} = k_\mathrm{off}^{-1}$ is the average lifetime of bonds between the polymers and the substrate. At steady state, $\rho_\text{p} = \rho_\mathrm{p0}\tau_\mathrm{off} /(\tau_\mathrm{off}  + \tau_\mathrm{on} )$, where $\rho_\text{p0} \approx 1/a^2$ denotes the density of all possible binding sites, $\tau_\mathrm{on}  = k_\mathrm{on} ^{-1}$ is the average time needed for a free floating polymer to bind to the surface again after unbinding. Substituting the steady state $\rho_\text{p}$ into the above expression for $f$, the kinetic friction reads
\begin{gather}
f = a^{-2} K^{\prime} v  \frac{\tau_\mathrm{off}^2}{(\tau_\mathrm{off} + \tau_\mathrm{on})}. \label{friction}
\end{gather}

{\bf Estimation of $\bs{\eta}$: } The thermodynamic picture of reversible binding implies that the stretching of a polymer would favor unbinding since the free energy of the bound state is increased by $E_\text{el} = (K^\prime v t)^2 / (2K_\mathrm{p})$, and the unbinding rate of a stretched polymer becomes
\begin{gather}
k_\mathrm{off} = k_\mathrm{off}^0 \exp(E_\text{el} /\kT) = k_\mathrm{on} \epsilon \exp\Big(\frac{{K^\prime}^2/K_p}{2\kT}(vt)^2\Big),
\end{gather}
where we define $\epsilon = \exp(-E_\text{adh}/\kT)$. We next calculate the average lifetime for bound polymers $\tau_\mathrm{off}$. Let $P(t)$ be the probability of remaining bound at time $t$ after initial binding. It is straightforward to show that $P$ follows $dP/dt = -k_\mathrm{off}(t) P $. Thus, $ P(t) = \exp\Big( - \int_0^t k_\mathrm{on} \epsilon \exp \Big(\frac{{K^\prime}^2/K_\mathrm{p}}{2\kT}(vt')^2\Big) dt' \Big),$ and $\tau_\mathrm{off}$ can be readily calculated by $\tau_\mathrm{off} = \int_0^\infty P(t) dt$. 

Let $\xi_T^\prime = (2\kT K_\mathrm{p})^{1/2}/K^{\prime}$ be the characteristic stretch of the polymer springs activated by thermal energy. We consider a dimensionless parameter $\alpha = v\tau_\mathrm{on}/\xi_T^\prime$ and a dimensionless variable $x = \epsilon t/\tau_\mathrm{on}$. To calculate $\tau_\mathrm{off}$ explicitly, one can expand the exponential terms in a series, and obtain $\epsilon \tau_\mathrm{off}/\tau_\mathrm{on} = \int_0^\infty \exp( - x - \sum_{n=1}^\infty\frac{(\alpha/\epsilon)^{2n}}{(2n+1)n!}x^{2n+1}) dx \equiv \phi(\alpha/\epsilon).$
Substituting this relation into (\ref{friction}), we obtain
\begin{gather}
f = \rho_{p0} K^\prime\xi_T^\prime \frac{\alpha \phi^2(\alpha/\epsilon)}{\epsilon(\epsilon + \phi(\alpha / \epsilon))}.
\end{gather}
In general, $f$ has a complex dependence on $v$ (or $\alpha$). However, in the limit $\alpha / \epsilon \ll 1$, i.e. when the binding/unbinding events happen so frequently that the change of activation energy, and its effect on the unbinding rate, can be neglected, we find that $\phi(\alpha / \epsilon) \to 1$, and we find that the friction force takes the form $f = \eta v$ with the drag coefficient
\begin{gather}
\eta = \frac{\rho_\mathrm{p0}K^\prime\tau_\mathrm{on}}{\epsilon(\epsilon+1)}\approx \epsilon^{-1}\rho_\mathrm{p0}K^\prime\tau_\mathrm{on},~~~~\text{for}~\epsilon\ll 1. \label{eq_eta_expression}
\end{gather}

{\bf Parameter estimation: } Finally, we estimate all the parameters in the above analysis  based on experimental results. First, for the effective spring constant of the polymer, $R_\mathrm{p} \sim 10$ nm, then $K_\mathrm{p} \sim 0.1$ pN$\cdot$nm$^{-1}$. For a substrate with shear modulus $G_\mathrm{s} = 10^3$ Pa, if $a \sim 3$ nm, then $K_\mathrm{s} \sim 10^{-2}$ pN$\cdot$nm$^{-1}$; thus, for the agar concentration we are interested in ($G_\mathrm{s} < 3 $ kPa), we expect $K_\mathrm{s} \ll K_\mathrm{p}$ and $K^\prime$ is dominated by $K_\mathrm{s}$ only. In this regime, $G_\mathrm{s}$ is related to $\eta$ by $\eta\propto G_\mathrm{s}$.

To estimate $\tau_\mathrm{on}$, we follow \cite{de1979scaling}, and assume each polymer blob thermally behaves as a hard sphere of radius $R_\mathrm{p}$ fluctuating around its equilibrium position, so $\tau_\mathrm{on} \approx \mu R_\mathrm{p}^3/\kT = (10^{-3}\text{ Pa} \cdot \text{s} \times 10^3 \text{ nm}^3)/(4 \text{ pN}\cdot\text{nm})\sim 10^{-6}\text{s}$, where $\mu$ denotes the viscosity of water. Next, the energetic factor $\epsilon$ can be estimated from the measurements of adhesion energy. From the microscopic picture, the energy required to separate the biofilm from the substrate is $\Gamma = \rho_\mathrm{b0}E_\text{adh}\tau_\mathrm{off}^0/(\tau_\mathrm{off}^0 + \tau_\mathrm{on}) = a^{-2}\kT \frac{(E_\text{adh}/\kT)}{1 + \exp(-E_\text{adh}/\kT)}.$  Experimentally, the measured  $\Gamma$ value is about $ 5 \text{ mN}\cdot\text{m}^{-1}$ \cite{yan2018bacterial}, which leads to $(E_\text{adh}/\kT)\approx 10$. However, because an order 1 difference in $(E_\text{adh}/\kT)$ can lead to 10-fold change in $\epsilon$, we estimate the range of $\epsilon$ to be $10^{-6}\sim 10^{-4}$ .

We then check whether our approximation in deriving the linear dependence on $v$, namely $\alpha \ll \epsilon$, holds true. The thermal length scale we defined, $\xi_T^{\prime}$ can be estimated to be $\xi_T^{\prime}\sim 100$ nm for $G_\mathrm{s} = 10^3$ Pa.  For a typical biofilm expansion velocity $v \sim 3\text{ $\mu$m/min} = 50 \text{ nm/s}$, one finds $\alpha  = v \tau_\mathrm{on} / \xi_T^{\prime} = 5\times10^{-7}<\epsilon$. Thus, the approximation is valid according to the estimated values. Finally, we can calculate the drag coefficient $\eta$ and compute the order of magnitude of the dimensionless control parameter $\xi=\frac{\eta (R_\mathrm{b0}/\tau_0)}{\Gb (H/R_\mathrm{b0})}$. Assuming the shear modulus of the film is $\Gb = 10^3 \text{ Pa}$, we obtain
$\eta \approx \epsilon^{-1} \rho_\mathrm{b0} K^\prime \tau_\mathrm{on} = (10^{4}\sim10^{6}) (0.1\text{ nm}^{-2})(10^{-2} \text{ pN/nm})(10^{-6} \text{s}) = 10^{4}\sim10^{6} \text{ Pa} \cdot\text{s}\cdot\mu\text{m}^{-1}, \text{ and}\\
\xi = \frac{\eta (R_\mathrm{b0}/\tau_0)}{\Gb (H/R_\mathrm{b0})} \approx \frac{(10^{4}\sim10^{6} \text{ Pa} \cdot\text{s}\cdot\mu\text{m}^{-1}) \times (3 \text{ $\mu$m/min})}{10^3\text{ Pa} \times (100\mu\text{m}/3\text{mm})} \approx 1\sim100 $. This estimation is also consistent with the value of $\xi$ we extracted by fitting the experimental velocity profile.

\section{\label{sec_three_stages} Analysis of Biofilm Expansion Dynamics}
\begin{figure}[t]
\centering
\includegraphics[width=.9\textwidth]{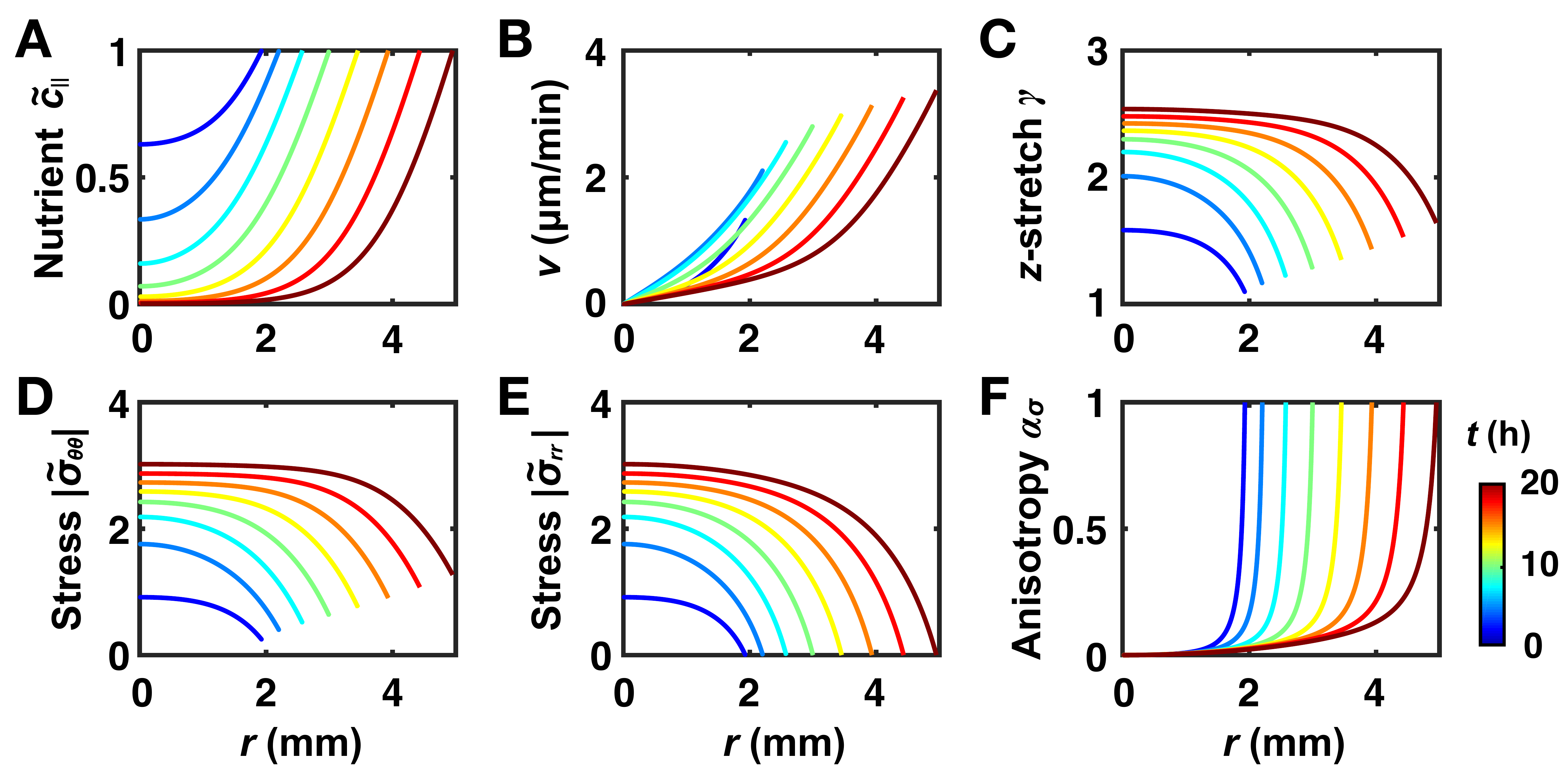}
\captionsetup{justification=justified,labelfont=bf,singlelinecheck=false}
\caption{
{\bf Spatiotemporal evolution of multiple fields in the chemo-mechanical model for biofilm expansion.} Time evolution of ({\bf A}) the reduced nutrient field $\tilde{c}_\parallel$, ({\bf B}) the radial velocity field $v$, ({\bf C}) the vertical stretch $\gamma$, ({\bf D}) the magnitude of the circumferential stress $\tilde{\sigma}_{\theta\theta}$, ({\bf E}) the magnitude of the radial stress $\tilde{\sigma}_{rr}$, and ({\bf F}) the stress anisotropy $\alpha_\sigma=(\tilde{\sigma}_{\theta\theta} - \tilde{\sigma}_{rr})/(\tilde{\sigma}_{\theta\theta} + \tilde{\sigma}_{rr})$ in the chemo-mechanical model for biofilm expansion. Simulation parameters are the same as in Figs. 2{\bf C} and {\bf D}. Color scale indicates time $t$.
}
\label{Fig:simulationsBiofilmExpansionDynamics}
\end{figure}

In both the simulations and the experiments, the expansion velocity of a biofilm shows three distinct stages (Figs.~2D and~\ref{Fig:simulationsBiofilmExpansionDynamics}). In the first stage, the interior remains immobile and only the edge of the biofilm moves outward. In the second stage, the entire biofilm undergoes uniform expansion and the radial velocity scales (almost) linearly with the radial coordinate, i.e. $v_r \propto r$. In the third stage, the expansion at the center slows down again due to nutrient limitation \cite{liu2015metabolic,srinivasan2019multiphase}. To understand the expansion kinematics in the first two stages, we hypothesize that nutrient limitation and non-uniform growth play a minor role in these early stages of biofilm development. Thus, we focus on analyzing the expansion dynamics of a biofilm that is undergoing uniform isotropic growth in the presence of surface friction. 

\subsection{Uniform biofilm growth}
Due to rotational symmetry, the force-balance equation (\ref{eq_ndim_fb_initial}) can be expressed in polar coordinates as
\begin{gather}
     J_\parallel\xi \Big(\frac{\partial \tilde r}{\partial  \tau}\Big)_{\tilde R_0} = \partial_{\tilde R_0}(\bs{\tilde S_{0,\parallel}})_{rr}+\frac{(\bs{\tilde S_{0,\parallel}})_{rr}-(\bs{\tilde S_{0,\parallel}})_{\theta\theta}}{\tilde R_0},\label{eq_force_balance}
\end{gather} 
where $J_\parallel = \partial (\tilde r^2)/\partial (\tilde R_0^2)$ measures the change of area from the initial to the current configuration and $(\bs{\tilde  S_{0,\parallel}})_{r\theta}=0$ due to symmetry. Note that due to the symmetry of the model, we do not explicitly distinguish between the radial and circumferential directions in the initial coordinates and those in the current coordinates.

Consider uniform and isotropic growth $\Fg_{,\parallel} \equiv e^\tau\mathbf{I}_\parallel$. For simplicity, we focus on discussing the behavior of an incompressible material whose dimensionless Cauchy stress is $\bs{\tilde \sigma}_\parallel = \Fep\Fep^T- \gamma^2 \mathbf{I}_\parallel$, where $\gamma=1/\Jep$ and the stress measure in the initial configuration is  $\bs{\tilde S_{0,\parallel}}=\bs{\tilde \sigma}_\parallel \cdot \Ft_\parallel^{-T} J_\parallel \gamma$.
Axisymmetry yields the following expressions for  $\Ft_\parallel$ and $\Fe_{,\parallel}$ in polar coordinates:
\begin{gather}
    \Ft_\parallel = \begin{pmatrix} \partial_{\tilde R_0} \tilde r & 0 \\ 0& \tilde r/\tilde R_0 \end{pmatrix},\text{ and  }\Fe_{,\parallel} =\Ft_\parallel
    \cdot\Fg_{,\parallel}^{-1} =  \begin{pmatrix} e^{-\tau}\partial_{\tilde R_0} \tilde r & 0 \\ 0& e^{-\tau}\tilde r/\tilde R_0 \end{pmatrix}, \label{eq_Fe_analysis}
\end{gather}
where the first (second) row and column denote the radial (circumferential) component. Substituting these expressions into the constitutive relation, the force-balance equation then becomes a partial differential equation (PDE) for the unknown function $\tilde r(\tilde R_0,\tau)$. It is difficult to analytically solve this equation. Nevertheless, to gain insight into the first two stages of biofilm expansion, we analyze the short time and long time expansion kinematics of Eq.~(\ref{eq_force_balance}).

\subsection{Stage I: Stress accumulation}
For the initial stage of biofilm growth, when the displacement $\tilde u$ is small compared to the biofilm radius, we can rewrite $\tilde r(\tilde R_0,\tau)$ as $\tilde R_0 + \tilde u(\tilde R_0,\tau)$ and expand the force-balance equation in powers of $\delta = \tilde u(\tilde R_0,\tau)/\tilde R_0 \ll 1$. This yields, to the leading order,
\begin{gather}
    \partial_{\tilde R_0}(\bs{\tilde S_{0,\parallel}})_{rr} = (3e^{6\tau}+1)\tilde u''+2e^{6\tau}(\tilde u/\tilde R_0)^\prime,\qquad \tilde R_0^{-1}\big[(\bs{\tilde S_{0,\parallel}})_{rr}-(\bs{\tilde S_{0,\parallel}})_{\theta\theta}\big] = (e^{6\tau}+1)(\tilde u'/\tilde R_0-\tilde u/\tilde R_0^2),\quad\text{ and }\\
    \xi \dot{\tilde u} = (3e^{6\tau} + 1)\big(\tilde u^{\prime\prime} + \tilde R_0^{-1}\tilde u^\prime-\tilde R_0^{-2}\tilde u\big)\label{Eq:StageI}.
\end{gather}
The boundary conditions are $\tilde u|_{\tilde R_0=0}\equiv 0$ and $(2\tilde u'+\tilde u/\tilde R_0)|_{\tilde R_0=1}\approx 3\tau$ for $\tau\ll1$, the latter deriving from the stress-free boundary condition at the edge ($(\bs{\tilde S_{0,\parallel}})_{rr}|_{\tilde R_0=1}=0$). 

In the early stage of biofilm expansion, the radial expansion velocity is negligible in the {\it interior} of the biofilm (denoted by zone I) because the internal stress is insufficient to drive the bulk of the biofilm to expand against friction. Clearly, $\tilde u = 0$ is one possible solution for the interior displacement field. By contrast, the {\it edge} of the biofilm is able to move outward (denoted by zone II) from the beginning due to the existence of a stress-free boundary. Thus, we seek for another solution in zone II (similar to the boundary layer theory in fluid mechanics) to connect the immobile core to the moving edge.

\begin{figure}[t]
\centering
\includegraphics[width=.7\textwidth]{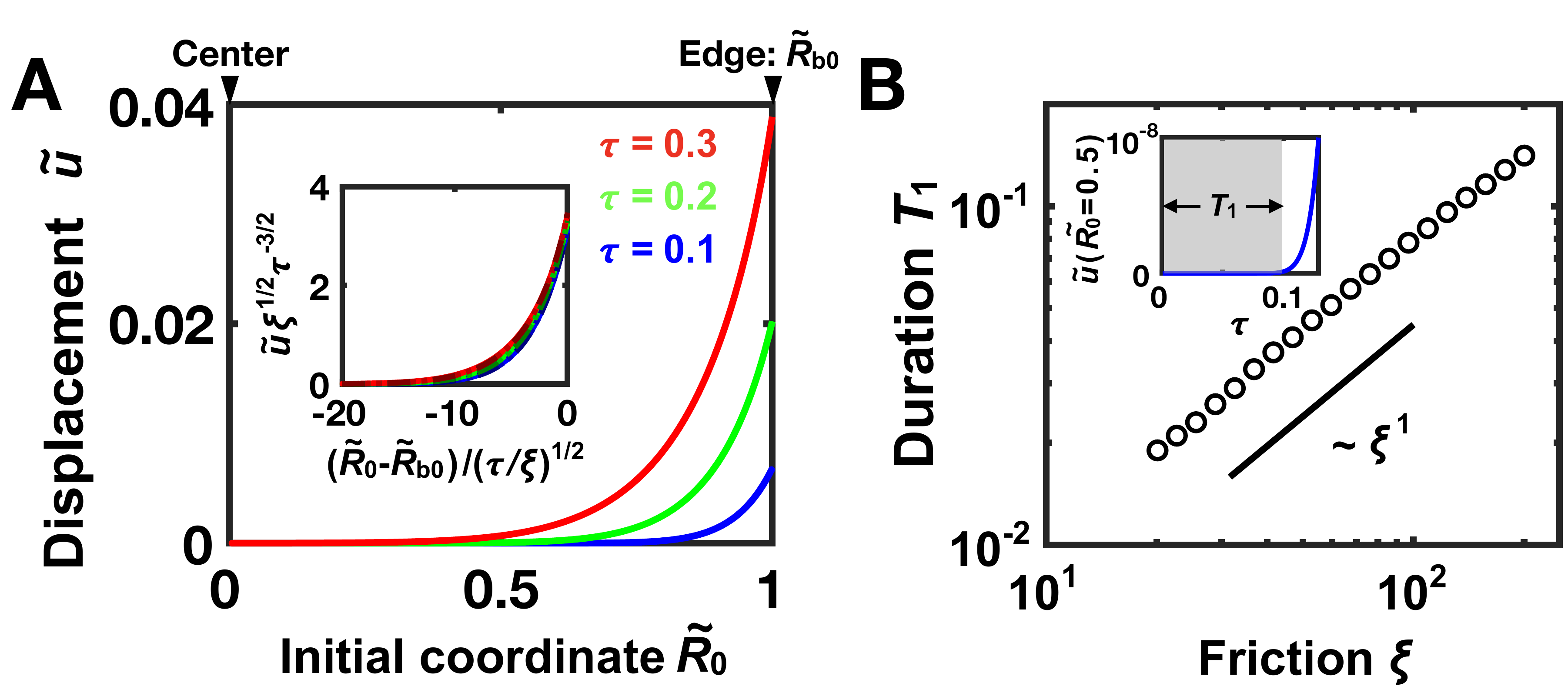}
\captionsetup{justification=justified,labelfont=bf,singlelinecheck=false}
\caption{
{\bf Biofilm expansion dynamics in the initial stage.} ({\bf A}) The radial displacement field $\tilde{u}$ plotted against the radial coordinate in the initial reference configuration, denoted by $\tilde{R}_\mathrm{0}$, at dimensionless time $\tau =$ 0.1 (blue), 0.2 (green), and 0.3 (red). $\tilde{R}_\mathrm{b0} = 1$ denotes the initial biofilm radius. The dimensionless friction parameter $\xi = 50$. Inset: the displacement fields $\tilde{u}$ computed for $\xi =$ 50 (solid curves), 100 (dotted curves), and 150 (dashed curves) at time $\tau =$ 0.1 (blue), 0.2 (green), and 0.3 (red) all collapsed onto a master curve corresponding to a self-similar solution $\tilde{u}(\tilde{R}_0)\approx\xi^{-1/2}\tau^{3/2}g(\frac{\tilde{R}_{B0}-\tilde{R}_0}{(\tau/\xi)^{1/2}})$, where $g$ is a function given by Eq. (\ref{Eq:gFunc}).
({\bf B}) The dimensionless time duration of the first stage of biofilm expansion $T_1$ (gray region), which ends when the material point at $\tilde{R}_0 = \tilde{R}_\mathrm{b0} /2 = 1/2$ starts to move outwards, i.e., the time at which $\tilde{u}(\tilde{R}_0 = 1) > 0$ (illustrated in the inset for $\xi = 100$), versus the dimensionless friction parameter $\xi$. The black line indicates a slope of 1 on a log-log scale.
}
\label{Fig:stageI}
\end{figure}

Analytically, in the short time limit, $\tau\ll 1$, the equation becomes $\xi \dot{\tilde u} = (4 + 18\tau)\big(\tilde u^{\prime\prime} + \tilde R_0^{-1}\tilde u^\prime-\tilde R_0^{-2}\tilde u\big)$, and the boundary at the edge $\tilde R_0 = 1$ is $(2\tilde u'+\tilde u)|_{\tilde R_0=1} = 3\tau$. We seek for solution in the form $\tilde u = \tau^n f(\frac{1-\tilde R_0}{\tau^m})$ where the shape function $f(x)$ is defined on $[0,+\infty)$ with $f(x\to\infty) \equiv 0$ to describe the velocity profile in the transition region. Let $x =(1-\tilde R_0)/\tau^m $. The time derivative of $\tilde u$ is given by
\begin{gather}
    \dot{\tilde u} = n\tau^{n-1} f(x) -m\tau^{n-1}xf'(x)\sim \tau^{n-1}.
\end{gather}
Similarly, we obtain that the R.H.S. of Eq.~(\ref{Eq:StageI}) is given by $4\tilde u'' = 4\tau^{n-2m}f''$ to leading order in $\tau$. Comparing the exponents of $\tau$ we find 
\begin{gather}
    m = \frac{1}{2}, \text{ and }\quad  \frac{4}{\xi } \frac{d^2 f}{d x^2} + \frac{x}{2}\frac{df}{dx} - nf = 0.
\end{gather}
The same analyses can be applied to the boundary condition at $\tilde R_0 = 1$, which yields $n = 3/2$, and $\frac{d f}{d x}|_{x=0} = -3/2$. Let $\bar{x} = \xi^{1/2} x$, and $g(\bar{x}) = \xi^{1/2} f(x)$. We find that the scaling function $g$ follows an ordinary differential equation independent of $\xi$, i.e.
\begin{gather}
    8\frac{d^2 g}{d \bar{x}^2} + \bar{x}\frac{d g}{d \bar{x}} - 3g = 0, \text{ and }\begin{cases}\frac{d g}{d \bar{x}}|_{\bar{x}=0} = -3/2, \\ \, \\ g(\bar{x}\to\infty) = 0.\end{cases} \label{Eq:gFunc}
\end{gather}
The exact solution of the ODE above contains an error function and is not of particular interest. However, from the definition $f(x) = \xi^{-1/2}g\big(\xi^{1/2}x\big)$, we can rewrite $\tilde u(\tilde R_0)$ as $\tilde u(\tilde R_0)\approx\xi^{-1/2}\tau^{3/2}g(\frac{1-\tilde R_0}{(\tau/\xi)^{1/2}})$. This means that the displacement/velocity profile assumes a self-similar solution. Furthermore, since the velocity field depends on $\tau$ only through $(\tau/\xi)^{1/2}$, this suggests that the time a biofilm spends in the first stage $T_{1}$ is proportional to $\xi$. Both the self-similar ansatz and the dependence of $T_1$ on $\xi$ are verified in the simulation,  where $T_1$ is defined as the time when $\tilde u(1/2)$ becomes larger than $10^{-12}$ (see Fig.~\ref{Fig:stageI}). To calculate the stress accumulation in stage I, we note that the inner part of the biofilm does not move, which, given the isotropic growth $\Fg = \lamg \mathbf{I}_\parallel$, implies a compensating isotropic in-plane deformation, i.e. $(\Fe_{,\parallel})_{rr} = (\Fe_{,\parallel})_{\theta\theta} \approx e^{-t}$. The Cauchy stress accumulates as $|\sigma_{rr}| = |\sigma_{\theta\theta}| \sim 6t$ when $t\ll 1$.

\subsection{Stage II: Isotropic expansion}
As the biofilm continues to grow, internal stresses build up, and the mobile region becomes larger. Ultimately, the entire biofilm expands outward.
In this stage, $\tilde r>\tilde R_0$ and both $\Frr$ and $\Ftt$ become larger than 1. Eq.~(\ref{eq_Fe_analysis}) implies that the accumulation of elastic strain is partially relaxed by biofilm expansion. Therefore, we expect that the stress will increase at a rate slower than material production. 

To understand the biofilm expansion kinematics in this regime, we focus on the long time asympoptic behavior of Eq.~(\ref{eq_force_balance}), i.e. $\tau \gg T_{1}$ and $\tau\gg 1$. Since $(\Fep)_{rr/\theta\theta}=\lamg^{-1}F_{\parallel,rr/\theta\theta}$, we expect that the isotropic stress $\gamma^2 = [(\Fep)_{rr}(\Fep)_{\theta\theta}]^{-2}$ dominates over the $\Fep\Fep^T$ term in Eq.~(\ref{eq_ndim_constitutive}) for the bulk of the biofilm for $\tau\gg 1$ provided that $\lamg^{-1}\ll 1$. Neglecting the contribution from $\Fep\Fep^T$, we derive the approximate force-balance equation to be
\begin{gather}
    \xi\dot{\tilde r} = \tilde R_0^{-1}\frac{e^{6\tau}}{\Frr^3\Ftt^3}(\Ftt^{-1}-\Frr^{-1}) - \Frr^{-1}\Ftt^{-1}\partial_{\tilde R_0}\Big(\frac{e^{6\tau}}{\Frr^3\Ftt^2}\Big),\label{eq_fb_longtime}
\end{gather}
where $\Frr = \partial_{\tilde R_0} \tilde r =\tilde r^{\prime}$ and  $\Ftt= \tilde r/\tilde R_0$. Eq.~(\ref{eq_fb_longtime}) is to be solved with the boundary conditions at the center $\tilde r|_{\tilde R_0=0} = 0$ and at the edge $\tilde r(\tilde r^\prime)^2|_{\tilde R_0=1} = e^{3\tau}$. Note that the boundary condition at the edge follows from $(\bs{\tilde S_\parallel})_{rr}=0$, where the $\Fep\Fep^T$ term is kept. Motivated by the observations in the simulations, we test a uniform expansion ansatz $\tilde r^\mathrm{B}(\tilde R_0,\tau) = (\beta_1 \tilde R_0 +\beta_2\tilde R_0^3)e^{\omega \tau}$ for the bulk part of the biofilm, assuming $\beta_2 \tilde R_0^2 \ll 1$ (the $\tilde R_0^2$ term is absent, because it would produce a constant term on the R.H.S. of Eq.~(\ref{eq_fb_longtime})). Note that the form of $\tilde r^\mathrm{B}$ naturally gives rise to the relation $\tilde v_r = \partial_\tau \tilde r^\mathrm{B} \propto \tilde r^\mathrm{B}$. Inserting the ansatz for $\tilde r^\mathrm{B}$ into Eq.~(\ref{eq_fb_longtime}), we find, to  leading order in $\tilde R_0$, $\xi \omega \beta_1 \tilde R_0 e^{\omega \tau} = 24\beta_2\beta_1^{-8}\tilde R_0e^{6 -7\omega \tau}$. Comparing the coefficient and the exponent leads to $\omega = 3/4$ and $\beta_2 = \frac{\xi\omega}{24}\beta_1^9= \frac{\xi\beta_1^9}{32}$. 

However, the ansatz $\tilde r^\mathrm{B}$ does not satisfy the boundary condition at the edge $\tilde R_0 = 1$, so we need to find another test solution near the edge. Consider an annulus $1 - \Delta \tilde R \leq \tilde R_0\leq 1$, where $\Delta \tilde R\ll 1$ denotes the width of the transition zone (to the bulk solution) at the edge. Since the radius of the biofilm is largely determined by the displacement in the bulk we expect that the edge solution will have the same exponent as the bulk solution. Specifically, we use the ansatz $\tilde r^\mathrm{E} = \bar{\beta_1} e^{3\tau/4}$ and $(\tilde r^\prime)^\mathrm{E} = \bar{\beta_1}^{-1/2} e^{9\tau/8}$ at the edge, the latter derived from the boundary condition. Here, the superscript E denotes functions evaluated at the boundary $\tilde R_0 = \tilde R_\mathrm{b0} = 1$. 

\begin{figure}[t]
\centering
\includegraphics[width=.6\textwidth]{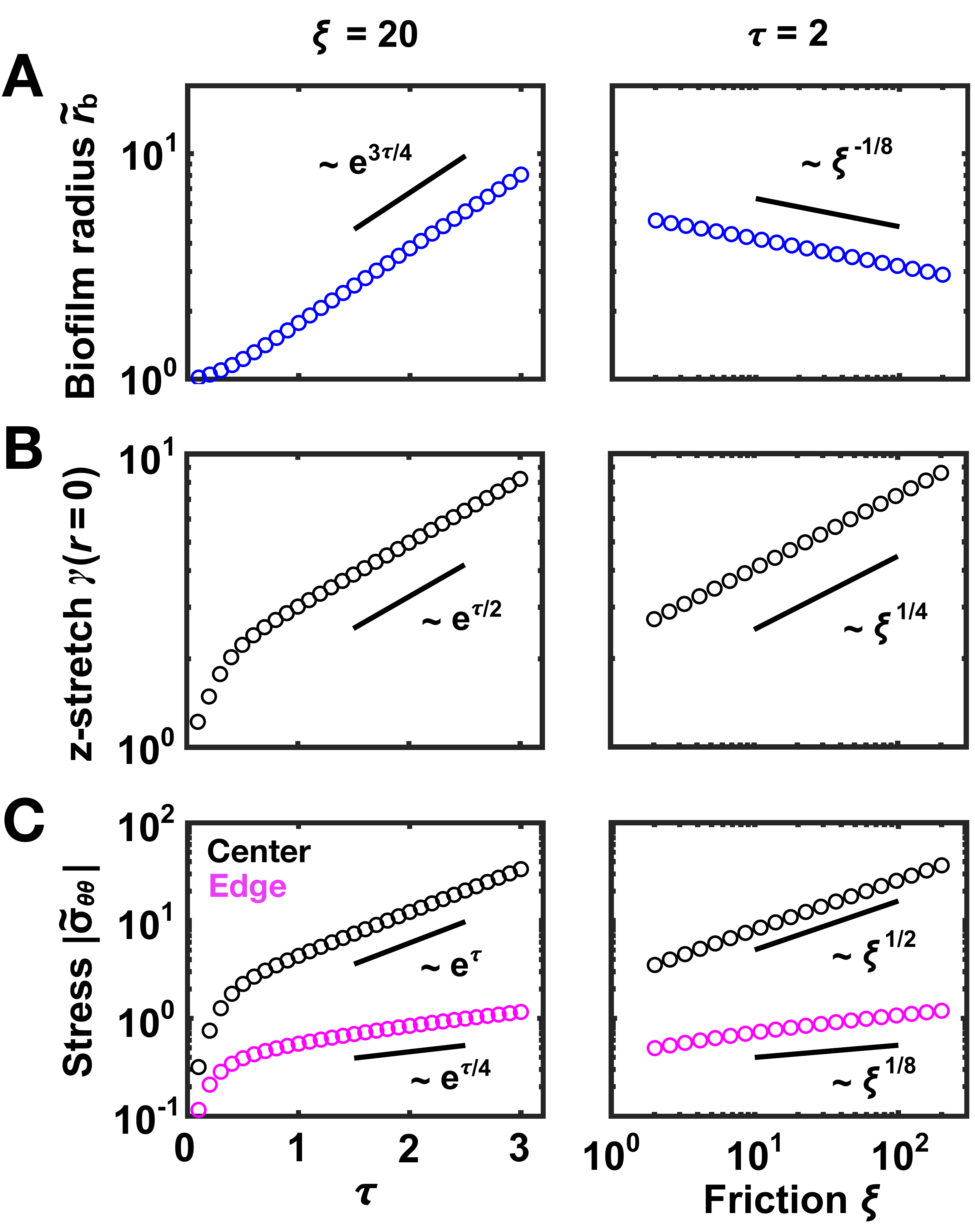}
\captionsetup{justification=justified,labelfont=bf,singlelinecheck=false}
\caption{
{\bf Validation of the scaling predictions for the second stage of biofilm expansion.}({\bf A}) The biofilm radius $r_\mathrm{b}$, ({\bf B}) the vertical stretch at the center of the biofilm $\gamma (r = 0)$, and ({\bf C}) the magnitude of the circumferential stress $\tilde{\sigma}_{\theta\theta}$ at the center (black) and edge (magenta) of the biofilm plotted versus the dimensionless time $\tau$ ({\it left}) and the dimensionless friction parameter $\xi$ ({\it right}). The black lines indicate the predicted slopes on log-lin ({\it left}) or log-log ({\it right}) scales. $\xi = 20$ in {\it left} and $\tau = 2$ in {\it right}. Simulation results with uniform growth rates are used to approximate the second stage of biofilm expansion, in which nutrient-dependent non-uniform growth plays a minor role. 
}
\label{Fig:stageII}
\end{figure}

To connect the two solutions, we leverage different continuity constraints at $\tilde R_0 = 1 - \Delta \tilde R$ to compute the undetermined coefficients. Assume $\Delta \tilde R$ takes the form $\Delta \tilde R = \overline{\Delta \tilde R}~\exp({\omega_\mathrm{E}\tau})$. First, the two solutions should give the same value of $\tilde r$ at the intersection, i.e. $\tilde r^\mathrm{B}|_{\tilde R_0 = 1-\Delta \tilde R} = \tilde r^\mathrm{E} - (\tilde r^\prime)^\mathrm{E} \Delta \tilde R$, from which one can obtain the edge time exponent $\omega_\mathrm{E} = -3/8$ and the prefactor $\overline{\Delta \tilde R}$ satisfies equation $\bar{\beta}_1 - \bar{\beta}_1^{-1/2}\overline{\Delta \tilde R} \approx  \beta_1 + \beta_2 = \beta_1 (1 + \frac{\xi\beta_1^8}{32})$. Additional equations for the undetermined coefficients $\bar{\beta}_1, {\beta}_1, \overline{\Delta \tilde R}$ are obtained by considering the continuity of the radial components of stress $(\bs{\tilde S_\parallel})_{rr}$, i.e. $(\bs{\tilde S_\parallel^\mathrm{B}})_{rr}|_{\tilde R_0 = 1-\Delta \tilde R} \approx (\bs{\tilde S_\parallel^\mathrm{E}})_{rr}|_{\tilde R_0 = 1} - \Delta \tilde R \,\partial_{\tilde R_0}({\bs S_\parallel^\mathrm{E}})_{rr}|_{\tilde R_0 = 1}$. Using the ansatz $\tilde r^{B}$ for the bulk we obtain the radial stress component $(\bs{\tilde S_\parallel^\mathrm{B}})_{rr}|_{\tilde R_0 = 1-\Delta \tilde R} \approx - \beta_1^{-5} e^{9 \tau /4}$. The boundary condition at the edge sets $(\bs{\tilde S_\parallel^\mathrm{E}})_{rr}|_{\tilde R_0 = 1}=0$. The derivative of the radial stress component at the edge $\partial_{\tilde R_0}({\bs S_\parallel^\mathrm{E}})_{rr}|_{\tilde R_0 = 1} = \frac{3}{4} \xi \overline{\beta_1}^{3/2} e^{21 \tau/8}$ is obtained by using the ansatz for the edge $\tilde r^{E}$ in Eq.~(\ref{eq_force_balance}) at $\tilde R_0 = 1$. For scaling analysis, we assume that all the prefactors have the power-law dependence on the control parameter $\xi$ 
, which yields $\bar{\beta_1}\sim \beta_1 \sim \xi^{-1/8}$ and $\overline{\Delta R} \sim \xi^{-3/16}$. In summary, we obtain the following scaling formulas,
\begin{align}
    \text{Bulk:}&\quad \tilde r^\mathrm{B}(\tilde R_0) \sim \xi^{-1/8}e^{3\tau/4} \tilde R_0, \quad  (\tilde r^\prime)^\mathrm{B} \sim \xi^{-1/8}e^{3\tau t/4},\\
    \text{Edge:}&\quad \tilde r^\mathrm{E} \sim \xi^{-1/8}e^{3\tau/4}, \quad\quad\quad\quad (\tilde r^\prime)^\mathrm{E} \sim \xi^{1/16}e^{9\tau/8},\\
    \text{Transition layer:}&\quad \Delta \tilde R \sim  \xi^{-3/16}e^{-(3/8)\tau}.
\end{align}
By substituting the relations above into the expressions for Cauchy stresses $\tilde \sigma_{rr}$, $\tilde \sigma_{\theta\theta}$ and the stretch factor $\gamma$, we find the following asymptotic solution: 
\begin{align}
    &\tilde\sigma^\mathrm{B}_{rr} \approx \tilde\sigma^\mathrm{B}_{\theta\theta} \sim \xi^{1/2}e^\tau ~ \text{in the bulk of the biofilm},\\
    &\tilde\sigma^\mathrm{E}_{\theta\theta}\sim \xi^{1/8}e^{\tau/4}\quad\quad \, \text{near the edge of the biofilm, and}\\
    &\gamma^\mathrm{B} \sim \xi^{1/4}e^{\tau/2} \quad \quad\  \text{for the increase of the biofilm thickness}.
\end{align}
All of the scaling relations derived above are verified by the simulations (Fig.~\ref{Fig:stageII}).

\subsection{Stage III: Nutrient-limited growth}
In the final stage of biofilm expansion, growth slows down in the central part of the biofilm due to nutrient limitation. 
We first argue that the resulting edge growth mode results in an approximately linear increase of biofilm radius with time. To see this, we consider the kinematics of growth of an incompressible flat biofilm, and neglect any nutrient-independent growth. Integrating the material conservation relation $\lamg^2 R_0 dR_0 = \gamma r dr$, we calculate the radial coordinate of a material point, labeled $R_0$ in the initial configuration, from
\begin{gather}
    r^2/2 = \int_0^{R_0}{\gamma}^{-1}{\lamg^2} R_0dR_0. \label{eq_area_cons}
\end{gather}
Taking the derivative of Eq.~(\ref{eq_area_cons}) with respect to time, and using Eq.~(\ref{eq_ndim_growth}) and the material conservation relation to rewrite the integral in the current coordinates, we obtain
\begin{gather}
    r\dot{r}= 2\int_0^r\phi(c_\parallel(\rho))\rho d\rho - \int_0^r\gamma^{-1}(\partial_\tau\gamma)\rho d\rho,
\end{gather}
which yields the velocity field in the growing biofilm. In particular, setting $r = r_\mathrm{b}(\tau)$, one finds how the biofilm radius $r_\mathrm{b}$ increases over time. In the edge growth regime, $\phi\approx 1$ for $r_\mathrm{b} - a_c<r <r_\mathrm{b}$ and $\phi\approx 0$ for $r<r_\mathrm{b} - a_c$, so the first term of the R.H.S. of the equation can be estimated to be $2a_cr_\mathrm{b}$. To estimate the upper bound on the second term of the R.H.S. of the equation, we consider a scenario where the local biofilm growth contributes solely to the increase of $\gamma$ with no in-plane expansion. In this limiting case, $\gamma^{-1}(\partial_t\gamma) =\partial_t(\ln\gamma) = \text{const}.$ in the active growing annulus. Therefore, the second integral should scale at most linearly with $r_\mathrm{b}$. Combining these results, we obtain $r_\mathrm{b}\dot{r}_\mathrm{b} \sim \text{constant} \times r_\mathrm{b}$, or $\dot{r}_\mathrm{b}\sim \text{constant}$. We conclude that edge growth yields a linear increase of biofilm radius with time.\footnote{One can apply a similar analysis to a model with a small amount of residual growth. This yields $\tilde r \sim \exp(\tilde{k}_\mathrm{r} \tau)$ where $\tilde{k}_\mathrm{r} \ll 1$. }

Next, we calculate the accumulation of circumferential stress at the edge in stage III. For simplicity, we consider an incompressible hyperelastic material, i.e. ${\bs{\tilde \sigma}}_\parallel=\Fep\Fep^T- \gamma^2 \mathbf{I}_\parallel$. The stress-free boundary condition $\sigma_{rr} = 0$ yields $(\Fep)_{rr}^2(\Fep)_{\theta\theta} = 1$, where we invoked the incompressibility condition $\gamma = (\Fep)_{rr}^{-1}(\Fep)_{\theta\theta}^{-1}$. Inserting this relation between $(\Fep)_{rr}$ and $(\Fep)_{\theta\theta}$ at the edge into the expression for circumferential stress, we find $\tilde \sigma_{\theta\theta} = (\Fep)_{\theta\theta}^2 - (\Fep)_{\theta\theta}^{-1}$.
Since nutrients are assumed to be always sufficient near the edge in our model, we have $\lamg|_{\tilde r=\tilde r_B} = e^\tau$, and hence ${\Fep}_{\theta\theta}|_{r=r_\mathrm{b}} = \lamg^{-1}|_{\tilde r=\tilde r_\mathrm{b}} \, \tilde r_\mathrm{b}/\tilde R_\mathrm{b0} \sim \tau e^{-\tau}$, where we have used the fact that $\tilde r_\mathrm{b}$ increases linearly with time. In the long time limit, $(\Fep)_{\theta\theta}^2 \ll (\Fep)_{\theta\theta}^{-1}$ at the edge, and the circumferential stress increases as $|\tilde \sigma_{\theta\theta}|_{\tilde r=\tilde r_\mathrm{b}}\sim e^\tau / \tau$ over time. This scaling relation was verified in simulations.

\section{Analysis of The Leading Angle}
In our chemo-mechanical model, we highlight the surface friction as a key factor in determining biofilm expansion kinematics and the spatiotemporal stress distribution. The shear force applied to the bottom of the biofilm also suggests that the leading angle increases with increasing friction/agar concentration. This is indeed the case in the experiments. To understand the quantitative relation between the biofilm leading angle and the kinematic friction, we focus on the growth and deformation of the wedge-shaped edge of a biofilm (whose width $\sim 100~\mu$m is much smaller than the biofilm radius $r_\mathrm{b}\sim 5~$mm). 

\begin{figure}[t]
\centering
\includegraphics[width=.8\textwidth]{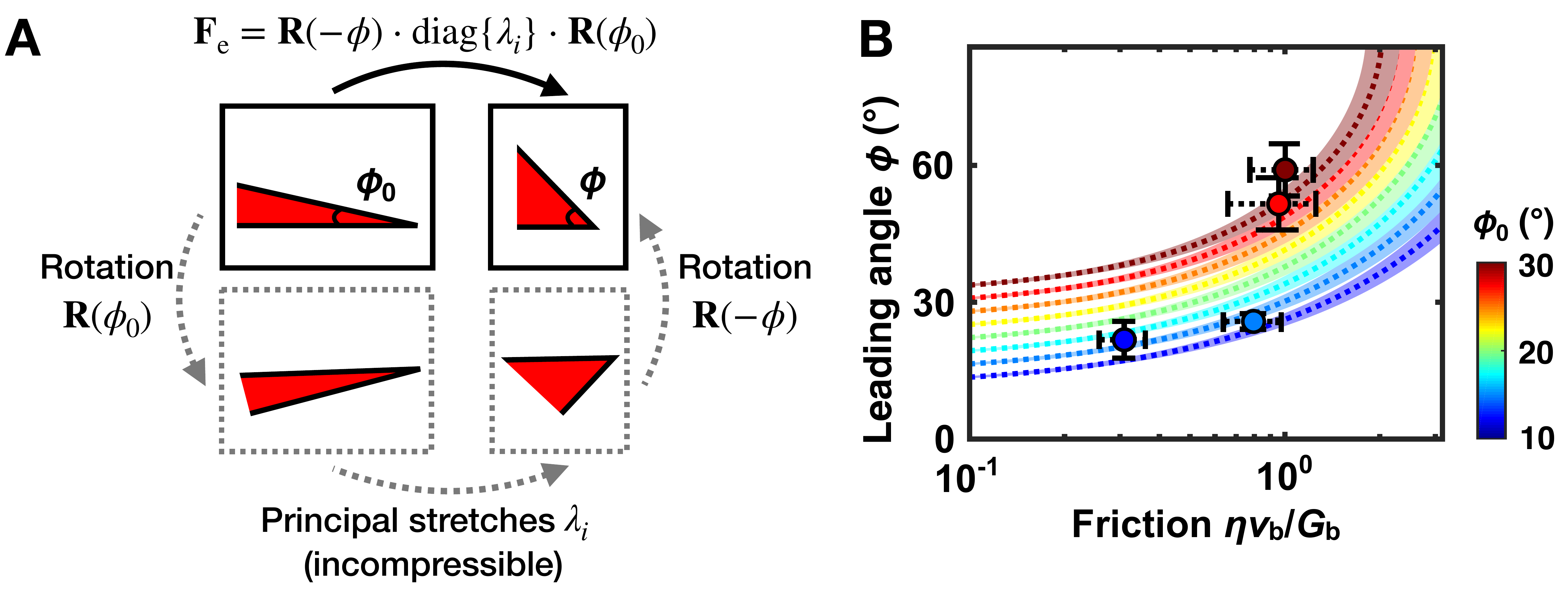}
\captionsetup{justification=justified,labelfont=bf,singlelinecheck=false}
\caption{
{\bf Higher surface friction increases the biofilm leading angle.} ({\bf A}) 2D schematic of the geometric interpretation for the decomposition of $\mathbf{F}_\mathrm{e}$. Surface friction leads to shear deformation and increases the leading angle of the biofilm from $\phi_0$ to $\phi$ ({\it upper row}, solid black arrow). This deformation, denoted by $\mathbf{F}_\mathrm{e}$ can be decomposed into the product of three simple operations (gray dashed arrows): $\mathbf{R}(\phi_0)$ and $\mathbf{R}(-\phi)$ describing rigid body rotations, and $\mathrm{diag}\{\lambda_i\}$ describing the principal shear deformation. ({\bf B}) Theoretical predictions for the biofilm leading angle $\phi$ as a function of the dimensionless friction $\eta v_\mathrm{b}/\Gb$ and initial angle $\phi_0$ (colorbar), where $v_\mathrm{b}$ is the expansion velocity at the edge of the biofilm and $\Gb$ is the biofilm shear modulus. This is the same plot as in Fig.~2E but with error estimation for the theory curves (shaded color bands). Theory predicts the relation between $\phi$, $\phi_0$, and $\eta v_\mathrm{b}/\Gb$, with an undetermined parameter $\mathbf{F}_\mathrm{e,\theta\theta}$ denoting the circumferential compression at the leading front of an expanding biofilm. The circumferential strain $\mathbf{F}_\mathrm{e,\theta\theta} - 1$ will first become increasingly negative due to growth-induced compression, and ultimately it will level off due to the wrinkling instability. We estimate the steady-state $\mathbf{F}_\mathrm{e,\theta\theta}$ to be $0.7\sim0.9$ \cite{yan2019mechanical}, which corresponds to the upper and lower boundaries of the theory color bands.
}
\label{Fig:wedge}
\end{figure}

We start with a 2D case, where an elastic wedge with initial angle $\phi_0$ is placed in the $x$-$z$ plane. The angle remains unchanged in the virtual configuration after isotropic growth.\footnote{Since here we focus on the very edge of the biofilm where the thickness of the biofilm has not yet reached a plateau, we consider the isotropic growth in the full space and ignore the oxygen/nutrient limitation.} A friction force is applied to the bottom surface $z=0$. The analyses below is restricted to the tip region of the wedge so that we can ignore spatial variations.\footnote{Although the analyses hold true up to a linear gradient.} The shear stress in the biofilm can be determined from the boundary condition at the bottom, i.e. $\sigma_{xz}|_{z = 0} = \eta v_\mathrm{b}$ where $v_\mathrm{b}$ is the velocity at the edge of the biofilm. Using the stress-free boundary condition $\bs{\sigma}\cdot \bs{n} = 0$ at the upper surface of the wedge, where $\bs{n}=(\sin \phi, \cos \phi)$ is the normal vector to the upper surface, we derive $\sigma_{xx} = -(\tan\phi)^{-1}\eta v_\mathrm{b}$ and $\sigma_{zz} = - (\tan\phi)\eta v_\mathrm{b}$, where $\phi$ is the leading angle measured in the current configuration. On the other hand, the stress should also be related to the deformation of the biofilm via the constitutive relation (incompressible) $\bs{\sigma} = \Gb \Fe\Fe^T -p\mathbf{I}$. The left Cauchy-Green deformation tensor $\mathbf{B} = \Fe\Fe^T$ is positive definite, and hence can be written as $\mathbf{B}_\text{e}=\mathbf{R}(\beta_1)\begin{pmatrix}\lambda_1^2 & 0 \\ 0& \lambda_1^{-2}\end{pmatrix}\mathbf{R}^T(\beta_1)$, where $\lambda_1$ denotes the principal stretch, $\mathbf{R}$ denotes the 2D rotation parameterized using a single variable $\beta$ denoting the rotation angle in the $x$-$z$ plane, i.e. $\mathbf{R}(\beta)=\begin{pmatrix}\cos\beta & -\sin\beta \\ \sin\beta & \cos\beta \end{pmatrix}$. Therefore, the deformation tensor $\Fe$ can be decomposed into $\Fe = \mathbf{R}(\beta_1)\begin{pmatrix}\lambda_1 & 0 \\ 0& \lambda_1^{-1}\end{pmatrix}\mathbf{R}(\beta_2)$. Expressing $\mathbf{B}_\text{e}$ and $\bs{\sigma}$ in terms of $\beta_1$ and $\lambda_1$, and comparing them with the expression of $\bs{\sigma}$ derived from the force-balance equation, we obtain
\begin{subequations}
\begin{equation}
    \sigma_{xz} = \Gb (\lambda_1^2 - \lambda_1^{-2})\cos\beta_1\sin\beta_1 = \eta v_\mathrm{b}, ~\text{and} \label{eq_angle_xz}
\end{equation}
\begin{equation}
    \sigma_{xx} - \sigma_{zz} = \Gb (\lambda_1^2 - \lambda_1^{-2})(\cos^2\beta_1 - \sin^2\beta_1) = \eta v_\mathrm{b} (\tan\phi-1/\tan\phi). \label{eq_angle_xx_zz}
\end{equation}
\end{subequations}
Dividing Eq.~(\ref{eq_angle_xx_zz}) by Eq.~(\ref{eq_angle_xz}), we obtain $(\tan\beta_1)^{-1} - \tan\beta_1 =\tan\phi - (\tan\phi)^{-1}$, from which we calculate $\beta_1 = -\phi$. Notice that the decomposition of $\Fe$ has a simple geometric interpretation (Fig.~\ref{Fig:wedge}) -- a rotation of $\beta_2$, followed by a principal stretch, followed by another rotation of $\beta_1$. Relating different geometric configurations yields $\beta_2 = \phi_0$ and $\lambda_1^{-2} = \tan\phi/\tan\phi_0$. Substituting for $\beta_1$ and $\lambda_1$ in terms of $\phi$ and $\phi_0$ in Eq. (\ref{eq_angle_xz}), we finally derive
\begin{gather}
    \tan^2\phi = \frac{\tan\phi_0 + \zeta}{1/\tan\phi_0 - \zeta},\quad \text{where } \zeta \equiv\eta v_\mathrm{b} /\Gb.\label{eq_angle_2D}
\end{gather}
In the absence of friction, i.e. $\zeta = 0$, this relation reduces to $\phi = \phi_0$. In the presence of friction, $\phi$ increases with $\phi_0$ for $\zeta\leq(\tan\phi_0)^{-1}$; if the friction is further increased the wedge will bulge out and tumble.

We can apply similar analyses to a 3D wedge in cylindrical coordinates. A correction from the principal deformation in the $\theta$ direction, $\lambda_\theta=\mathbf{F}_{\textrm{e},\theta\theta}$, will be introduced, and $\zeta$ should be replaced by $\lambda_\theta\zeta$ in Eq. (\ref{eq_angle_2D}) to give the final result in 3D. Although the specific value of $\lambda_\theta$ can only be obtained by solving the coupled mechanical growth equations, we note that $\lambda_\theta$ is bounded by the critical strain of mechanical instability, so we estimate $1 > \lambda_\theta \gtrsim 0.7$. Within this regime, the effect of $\lambda_\theta$ on the prediction of $\phi$ is relatively small (Fig.~\ref{Fig:wedge}).

\section{Origin of stress anisotropy}
\begin{figure}[t]
\centering
\includegraphics[width=1\textwidth]{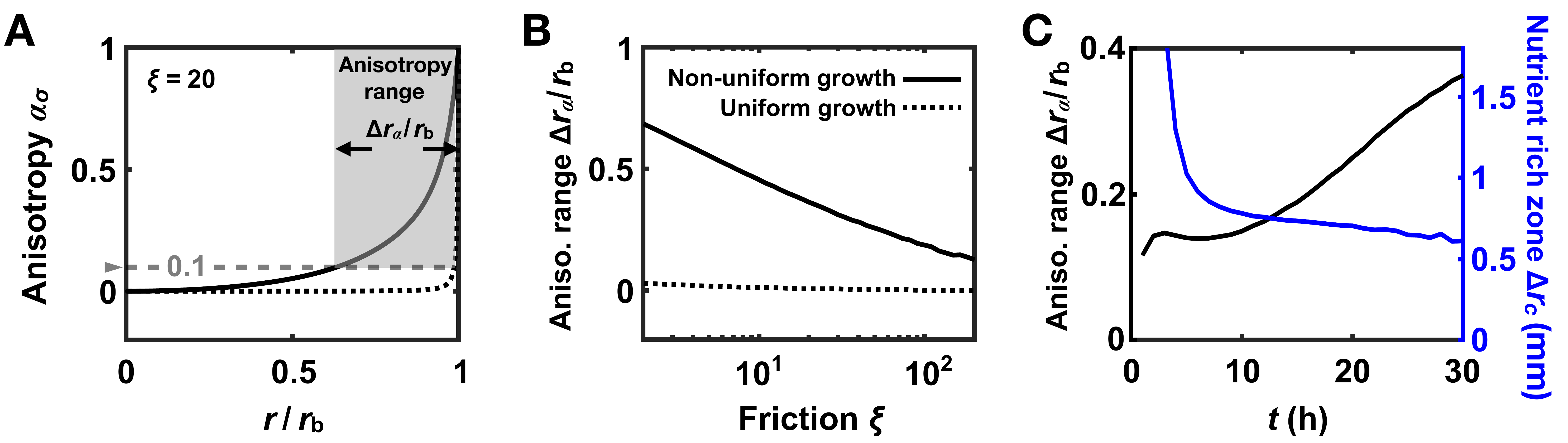}
\captionsetup{justification=justified,labelfont=bf,singlelinecheck=false}
\caption{
{\bf Non-uniform biofilm growth results in anisotropic stress.} ({\bf A}) Stress anisotropy $\alpha_\sigma$, defined as the difference divided by the sum of the two principal stresses, versus the radial coordinate $r$ normalized by the biofilm radius $r_\mathrm{b}$, and ({\bf B}) anisotropy range $\Delta r_\alpha/r_\mathrm{b}$, defined as the normalized range of the biofilm region where $\alpha_\sigma> 0.1$ (illustrated by gray shaded area in {\bf A}), versus the dimensionless friction parameter $\xi$, plotted for the model with non-uniform nutrient-limited growth (solid curve) and uniform growth (dotted curve). Simulation results are shown at a specific time $t = 30$ h. ({\bf C}) Anisotropy range $\Delta r_\alpha/r_\mathrm{b}$, and the width of the nutrient rich zone ($\tilde{c} > 0.5$) $\Delta r_c$, plotted versus time $t$. $\xi = 20$ was used for the simulations in {\bf A} and {\bf C}.
}
\label{Fig:nonuniformGrowth}
\end{figure}

Stress anisotropy plays an important role in determining the morphology of wrinkles. It was previously shown that, for anisotropic stresses, wrinkles are oriented orthogonal to the direction of maximum compressive stress, while for isotropic stresses, wrinkles form zigzag herringbone-like pattern \cite{audoly2008buckling, audoly2008buckling2,cai2011periodic}. In this section, we discuss how nonhomogeneous isotropic growth could result in stress anisotropy. 
As already discussed in Sec. \ref{sec_three_stages}, if the biofilm undergoes homogeneous isotropic growth, the isotropic stress will dominate in the bulk part of the biofilm, i.e. $(\mathbf{ S_{0,\parallel}})_{rr}\approx (\mathbf{S_{0,\parallel}})_{\theta\theta}$. In contrast, non-homogeneous growth could give rise to anisotropic stress in a much larger region of biofilm. We also verify this argument by quantifying the stress anisotropy in the simulation (see Fig.~\ref{Fig:nonuniformGrowth}).

To see why non-homogeneous growth leads to anisotropic stress, one can simply check the stress state of a biofilm if there is no radial elastic deformation, i.e. $r = \int_0^{R_0} \lamg(R_0')dR_0'$ where $ \lamg$ is the growth factor. We confirm that indeed $(\Fep)_{rr} = \lamg^{-1}\partial r/\partial R_0 = 1$ in this state. The circumferential deformation in this state can be calculated by $(\Fep)_{\theta\theta} = (r/ R_0) / \lamg(R_0)  =  \langle \lamg \rangle_{R_0} / \lamg(R_0)$ where $\langle \lamg \rangle_{R_0} = R_0^{-1}\int_0^{R_0}\lamg(R_0')dR_0'$ denotes the average $\lamg$ up to some reference radius $R_0$. Thus, if $\lamg$ in an increasing function of $R_0$, i.e. the production of biofilm material is faster at the edge than at the center (as is the case in {\it V. cholerae}), then $\langle \lamg \rangle_{R_0}  < \lamg (R_0)$ and the biofilm will be under circumferential compression. Although the true stress state is also affected by the boundary condition and force balance as a whole, this simple analysis captures the basic physical picture of how stress anisotropy is generated: compression preferentially in one direction ensures the compatibility of the material that would otherwise break apart due to non-homogeneous growth.

This analysis also immediately points out another possibility of anisotropic stress if the non-homogeneous growth pattern is reversed. If $\lamg$ is an decreasing function of $R_0$, i.e. matrix production is faster in the center of the biofilm than at the rim, then we expect a region where the radial compressive stress is larger than the circumferential stress. Such matrix production pattern is indeed found in other bacterial biofilm formers. See the main text for further discussion of this point.

\section{Coarse-grained Model of Wrinkle Formation\label{sec_wrinkle}}
\subsection{Plane-stress wrinkling model\label{sec_wrinkle_model}}
To investigate how internal mechanical stress shapes the 3D pattern of wrinkles, we develop a plane-stress wrinkling model. Following \cite{holland2017instabilities}, we decompose the elastic deformation tensor $\Fe$ of a wrinkled biofilm into two parts -- a planar compression $\Fe^0$ and a small wrinkling deformation $\Fe^\mathrm{w}$ that describes the out-of-plane undulation. Specifically,
\begin{gather}
\Fe = \Fe^\mathrm{w}\Fe^0 = 
\begin{pmatrix}
\mathbf{I}+\nablap \bs{u} & \bs{u}_{,z}\\
(\nablap w)^T & 1 + w_{,z}
\end{pmatrix}
\begin{pmatrix}
\Ainit & \mathbf{0}\\
\mathbf{0} & \gamma^0
\end{pmatrix},\label{Wrinkle_DefTensor}
\end{gather}
where $\bs{u}$ and $w$ denote, respectively, the in-plane and out-of-plane displacement due to wrinkling. Here $\nablap$ denotes the in-plane derivatives with respect to the spatial coordinates in the deformed flat-film configuration (after $\Fe^0$), which we refer to as $\bs{x}^\prime$ in the following. In this section, we use simplified notation for derivatives of $\bs{u}$ and $w$ -- subscript ``$,ij$'' denoting $\frac{\partial^2}{\partial x_i'\partial x_j'}$ $(i,j = 1,2)$ and subscript ``$,z$'' denoting $\frac{\partial}{\partial z}$. For small deformations and moderate rotations of thin plates it is known that derivatives scale as $\nablap w \sim \bs{u}_{,z} \sim \delta_h$ and $\nablap\bs{u} \sim w_{,z}\sim \delta_h^2$, \cite{audoly2010elasticity} where $\delta_h = h/r_\mathrm{b}\ll 1$ is the aspect ratio of thin biofilm. In our derivations below, we keep all the terms to leading orders in $\delta_h$.

First, according to the plane-stress assumption, we set the normal components of the Cauchy stress to vanish at the lowest order, i.e. $\boldsymbol{\sigma}\cdot\mathbf{n} = \mathbf{0}+  \mathcal{O}(\delta_h^3)$, where $\mathbf{n} \approx (\nablap w, 1)$ denotes the normal to the wrinkled configuration. Substituting Eq. (\ref{Wrinkle_DefTensor}) into the expression for Cauchy stress $\Je\boldsymbol{\sigma} = \Gb\Fe\Fe^T +(\lambdab\ln \Je-\Gb)\mathbf{I}$, we find, to the lowest order in $\delta_h$,
\begin{gather}
    \bs{u}_{,z} = -\nablap w \label{u3_Deriv},\\
    w_{,z} = -\frac{\lambdab}{\lambdab+2\Gb(\gamma^0)^2}\tr(\nablap \bs{u}) - \frac{\lambdab + \Gb (\gamma^0)^2}{\lambdab + 2\Gb(\gamma^0)^2}\lVert\nablap w\rVert^2.
\end{gather} 
Eq.~(\ref{u3_Deriv}) implies that the in-plane displacement takes the form $\bs{u} = \bs{u}_\mathrm{m}-z^\prime \gradw$ where $\bs{u}_\mathrm{m}$ denotes the in-plane displacement of the mid-plane of the biofilm. Note that $\bs{u}_\mathrm{m}$ and $w$ are functions of the two in-plane coordinates $x_1^\prime$ and $x_2^\prime$, and hence $\nablap \bs{u} = \gradv - z^\prime \nablap\nablap w$.

Next, we calculate the total elastic energy of the system in the wrinkled configuration. The elastic energy stored in the biofilm can be calculated from
\begin{gather}
    \Pi^\mathrm{biofilm} =\int \Psi dV =J_\mathrm{e,0}^{-1}\int ds^\prime \int_{-\gamma^0 H /2}^{\gamma^0 H/2} \Psi dz^\prime,
\end{gather}
where $J_\mathrm{e,0} =\det(\Fe^0)$. We shall leave out the full details of derivation here, but add several comments. All the terms in $\Psi$ (which do not average to zero over $z^\prime$) can be classified into three categories: (1)~terms independent of $\bs{u}_\mathrm{m},w$; (2)~terms depending on $\bs{u}_\mathrm{m},w$, but independent of $z^\prime$; (3)~terms proportional to $(z^\prime)^2$. The first terms correspond to the elastic energy stored in the pre-strained configuration, the second terms describe the release of compression energy $\propto H$, and the integration of the third terms give rise to the bending energy $\propto H^3$. To facilitate the discussion, we focus on incompressible material, i.e. $\lambdab\to \infty$. In this case, the elastic energy $\Pi^\mathrm{biofilm} = \int \Psi_\parallel ds^\prime $, with the areal energy density $\Psi_\parallel$ given by
\begin{gather}
\Psi_\parallel = \Gb\gamma^0 H(f_\mathrm{pre-stress}+f_\mathrm{strain,lin}+f_\mathrm{strain,nlin}) + \frac{\Gb(\gamma^0)^3H^3}{12} f_\mathrm{bending}, \label{area_energy_density_1}
\end{gather}
where
\begin{subequations}
\label{area_energy_density_2}
\begin{equation}
f_\mathrm{pre-stress} = \tr(\Bo^0) + \left(\gamma^0\right)^2-3,
\end{equation}
\begin{equation}
 f_\mathrm{strain,lin} = \tr\Big(\mathrm{sym}\big[(\gradv)\Bo^0\big]\Big) - (\gamma^0)^2\tr\Big(\gradv\Big) +  \frac{1}{2}\Big(\gradw \cdot \Bo^0 \cdot\gradw - (\gamma^0)^2\lVert\gradw\lVert^2\Big),
\end{equation}
\begin{equation}
\begin{split}
 f_\mathrm{strain,nlin} = & \frac{1}{2}(\gamma^0)^2\Big(\tr(\gradv) + \lVert\gradw\lVert^2 \Big)^2 + \frac{1}{2}\tr\Big(\gradv\cdot\Bo^0\cdot(\gradv)^T\Big) \\ &+\frac{1}{2}(\gamma^0)^2\tr\Big(\gradv \cdot \gradv\Big) + \frac{1}{2}(\gamma^0)^2\Big(\tr\big(\gradv\big)\Big)^2 + (\gamma^0)^2(\gradw)^T\cdot\gradv\cdot\gradw,
 \end{split}
\end{equation}
\begin{equation}
\mathrm{and}\quad f_\mathrm{bending} = (\gamma^0)^2\Big[\tr(\nablap \gradw)\Big]^2 +\frac{1}{2}(\gamma^0)^2 \tr\Big(\nablap\gradw\cdot \nablap\gradw\Big) +  \frac{1}{2}\tr\Big((\nablap \gradw)\cdot\Bo^0\cdot(\nablap \gradw)\Big).
\end{equation}
\end{subequations}
Here, we separated the energy costs of stretching according to the scaling with the aspect ratio $\delta_h$ as $f_\mathrm{pre-stress}\sim\delta_h^0$, $ f_\mathrm{strain,lin}\sim\delta_h^2$, and $ f_\mathrm{strain,nlin}\sim\delta_h^4$.
We denote by $\mathrm{sym}[\cdot]$ the symmetric part of a tensor. $\Bo^0 = \Ainit(\Ainit)^T$ denotes the in-plane components of the left Cauchy-Green deformation tensor.

The elastic energy of the deformed elastic substrate can be formally written as $\Pi^\mathrm{sub} = \int \frac{1}{2}K_\mathrm{s} w^2 ds^\prime$, where $K_\mathrm{s}$ is an effective spring constant, and could be a function of the wavenumber $\kw$ of the wrinkling profile depending on the substrate model. Some examples are as follows \cite{huang2005nonlinear,lejeune2016tri}: (1)Winkler foundation: $K_\mathrm{s}$  = constant; (2) infinitely large elastic substrate (incompressible): $K_\mathrm{s} = 2\Gs \kw$, where $\Gs$ is the shear modulus of the substrate; (3) composite substrate in the tri-layer model (incompressible): $K_\mathrm{s}=\frac{4\Gs \kw}{2 + \kw h_\mathrm{i}(\Gs/G_\mathrm{i} -1)}$, where $h_\mathrm{i}$ is the thickness of the intermediate layer and $\Gs/G_\mathrm{i}$ denotes the stiffness ratio between the substrate and the intermediate layer.

\subsection{Stress relaxation}
One direct effect of forming wrinkles is the relaxation of the compressive stress in the biofilm. To see this, we calculate the change of the in-plane components of the Cauchy stress before and after wrinkling occurs. 
Substituting Eq.~(\ref{Wrinkle_DefTensor}) into the general expression of Cauchy stress $\tilde{\bs{\sigma}} = (\Fe\Fe^T) - \tilde p\mathbf{I}_\parallel$, we obtain, to order $\delta_h^2$,
\begin{subequations}
\begin{align}
    \tilde{\bs{\sigma}}_\parallel &= \Bo^0 +  \gradv \cdot \Bo^0 + \Bo^0\cdot\left(\gradv\right)^T + (\gamma^0)^2\,\gradw\left(\gradw\right)^T  - \tilde p\mathbf{I}_\parallel, \\
    \tilde{\bs{\sigma}}_{\parallel,z} & = \Bo^0 \cdot \gradw - (\gamma^0)^2 \gradw,\\
    \tilde{\bs{\sigma}}_{z,\parallel} & = (\gradw)^T \cdot \Bo^0  - (\gamma^0)^2 (\gradw)^T,\\
    \tilde \sigma_{zz} &=(\gamma^0)^2+2 (\gamma^0)^2 w_{,z} + (\gradw)^T \cdot \Bo^0 \cdot \gradw - \tilde p,
\end{align}
\label{post_wrinkle_stress}%
\end{subequations}
where stresses are evaluated at the midline of the biofilm.
The isotropic stress $\tilde p$ can be determined again from the plane-stress assumption $\boldsymbol{\sigma}\cdot\mathbf{n}= 0$ as before, which yields
\begin{gather}
    p = (\gamma^0)^2\big[ 1 - \lVert\gradw\rVert^2 - 2\tr(\gradv) \big]. 
\end{gather}

Taken together, we rewrite the stress in the wrinkling configuration as
\begin{gather}
    \tilde{\bs{\sigma}}_\parallel = \underbrace{\Big[\Bo^0- (\gamma^0)^2 \mathbf{I}_\parallel\Big] }_{\tilde{\bs{\sigma}}_\parallel^0} + \underbrace{\left[(\gamma^0)^2\left(\gradw\cdot\big(\gradw\big)^T+\lVert\gradw\rVert^2\mathbf{I}_\parallel\right)+ \gradv \cdot \Bo^0 + \Bo^0\cdot\left(\gradv\right)^T +  2(\gamma^0)^2\tr(\gradv)\mathbf{I}_\parallel\right] }_{\Delta \tilde{\bs{\sigma}}_\parallel}. \label{eq_stress_relaxation}
\end{gather}
Clearly, the first term recovers the initial pre-stress applied to the biofilm in the deformed flat configuration, and the second term denotes the stress relaxation due to in-plane stretch ($\bs{u}_\mathrm{m}$ terms) and out-of-plane displacement ($w$ terms).

\subsection{A coarse-grained description of post-wrinkling evolution}
So far, our model only considers a flat circular biofilm expanding on agar. The solutions are axisymmetric, and therefore the problem is essentially 1D under the plane-stress assumption. On the other hand, the formation of wrinkles breaks the symmetry. To study the effect of mechanical instability, one could use a vectorial displacement field $\bs{x}-\bs{x}^\prime = (\bs{u},w)$ to describe the wrinkling profile and perform direct numerical simulations to solve the full growth-instability problem. However, this would be computationally expensive. 

Here, we propose to study the wrinkling dynamics by means of a coarse-grained model, which maintains axisymmetry and, at the same time, captures the most important features of the wrinkling instability. The idea is to conceptually cut the film into patches, which are small compared to the biofilm radius $r_\mathrm{b}$, but larger than the wavelength of wrinkles. The amplitude and shape of wrinkles are assumed to be constant within patches and to vary slowly over distances much larger than the wavelength of wrinkles, which slow variation is described with respect to the coarse-grained coordinates $\bs{x}^c\equiv(r^c,\theta^c)$. Below, we start by coarse-graining the elastic energy in Eq.~(\ref{area_energy_density_1}) over such patches. Next, we compute the amplitude and the shape of wrinkles by minimizing the coarse-grained elastic energy. In turn, this information is used to calculate the coarse-grained stresses of the wrinkled configuration given by Eq.~(\ref{eq_stress_relaxation}).
Throughout this section we assume that the expansion of the biofilm is much slower than the mechanical relaxation, and hence, the biofilm morphology evolves quasi-statically.

To describe the undulation of wrinkles within each coarse-grained patch, we use an ansatz that was previously proposed for herringbone patterns \cite{cai2011periodic,audoly2008buckling} 
\begin{gather}
    w(x_1^\prime,x_2^\prime) = A\cos(k x_1^\prime -\beta_1 \cos \beta_2 k x_2^\prime) \approx  A\Big[\cos(k x_1^\prime) + \beta_1\sin(k x_1^\prime)\cos(\beta_2k x_2^\prime)\Big], \label{herringbone}
\end{gather}
where $x_1^\prime$ and $x_2^\prime$ denote the coordinates along the two principal axes within the patch, $A$ is the amplitude of undulations, $k$ is the wave number of wrinkles in the $x_1^\prime$ direction, and $\beta_1/k$ and $\beta_2 k$ denote, respectively, the amplitude and the wave number of the in-plane wiggles (see Fig.~4B in the main text). Here, we assumed that $\beta_1\ll 1$ and that $k, A,\beta_1,\beta_2$ are constant over the coarse-grained patch. Following the analysis in Sec.~\ref{sec_three_stages}, we assume that the principal axes are in the radial and circumferential directions. Thus, we choose $x_1^\prime = r^\mathrm{c}\theta^\prime$ and $x_2^\prime = r^\prime$, and write the planar compression $\Fe^0\equiv \mathrm{diag}(a_{11}^0,a_{22}^0, \gamma^0)$ as a diagonal matrix, where $\gamma_0 = 1/(a_{11}^0a_{22}^0)$ assuming incompressibility of the biofilm. The slow variation of wrinkle patterns over the biofilm is given by the amplitude $A(\bs{x}^\mathrm{c})$ and by the orientation/shape parameters $\beta_1(\bs{x}^\mathrm{c})$ and $\beta_2(\bs{x}^\mathrm{c})$, which are obtained via the minimization of the coarse-grained areal elastic energy density. Note that the ansatz for the herringbone pattern in the above Eq.~(\ref{herringbone}) is a perturbative expansion around the stripe state ($\beta_1=0$) that would result from the uniaxial compression in the $x_1^\prime$ direction. The assumption $\beta_1\ll 1$ is thus violated for the zigzag herringbone pattern resulting from the isotropic compression. Below we discuss how to address this limitation to obtain meaningful results for the isotropic compressive stresses as well.

The coarse-graining of the elastic energy stored in a compressed biofilm $\Psi_\parallel^\mathrm{biofilm}$ given in Eqs.~(\ref{area_energy_density_1}, \ref{area_energy_density_2}) is done in two steps. First we average the terms that include only the out-of-plane displacements $w(x'_1,x'_2)$ by using the ansatz from Eq.~(\ref{herringbone}). For example, averaging the terms directly related to $w$ in $f_\mathrm{strain,lin}$ yields
\begin{align}
    \langle f_\mathrm{strain,lin}^{(w)}\rangle &= \frac{1}{2} \Bigg\langle \Big[(a_{11}^0)^2 w_{,1}^2 + (a_{22}^0)^2 w_{,2}^2\Big]\Bigg\rangle - \frac{1}{2}(\gamma^0)^2 \Bigg\langle  \Big(w_{,1}^2 + w_{,2}^2\Big)\Bigg\rangle \notag, \\ & = \frac{1}{4}\Big[(a_{11}^0)^2 - (\gamma^0)^2\Big](kA)^2 +\frac{1}{8}\Big[(a_{11}^0)^2 - (\gamma^0)^2\Big](kA)^2 (\beta_1)^2 + \frac{1}{8}\Big[(a_{22}^0)^2 - (\gamma^0)^2\Big](kA)^2 (\beta_1\beta_2)^2,
\end{align}
where $w_{,i}=\partial_{x_i'} w~(i = 1,2)$, and averaging is defined as $\langle\dots\rangle = (\Delta s^\prime)^{-1}\int \dots dx_1^\prime dx_2^\prime$, where $\Delta s^\prime$ is the projected area of the coarse-grained biofilm patch on to the horizontal plane. Similar calculations can be done for the rest of the terms in the elastic energy density $\Psi_\parallel^\mathrm{biofilm}$ involving only terms with $w$.

In the second step of the coarse-graining of the elastic energy $\Psi_\parallel^\mathrm{biofilm}$, we discuss how to deal with terms that include the in-plane displacements $\bs{u}_\mathrm{m}$. We start by minimizing the energy density in Eqs.~(\ref{area_energy_density_1}, \ref{area_energy_density_2}) with respect to $\bs{u}_\mathrm{m}$  for fixed out-of-plane displacements $w$. This yields
\begin{subequations}
    \label{v_eq_all}
    \begin{equation}
        \Big[3(\gamma^0)^2 + (a_{11}^0)^2\Big]u_{\text{m}1,11} + 3(\gamma^0)^2 u_{\text{m}2,21}+ (a_{22}^0)^2 u_{\text{m}1,22} + \frac{3}{2}(\gamma^0)^2\lVert\gradw\rVert^2_{,1} + (\gamma^0)^2(w_{,1})(w_{,11} + w_{,22}) = 0 \label{v_eq_a},
    \end{equation}
    \begin{equation}
        \Big[3(\gamma^0)^2 + (a_{22}^0)^2\Big]u_{\text{m}2,22} + 3(\gamma^0)^2 u_{\text{m}1,12} + (a_{11}^0)^2 u_{\text{m}2,11} + \frac{3}{2}(\gamma^0)^2\lVert\gradw\rVert^2_{,2} + (\gamma^0)^2(w_{,2})(w_{,11} + w_{,22}) = 0. \label{v_eq_b}
    \end{equation}
\end{subequations}
The above partial differential equations for the in-plane-displacements $\boldsymbol{u}_\mathrm{m}$ can be solved with the help of Fourier transforms, which yield
\begin{subequations}
\label{v_solution_lin}
\begin{equation}
    \begin{split}
    u_{\text{m}1} = & \frac{4 + \beta_1^2(-2+\beta_2^2)}{8\big[(a_{11}^0)^2 + 3(\gamma^0)^2\big]}kA^2\sin(2kx_1^\prime) + \frac{\beta_1 (\gamma^0)^2}{2(a_{22}^0)^2}\cos(\beta_2kx_2^\prime)\\
    & - \frac{\beta_1^2(1+\beta_2^2)(\gamma^0)^2}{4\Big((a_{11}^0)^2 + 3(\gamma^0)^2 + \big[(a_{22}^0)^2 + 3(\gamma^0)^2\big]\beta_2^2  \Big)}kA^2\sin(2k x_1^\prime)\cos(2\beta_2kx_2^\prime)\\
    & -\frac{\beta_1(\gamma^0)^2\Big[(8+\beta_2^2)\big[4(a_{11}^0)^2 + (a_{22}^0)^2\beta_2^2\big] + 3(\gamma^0)^2\beta_2^4\Big]}{2\Big[4(a_{11}^0)^2 + (a_{22}^0)^2 \beta_2^2\Big]\Big[4(a_{11}^0)^2 +12(\gamma^0)^2 + \big[(a_{22}^0)^2 + 3(\gamma^0)^2 \big] \beta_2^2\Big]} kA^2\cos(2k x_1^\prime)\cos(\beta_2kx_2^\prime),~\text{and}
    \end{split}
\end{equation}
\begin{equation}
    \begin{split}
    u_{\text{m}2} = & \frac{\beta_1^2\big(-1+2\beta_2^2\big)(\gamma^0)^2}{8\beta_2\big[(a_{22}^0)^2 + 3(\gamma^0)^2\big]}kA^2\sin(2\beta_2kx_2^\prime) \\
    &+ \frac{\beta_1\beta_2(\gamma^0)^2\Big[8(a_{11}^0)^2 + \beta_2^2\big[2(a_{22}^0)^2 - 3(\gamma^0)^2\big]\Big]}{\Big[4(a_{11}^0)^2 + (a_{22}^0)^2 \beta_2^2\Big]\Big[4(a_{11}^0)^2 +12(\gamma^0)^2 + \big[(a_{22}^0)^2 + 3(\gamma^0)^2 \big] \beta_2^2\Big]}kA^2\sin(2k x_1^\prime)\sin(\beta_2kx_2^\prime)\\
    & -\frac{\beta_1^2\beta_2(1+\beta_2^2)(\gamma^0)^2}{4\Big[(a_{11}^0)^2 + 3(\gamma^0)^2 + \big[(a_{22}^0)^2 + 3(\gamma^0)^2\big]\beta_2^2 \Big]} kA^2\cos(2k x_1^\prime)\sin(2\beta_2kx_2^\prime).
    \end{split}
\end{equation}
\end{subequations}
The above expressions for $\bs{u}_\mathrm{m}$ can then be inserted into the elastic energy density $\Psi_\parallel^\mathrm{biofilm}$ in Eqs.~(\ref{area_energy_density_1}, \ref{area_energy_density_2}) and averaged over the patch as described above. For example,
\begin{align}
   \Big\langle \tr(\gradv)\Big\rangle &= \langle u_{\text{m}1,1} + u_{\text{m}2,2}\rangle = 0,\text{ and}\\ 
   \Big\langle \tr(\gradv\cdot\Bo^0\cdot\gradv^T)\Big\rangle &= (a_{11}^0)^2\Big(\langle u_{\text{m}1,1}^2 \rangle+ \langle u_{\text{m}2,1}^2 \rangle\Big) + (a_{22}^0)^2\Big(\langle u_{\text{m}1,2}^2 \rangle +\langle u_{\text{m}2,2}^2 \rangle\Big) \notag \\ 
   & \to \frac{1}{32}(kA)^4 
   + \frac{32+176\beta_2^2 + 79\beta_2^4 + 13\beta_2^6}{64\big(4+\beta_2^2\big)^2} (kA)^4\beta_1^2 
   + \frac{7-4\beta_2^2 + 7\beta_2^4}{512}(kA)^4\beta_1^4.
\end{align}
For brevity, we report only results in the small deformation limit, i.e. $a_{11}^0\to 1$, and $a_{22}^0\to 1$. Detailed expressions for the moderate deformations $a_{11}^0, a_{22}^0$ can be found in the Mathematica notebook available on Github (\hyperlink{https://github.com/f-chenyi/biofilm-mechanics-theory}{https://github.com/f-chenyi/biofilm-mechanics-theory}). In a similar way we average all terms in the elastic energy density $\Psi_\parallel^\mathrm{biofilm}$ in Eqs.~(\ref{area_energy_density_1}, \ref{area_energy_density_2}) by using expressions for the in-plane-displacements $\bs{u}_\mathrm{m}$ in Eq.~(\ref{v_solution_lin}) and for the out-of-plane displacements $w$ in Eq.~(\ref{herringbone}). The resulting averaged energy density of the biofilm is given and discussed in the next section below.

Finally, we also compute the elastic energy of deforming the substrate. As discussed in Refs.~\cite{audoly2008buckling,audoly2008buckling2} the elastic energy of the substrate can be written as
\begin{equation}
\langle\Psi_\parallel^\mathrm{sub}\rangle =  \frac{1}{\Delta s'}\int \mathrm{d}k_1\mathrm{d}k_2\  \frac{1}{2} K_\mathrm{s}(||k||)\ ||\hat{w}(k_1,k_2)||^2,
\end{equation}
where $||k|| = \big(k_1^2+k_2^2\big)^{1/2}$ and we defined the Fourier transform
\begin{equation}
\hat{w}(k_1,k_2) = \frac{1}{2\pi}\int_{\Delta s'}\mathrm{d}x_1'\mathrm{d}x_2'\  w(x_1',x_2')\ \mathrm{exp}\Big[-i(k_1x_1'+k_2x_2')\Big].
\end{equation}
Note that the results differ between different substrate models. For the Winkler foundation with elastic constant $K_s$ one obtains $\langle \Psi_\parallel^\mathrm{sub} \rangle = \frac{1}{4}K_sA^2(1+\beta_1^2/2)$, while for an infinitely thick incompressible elastic substrate with shear modulus $\Gs$ we find $\big\langle \Psi_\parallel^\mathrm{sub} \big\rangle = \frac{1}{2}\Gs kA^2 \big(1 + \frac{\beta_1^2}{2}\sqrt{1+\beta_2^2}\big)$. Here, we treat the agar as an infinitely thick elastic substrate.

\subsection{Summary and discussion}
In the small deformation limit ($a_{11}^0\to 1$, and $a_{22}^0\to 1$), the total averaged free energy density $\langle\Psi_\parallel^\mathrm{total}\rangle$ that includes the energy density of the deformed biofilm and the substrate can be written as~\cite{audoly2008buckling}
\begin{equation}
\begin{split}
    \frac{\langle\Psi_\parallel^\mathrm{total}\rangle}{\Gb H} = &-\frac{1}{4}\underbrace{\Bigg(|\tilde{\sigma}_{11}^0| + \frac{1}{2}|\tilde{\sigma}_{11}^0|\beta_1^2 + \frac{1}{2}|\tilde{\sigma}_{22}^0|\beta_1^2\beta_2^2\Bigg)(kA)^2 + \frac{1}{8}\Bigg(1 + f_a(\beta_2)\beta_1^2 + f_b(\beta_2)\beta_1^4\Bigg)(kA)^4}_\text{biofilm stretching~energy} \\ 
    & + \underbrace{\frac{(kH)^2}{12}\Bigg(1 + \frac{1}{2}\beta_1^2(1+\beta_2^2)^2\Bigg)(kA)^2}_\text{biofilm bending energy} 
    + \underbrace{\frac{\Gs}{2\Gb (kH)}\Big(1 + \frac{1}{2}\beta_1^2\sqrt{1+\beta_2^2}\Big)(kA)^2}_\text{substrate energy},
\end{split}
\label{eq_wrinkle_energy}
\end{equation}
where $\tilde{\sigma}_{\alpha\alpha}^0 = (a_{\alpha\alpha}^0)^2 - (\gamma^0)^2$ denotes the in-plane pre-stress, and $f_a(\beta_2) = \frac{64 + 64\beta_2^2 + 23\beta_2^4 + 2\beta_2^6}{4(4+\beta_2^2)^2}$ and $f_b(\beta_2) = (11 + 8 \beta_2^2 + 11\beta_2^4)/32$ are functions of $\beta_2$ only. Here, we assumed that stresses are compressive ($\tilde{\sigma}_{11}^0,\tilde{\sigma}_{22}^0\le0$).  Note that we have omitted the constant terms corresponding to the elastic energy stored in the pre-stressed flat configuration.

Next we comment on the form of the energy density in Eq.~(\ref{eq_wrinkle_energy}). By considering only terms to the order $(\beta_1)^0$, we recover the energy density 
\begin{gather}
    \frac{\langle\Psi_\parallel^\mathrm{total}\rangle}{\Gb H} =  \Bigg(-\frac{1}{4} |\tilde{\sigma}_{11}^0| + \underbrace{\Big[\frac{G_s}{2\Gb(kH)}+ \frac{(kH)^2}{12}\Big]}_{f_{\sigma}(k)}\Bigg) (kA)^2 + \frac{1}{8}(kA)^4
    \label{Eq:energy_1D_wrinkle}
\end{gather}
 that corresponds to 1D wrinkles for uniaxial compression ($\tilde{\sigma}_{11}^0 < 0$, $\tilde{\sigma}_{22}^0\approx 0$).~\cite{huang2005nonlinear}
The minimum of $f_\sigma(k)$ determines the critical stress for the formation of wrinkles, i.e. $\tilde{\sigma}_c = 4 \min_k f_\sigma(k)=(3 G_s/\Gb)^{2/3}$, and the corresponding argument $k_w=H^{-1} (3 G_s/\Gb)^{1/3}$ determines the  wavelength of wrinkles $\lambda=2\pi/k_w$. Consequently, when $|\tilde{\sigma}_{11}^0| < \tilde{\sigma}_c$, the coefficient of $(kA)^2$ term in Eq.~(\ref{Eq:energy_1D_wrinkle}) is positive, which is minimized for the amplitude $A=0$, i.e., a flat biofilm. The bifurcation from the planar to the wrinkled state occurs when the compressive stress $|\tilde{\sigma}_{11}^{0}|$ surpasses the critical stress $\tilde{\sigma}_\mathrm{c}$,  which, in the small deformation limit, corresponds to a phase boundary $|\epsilon_{11}^0| + |\epsilon_{22}^0|/2 = \tilde{\sigma}_\mathrm{c}/4$, where we expressed  compressions $a_{ii}^0=1+\epsilon_{ii}^0$ in terms of strains $\epsilon_{ii}^0$. The amplitude of wrinkles grows as $k_w A=\sqrt{|\tilde{\sigma}_{11}^{0}|-\tilde{\sigma}_\mathrm{c}}$.

\begin{figure}[t]
\centering
\includegraphics[width=.7\textwidth]{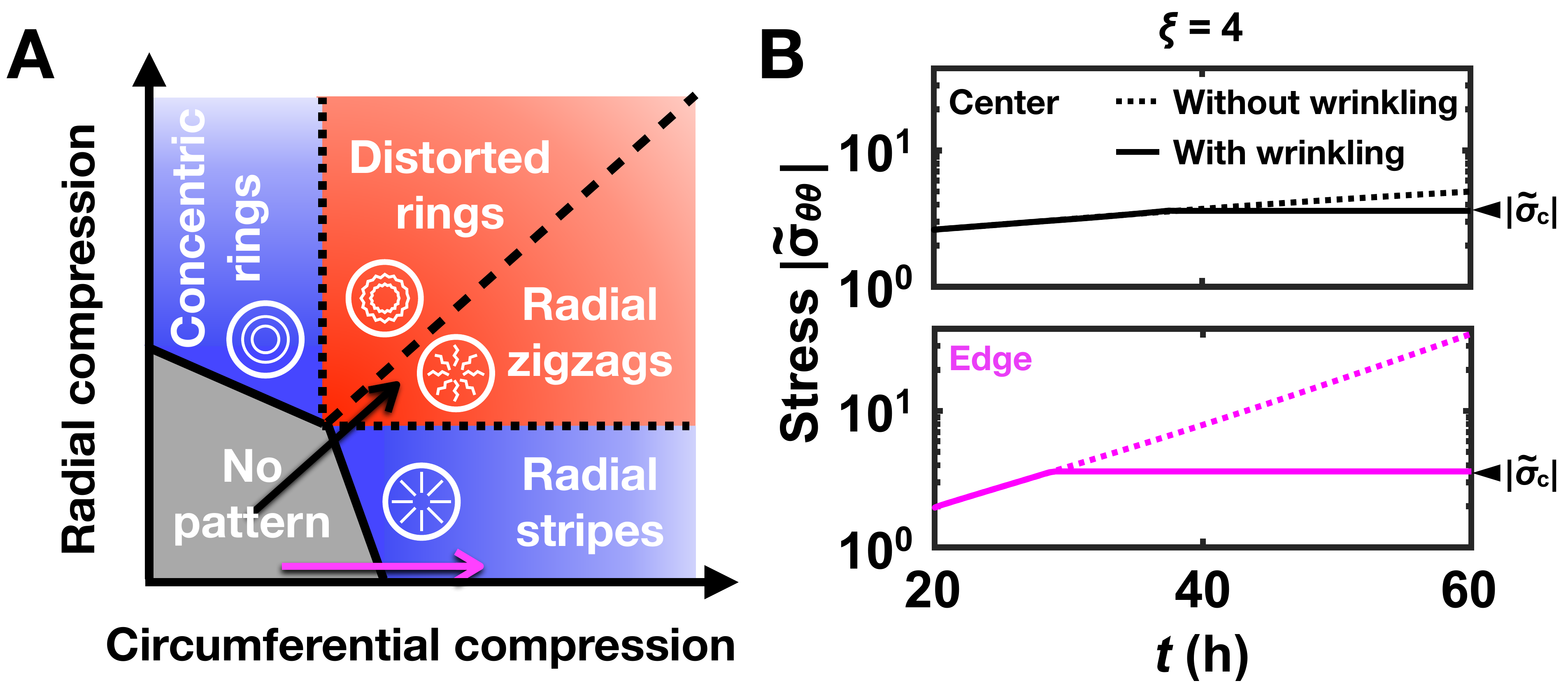}
\captionsetup{justification=justified,labelfont=bf,singlelinecheck=false}
\caption{
{\bf Mechanical characterization of biofilm wrinkling instability.}  ({\bf A}) Schematic phase diagram of the coarse-grained model for biofilm wrinkling instability.  Axes correspond to the circumferential compressive strain, measured by $|\epsilon^0_{\theta\theta}| = 1-(\mathbf{F}_\mathrm{e,\parallel}^0)_{\theta\theta}$, and the radial compressive strain,  measured by $|\epsilon^0_{rr}| = 1-(\mathbf{F}_\mathrm{e,\parallel}^0)_{rr}$, in the pre-stressed state. For the lower portion of the phase diagram, i.e. $|\epsilon^0_{rr}| < |\epsilon^0_{\theta\theta}|$, the biofilm transitions from the planar state (no pattern) to a wrinkled state when $|\tilde{\sigma}_{\theta\theta}^{0}|$ surpasses the critical stress $|\tilde{\sigma}_\mathrm{c}|$, which, in linear elasticity, corresponds to a phase boundary $|\epsilon_{\theta\theta}^0| + |\epsilon_{rr}^0|/2 = |\tilde{\sigma}_\mathrm{c}|/4$ (black solid line). This primary instability results in a radial stripe pattern. The secondary transition from radial stripes to radial zigzags occurs when $|\epsilon_{rr}^0| > |\tilde{\sigma}_\mathrm{c}|/6$ (linear elasticity phase boundary, black dotted line). Similarly, for the upper portion of the phase diagram, the modeled biofilm will transition from the planar state to a wrinkled state with concentric rings then to a wrinkled state with distorted rings. The black and magenta arrows illustrate the evolution of the center and the edge of the modeled biofilm, respectively.
({\bf B}) Magnitude of the compressive stress $\tilde{\sigma}_{\theta\theta}$ at the center (black, {\it top}) and edge (magenta, {\it bottom}) of the modeled biofilm versus time $t$, plotted for the chemo-mechanical model with (solid curves) and without (dotted curves) permitting wrinkling. Simulation parameters as in Fig. 4{\bf C}.
\label{Fig:phaseDiagram}
}
\end{figure}

Going beyond the leading order in $\beta_1$ and minimizing the Landau-Ginzburg-type free energy in Eq.~(\ref{eq_wrinkle_energy}) with respect to the amplitude $A$ and the wave number $k$ leads to
\begin{subequations}
\begin{gather}
    k_w^* = H^{-1} \left(\frac{3 G_s (1 + \frac{1}{2}\beta_1^2\sqrt{1+\beta_2^2})}{\Gb (1 + \frac{1}{2}\beta_1^2(1+\beta_2^2)^2)}\right)^{1/3}, \label{eq_wrinkle_k}
\end{gather}
\begin{gather}
    (k_w^* A^*)^2 \approx \frac{\displaystyle \Big(|\tilde{\sigma}_{11}^0| -\tilde{\sigma}_c + \tilde{\Delta} \beta_1^2\Big)}{1 + f_a(\beta_2)\beta_1^2 + f_b(\beta_2)\beta_1^4}. \label{eq_wrinkle_kA}
\end{gather}
\end{subequations}
In Eq.~(\ref{eq_wrinkle_kA}) we expanded the numerator to the second order term $\beta_1^2$  and introduced $\tilde{\Delta} = |\tilde{\sigma}_{11}^0|/2 + |\tilde{\sigma}_{22}^0|\beta_2^2/2 -\tilde{\sigma}_c\big(1+2\beta_2^2 + \beta_2^4 + 2\sqrt{1+\beta_2^2} \big)/6$. The bifurcation from the planar to the wrinkled state occurs when $|\tilde{\sigma}_{11}^{0}| + \tilde{\Delta} \beta_1^2 > \tilde{\sigma}_\mathrm{c}$. In this case the minimized energy becomes 
\begin{gather}
    \left.\frac{\langle\Psi_\parallel^\mathrm{total}\rangle}{\Gb H}\right|_{A=A^*,k=k_w^*} = - \frac{(|\tilde{\sigma}_{11}^{0}| - \tilde{\sigma}_c+\tilde{\Delta}~\beta_1^2)^2}{1+f_a \beta_1^2 + f_b \beta_1^4}. \label{eq_wrinkle_energy_A}
\end{gather}
We then solve for the optimal value of $\beta_1$ by minimizing the free energy density in the above equation, which yields solutions $\beta^*_1=0$ and  
\begin{gather}
    (\beta_1^*)^2 = \frac{(|\tilde{\sigma}_{11}^{0}| - \tilde{\sigma}_c)f_a(\beta_2) - 2\tilde{\Delta}}{f_a(\beta_2)\,\tilde{\Delta} - 2 (|\tilde{\sigma}_{11}^{0}| - \tilde{\sigma}_c) f_b(\beta_2) }\approx \frac{8 (2 |\tilde{\sigma}_{22}^{0}| - |\tilde{\sigma}_{11}^{0}| - \tilde{\sigma}_c)}{3(|\tilde{\sigma}_{11}^{0}| - \tilde{\sigma}_c)}\  \beta_2^2, \label{eq_beta_1}
\end{gather}
where the above result was expanded to the lowest order in $\beta_2$. In the relevant regime $(|\tilde{\sigma}_{11}^{0}| > \tilde{\sigma}_c)$,
 the secondary bifurcation from the stripe pattern ($\beta_1^*=0$) to the zigzag pattern ($\beta_1^*>0$) happens when the compressive stress $\tilde{\sigma}_{22}^{0}$ is sufficiently large, i.e., when $|\tilde{\sigma}_{22}^{0}| > |\tilde{\sigma}_{11}^{0}|/2 + \tilde{\sigma}_c/2$. This corresponds to the phase boundary $|\epsilon_{22}^0| = \tilde{\sigma}_\mathrm{c}/6$, which is consistent with previous studies on herringbone patterns~\cite{audoly2008buckling}. A schematic phase diagram is presented in Fig.~\ref{Fig:phaseDiagram}.
 
In principle, the expression for $\beta_1^*$ in Eq.~(\ref{eq_beta_1}) can then be substituted into Eq.~(\ref{eq_wrinkle_energy_A}) for the minimized energy density, which can then be expressed as a function of $\beta_2$. As shown in Ref.~\cite{audoly2008buckling}, the optimal value $\beta_2^*$ that minimizes the resulting energy density scales as $\epsilon^{1/2}$ near the onset of the secondary bifurcation, where $\epsilon = |\epsilon_{22}^0| - \tilde{\sigma}_c/6$. However, as shown below this would result in anisotropic coarse grained stresses $\langle \tilde{\bs{\sigma}}_\parallel \rangle$ in the wrinkled configuration  in Eq.~(\ref{eq_stress_relaxation}) in response to isotropic compression $\tilde{\sigma}_{11}^0 = \tilde{\sigma}_{22}^0$. To overcome this issue, we do not obtain the value of parameter $\beta_2$ via the energy minimization, but rather we fix the value to $\hat{\beta}_2$, such that the coarse grained  stresses $\langle \tilde{\bs{\sigma}}_\parallel \rangle$ in the wrinkled configuration remain isotropic in response to isotropic compression.

From Eq.~(\ref{eq_stress_relaxation}) for the stresses $\tilde{\bs{\sigma}}_\parallel$ in the wrinkled configuration, we see that the in-plane displacements $\bs{u}_\mathrm{m}$ appear only in linear terms that average out to zero in the coarse-grained description. The out-of-plane displacements $w$ appear in quadratic terms $\gradw \cdot (\gradw)^T$ and $\lVert\gradw \rVert^2 \mathbf{I}$ that result in the anisotropic and isotropic stress relaxation, respectively. Their averages over the coarse-grained patches read
\begin{subequations}
\begin{gather}
    \langle \gradw \cdot (\gradw)^T \rangle = 
    \begin{pmatrix}
        \frac{1}{2}(kA)^2\Big(1 + \frac{\beta_1^2}{2} \Big) & 0\\
        0 & \frac{1}{4}(\beta_1\beta_2)^2(kA)^2
    \end{pmatrix},
\end{gather}
\begin{gather}    
    \langle \lVert\gradw \rVert^2 \mathbf{I}_\parallel \rangle = \Big[\frac{1}{2}(kA)^2\Big(1 + \frac{\beta_1^2}{2}\Big) + \frac{1}{4}(\beta_1\beta_2)^2(kA)^2 \Big]
    \mathbf{I}_\parallel.
\end{gather}
\end{subequations}
Therefore, the coarse-grained stresses in the wrinkled configuration become
\begin{subequations}
\label{eq_stress_relaxation_A}
\begin{gather}
\langle\tilde{\sigma}_{11}\rangle =  - |\tilde{\sigma}_{11}^0| + (kA)^2\Big(1 + \frac{\beta_1^2}{2} + \frac{\beta_1^2\beta_2^2}{4}\Big),
\end{gather}
\begin{gather}
\langle\tilde{\sigma}_{22}\rangle =  - |\tilde{\sigma}_{22}^0| + \frac{1}{2}(kA)^2\Big(1 + \frac{\beta_1^2}{2} + \beta_1^2\beta_2^2\Big). \end{gather} 
\end{subequations}
By requiring that the coarse-grained stresses in the wrinkled state remain equal, i.e. $\langle\tilde{\sigma}_{11}\rangle = \langle\tilde{\sigma}_{22}\rangle$, in response to the isotropic compression ($\tilde{\sigma}_{11}^0 = \tilde{\sigma}_{22}^0$), we obtain the condition $\beta_1^2 = 2/(\beta_2^2 - 1)$. Comparing this condition with Eq.~(\ref{eq_beta_1}) for $\beta_1^*$ that minimizes the energy density, we obtain a self-consistent equation $2/((\hat{\beta}_2)^2 - 1)\approx 8/3 (\hat{\beta}_2)^2$ with solution $\hat{\beta}_2 \approx \sqrt{3/2}$.
Moreover, we observe that by inserting the optimal values of $k_w^* A^*$, $\beta_1^*$, and $\hat{\beta}_2$ into Eq.~(\ref{eq_stress_relaxation_A}) for the coarse-grained stresses in the wrinkled state, we find that $\langle\tilde{\sigma}_{11}\rangle \equiv - \tilde{\sigma}_\mathrm{c}$ after the primary bifurcation, and $\langle\tilde{\sigma}_{22}\rangle \equiv - \tilde{\sigma}_\mathrm{c}$ after the secondary bifurcation. Thus, the wrinkling instability prevents the compressive stresses from rising above the critical stress by releasing the compressing mechanical strain via the out-of-plane deformations~(see Fig.~\ref{Fig:phaseDiagram}).

The results presented above can be represented in terms of the dimensionless amplitude of wrinkles $\tilde{A} \equiv k_w^* A$ and in terms of the dimensionless shape parameter $\tilde{S}^2 \equiv (\beta_1\hat{\beta}_2)^2/6$, which is $\tilde{S}=0$ for striped wrinkles and $\tilde{S}>0$ for zigzag wrinkles ($\tilde{S}=1$ for zigzag wrinkles due to isotropic compression). The energy density in Eq.~(\ref{eq_wrinkle_energy}) can be rewritten in terms of the two parameters $\tilde{A}$ and $\tilde{S}$. For the optimal wavelength of wrinkles $k_w^*$ in Eq.~(\ref{eq_wrinkle_k}), the energy density to the leading order becomes
\begin{gather}
\frac{4\langle\Psi_\parallel^\mathrm{total}\rangle}{\Gb H} \approx - \Big((|\tilde{\sigma}_{11}^{0}| - \tilde{\sigma}_\mathrm{c})(1+3 b\tilde{S}^2) + 3  (|\tilde{\sigma}_{22}^{0}| - \tilde{\sigma}_\mathrm{c})\tilde{S}^2\Big)\tilde{A}^2 + \frac{1}{2}\Big(1 + (6 b + 3) \tilde{S}^2 \Big)\tilde{A}^4,
\label{eq:EnergyWrinkleMain}
\end{gather}
where $b=(\hat{\beta}_2)^{-2}=2/3$ and $\tilde{\sigma}_c =(3 G_s/\Gb)^{2/3}$ denotes the normalized critical stress for wrinkling as discussed above.
Similarly, we rewrite the stress relaxation due to wrinkles in Eq.~(\ref{eq_stress_relaxation_A}) in terms of the two parameters $\tilde{S}$ and $\tilde{A}$ as
\begin{gather}
    \langle\tilde{\sigma}_{11}\rangle =  - |\tilde{\sigma}_{11}^0| + \tilde{A}^2 \Big[ 1+(3 b  + 3/2) \tilde{S}^2 \Big], ~\mathrm{and}~ \langle\tilde{\sigma}_{22}\rangle =  - |\tilde{\sigma}_{22}^0| + \tilde{A}^2\Big[1/2+(3b/2  + 3) \tilde{S}^2 \Big].
    \label{eq:StressWrinkleMain}
\end{gather}
The above Eqs.~(\ref{eq:EnergyWrinkleMain}) and (\ref{eq:StressWrinkleMain}) correspond to Eqs. [4] and [5] quoted in the main text, respectively.

Using the coarse-grained model discussed above, the coarse-grained force-balance equation after the wrinkling instability occurs is given by
$
     \tilde{\bs{\nabla}}_\parallel\cdot \langle\tilde{\mathbf{N}}_\parallel \rangle = \xi q(k_w^* A^*,\beta_1^*,\hat{\beta}_2)\tilde{\bs{v}}_\parallel,
$
where $q(k_w^* A^*,\beta_1^*,\hat{\beta}_2) = \langle \Vert (-w_{,1},-w_{,2},1)\Vert \rangle \approx  1 + \frac{1}{4}(k_w^* A^*)^2\big(1+\frac{1}{2}(\beta_1^*)^2(1+(\hat{\beta}_2)^2)\big)$ accounts for the increase of the contact area between the wrinkled biofilm and the substrate. The average stress is defined as  $\langle\tilde{\mathbf{N}}_\parallel \rangle = (\Delta s^\prime)^{-1}\int \tilde{\bs{\sigma}}_\parallel dz \,ds^\prime$, where the integral $\int\!dz$ is taken along the $z$ direction in the current deformed configuration and the integral $\int\! ds'$ is over the reference patch coordinates. Note that $dz = dz_n / \cos\theta_{nz}$ and $ds^\prime = ds \cos\theta_{nz}$ where $z_n$ denotes the coordinate along the normal to the biofilm surface, $ds$ denotes the area of the biofilm patch whose projected area on the $x-y$ plane is $ds^\prime$, and $\theta_{nz}$ denotes the angle between the normal to the wrinkled biofilm surface and the $z$ direction. Because $dz ds'=dz_n ds_n$, we can rewrite $\langle\tilde{\mathbf{N}}_\parallel \rangle$ as $\langle\tilde{\mathbf{N}}_\parallel \rangle = (\Delta s^\prime)^{-1}\int \tilde{\bs{\sigma}}_\parallel dz_n ds =(\Delta s^\prime)^{-1}\int \tilde{\bs{\sigma}}_\parallel dz^\prime ds^\prime $, where we used the incompressibility of the biofilm, $dz_n ds = dz^\prime ds^\prime$, to convert the integration to coordinates $dz'$ and $ds'$ in the deformed flat-film configuration. Thus we obtain $\langle\tilde{\mathbf{N}}_\parallel \rangle = \gamma^0 \langle \tilde{\bs{\sigma}}_\parallel\rangle$, where the average stress $\langle\tilde{\bs{\sigma}}_\parallel\rangle$ is given by Eq.~(\ref{eq_stress_relaxation_A}).  

In summary, the force-balance equation for the biofilm can be written as
\begin{gather}
     \tilde{\bs{\nabla}}_\parallel\cdot(\gamma^0 \langle \tilde{\bs{\sigma}}_\parallel)\rangle  = \xi q(k_w^* A^*,\beta_1^*,\hat{\beta}_2)\tilde{\bs{v}}_\parallel,
     \label{eq:forceBalanceWrinkled}
\end{gather}
where $q = 1 + \frac{1}{4}(k_w^* A^*)^2\big(1+\frac{1}{2}(\beta_1^*)^2(1+(\hat{\beta}_2)^2)\big) = 1 + \frac{1}{4}\tilde{A}^2\big[1+3(b+1)\tilde{S}^2\big]$, and the stress $\langle \tilde{\bs{\sigma}}_\parallel \rangle = \tilde{\bs{\sigma}}_\parallel^0 + \Delta \tilde{\bs{\sigma}}_\parallel (k_w^* A^*,\beta_1^*,\hat{\beta}_2)$ is given by Eq.~(\ref{eq_stress_relaxation_A}) and (\ref{eq:StressWrinkleMain}) with the pre-stress  given by $\tilde{\bs{\sigma}}_\parallel^0 = \Fep^0(\Fep^0)^T - (\gamma^0)^2\mathbf{I}_\parallel$. Before the wrinkling instability occurs, the amplitude is $A^*=0$. Thus $q=1$, and $\langle \tilde{\bs{\sigma}}_\parallel \rangle =  \tilde{\bs{\sigma}}^0_\parallel$, and the force-balance Eq.~(\ref{eq:forceBalanceWrinkled}) reduces to Eq.~(\ref{eq_ndim_force_balance}) in Sec.~\ref{sec:ChemoMechanicalModel}. In order to simulate the evolution of the wrinkled biofilm, we solve the governing equations (\ref{eq_ndim_nc}) -- (\ref{eq_ndim_constitutive}) defined in Sec.~\ref{sec:ChemoMechanicalModel}, where the force-balance equation is replaced with Eq.~(\ref{eq:forceBalanceWrinkled}). Note that the evolution of the wrinkle pattern is fully coupled to the dynamics of nutrients and biofilm expansion.

\begin{figure}[t]
\centering
\includegraphics[width=\textwidth]{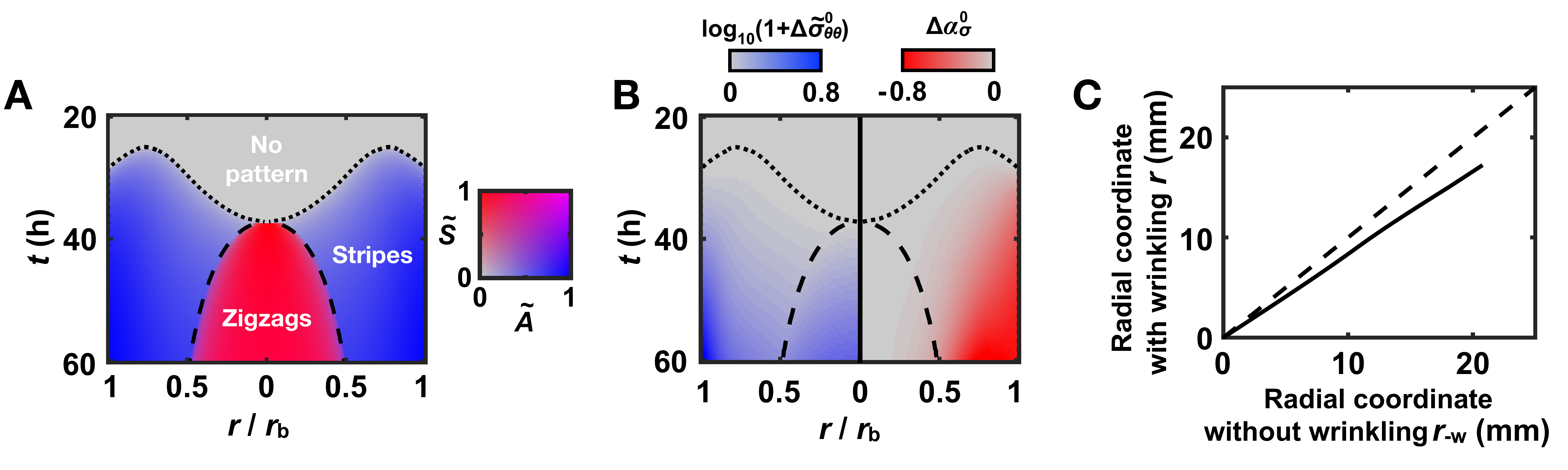}
\captionsetup{justification=justified,singlelinecheck=false,labelfont=bf}
\caption{
{\bf Formation of wrinkles slows biofilm expansion and affects the subsequent formation of the biofilm wrinkle patterns.} ({\bf A,~B}) Kymograph representations of ({\bf A}) biofilm wrinkling pattern formation ($\tilde{A} \equiv k_w^* A$ is the dimensionless amplitude of wrinkles and $\tilde{S} \equiv \sqrt{(\beta_1\hat{\beta}_2)^2/6}$ is the dimensionless shape parameter, see text for details), and ({\bf B}) differences in circumferential pre-stress (denoted by $\Delta \tilde{\sigma}_{\theta\theta}^0$, {\it left}) and in pre-stress anisotropy (denoted by $\Delta \alpha_\sigma^0$, {\it right}) between the chemo-mechanical model with and without wrinkling (denoted by $\mathrm{-w}$), i.e., $\Delta\tilde{\sigma}_{\theta\theta}^0\equiv |\tilde{\sigma}_{\theta\theta}^0| - |\tilde{\sigma}_{\theta\theta,\mathrm{-w}}^0|$ and $\Delta \alpha_\sigma^0 = \alpha_\sigma^0 - \alpha_{\sigma,-w}^0$, where the pre-stress anisotropy is defined as $\alpha_\sigma^0 = (\tilde{\sigma}_{\theta\theta}^0 - \tilde{\sigma}_{rr}^0)/(\tilde{\sigma}_{\theta\theta}^0 + \tilde{\sigma}_{rr}^0)$. Axes correspond to time $t$, and the radial coordinate normalized by the biofilm radius $r/r_\mathrm{b}$. Color code of $\bf{A}$ is the same as in Fig. 4{\bf C}. The color scale in {\bf B} {\it left} shows $(1+\Delta \tilde{\sigma}_{\theta\theta})$ on a logarithmic scale for visualization of both small and large values of $\Delta \tilde{\sigma}_{\theta\theta}$. ({\bf C}) The radial coordinate $r$ of a modeled biofilm with wrinkles, plotted against that of a modeled biofilm when wrinkling is not permitted (denoted by $r_{-\mathrm{w}}$), at the designated time (solid curve). The dashed black line $r = r_{-\mathrm{w}}$ is provided as a guide to the eye. Simulation parameters as in Fig. 4{\bf C}. 
}
\label{Fig:kymographs}
\end{figure}

According to our model, compared to the case where the wrinkling instability is prevented (force-balance Eq.~(\ref{eq_ndim_force_balance})), the expansion of the wrinkled biofilm is slowed, the pre-stress anisotropy is reduced, and the magnitude of compressive true stress is reduced while that of the pre-stress is increased (Fig. ~\ref{Fig:phaseDiagram} and \ref{Fig:kymographs}). Thus, our model suggests that wrinkling due to a growth-induced mechanical instability feeds back and further influences biofilm expansion and pattern formation by modifying the distribution of internal stress.

\section{\label{sec_plasticity}Supplementary Discussion: biofilm material properties}

We have focused on the question of how stresses accumulate spatiotemporally in a growing biofilm. For simplicity, we used a hyperelastic model to describe the biofilm material. However, our previous measurements show that the material properties of experimental biofilms are much more complicated, and very closely resemble those of a hydrogel~\cite{yan2018bacterial}. For example, the biofilm material starts to yield with a dramatic decrease in elastic modulus $\Gb$ above a critical strain $\epsilon_\mathrm{Y} \approx 10\%$. The internal stress predicted by our simple elastic model is larger than the yield stress.  Thus, we suspect that some yielding of the material likely occurs as the biofilm grows.

To explore the effect of yielding on biofilm expansion dynamics, we implement some of the simplest plastic material models to study how the stresses build up before mechanical instability occurs. We adopt the von Mises yield criterion which suggests that yielding of material begins when the second invariant of the deviatoric stress $\tilde{\mathbf{s}}$, i.e. $\sige = (\frac{3}{2}\tilde{\mathbf{s}}:\tilde{\mathbf{s}})^{1/2}$, reaches a critical value. 

{\bf Incompressible hyperelastic material}: 
We start by rewriting the mechanical stress as the sum of a deviatoric stress and a hydrostatic stress. Let $\mathbf{E} = \Fe\Fe^T - \frac{1}{3}\mathrm{tr}\Big(\Fe\Fe^T\Big)\mathbf{I}$ be the {\it deviatoric deformation tensor}. The dimensionless stress $\tilde{\boldsymbol{\sigma}}$ can be reorganized as $\tilde{\boldsymbol{\sigma}} = \tilde{\mathbf{s}} - \tilde{p}\mathbf{I}$ where $\tilde{\mathbf{s}} = \mathbf{E}$ denotes the deviatoric stress, and $\tilde{p}$ is the hydrostatic pressure $\tilde{p} = \gamma^2 - \tr(\Fe\Fe^T)/3$.

\begin{figure}[t]
\centering
\includegraphics[width=.7\textwidth]{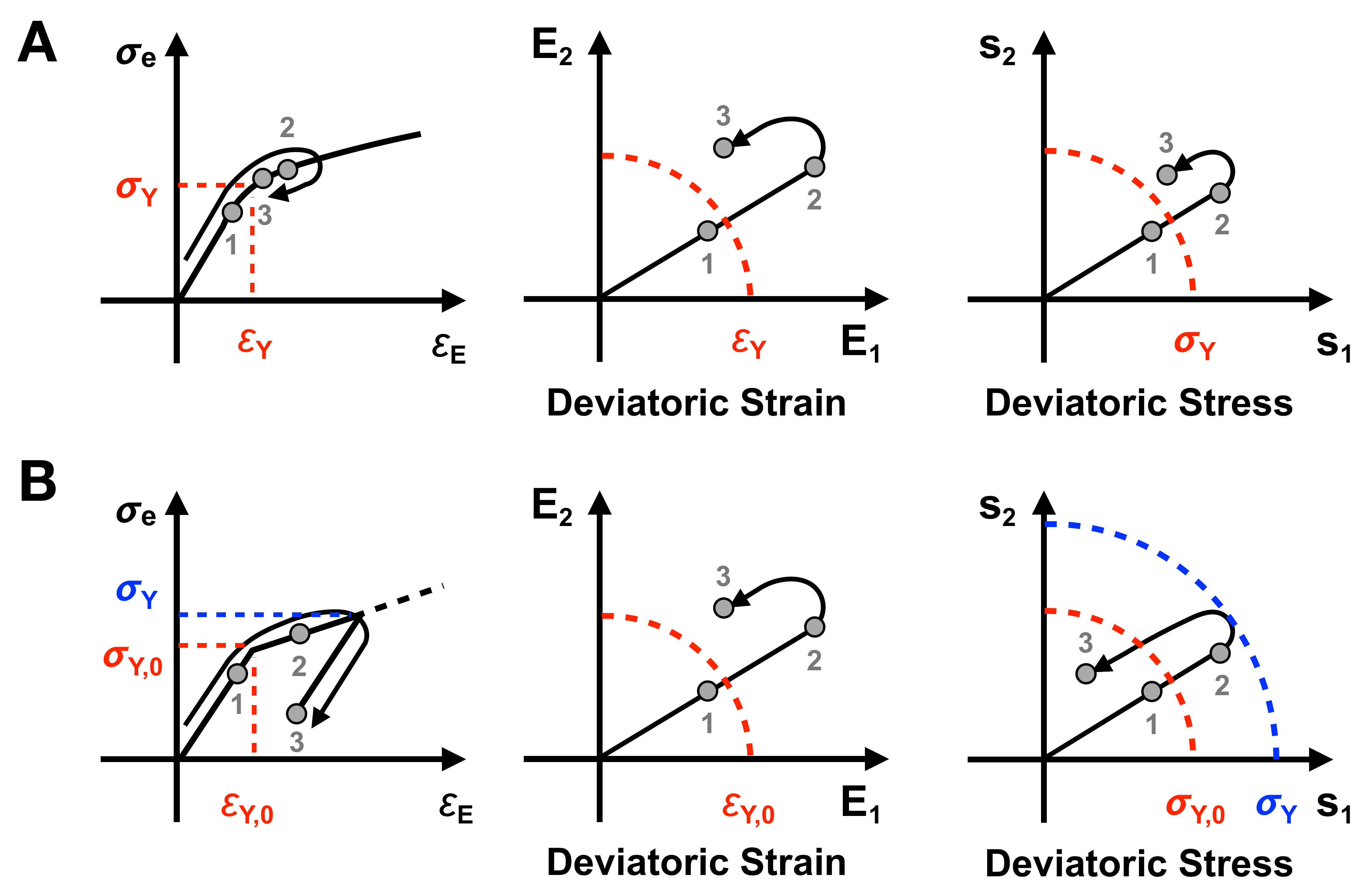}
\captionsetup{justification=justified,labelfont=bf,singlelinecheck = false}
\caption{
{\bf Schematic illustration for elastoplastic constitutive models of biofilm materials.} Schematics of scalar ({\it left}) and tensorial ({\it middle} and {\it right}) stress-strain diagrams for ({\bf A}) deformation theory of elasto-plasticity and ({\bf B}) von Mises theory of elasto-plasticity. For both {\bf A} and {\bf B},  the initial yielding surface is illustrated by red in the axis/plane of deviatoric strain (denoted by $\epsilon_\mathrm{E}$ or $\mathbf{E}$) and deviatoric stress (denoted by $\sigma_\mathrm{e}$ or $\mathbf{s}$). Gray dots with designated numbers represent three successive states during a hypothetical loading/unloading process (black curves with arrows). In {\bf B}, von Mises theory evolves the yielding surface (illustrated by blue dashed circles) and is able to capture hysteresis due to plasticity upon loading and unloading.
}
\label{Fig:elastoplasticModels}
\end{figure}
{\bf Deformation theory of plasticity}:
The deformation theory of plasticity attempts to develop a one-to-one stress-strain relationship (Fig.~\ref{Fig:elastoplasticModels}) \cite{lubarda2000deformation}. One common power-law deformation theory of plasticity assumes that the total deviatoric deformation can be decomposed into an elastic part and a plastic part, i.e. $\mathbf{E} = \mathbf{E}^\mathrm{e} + \mathbf{E}^\mathrm{p}$, where the elastic strain is $\mathbf{E}^\mathrm{e} = \tilde{\mathbf{s}}$ and the plastic strain depends nonlinearly on $\tilde{\mathbf{s}}$,
\begin{gather}
\mathbf{E}^\mathrm{p} = \Big(\frac{\sige}{\sigyo}\Big)^n\tilde{\mathbf{s}},
\end{gather}
where $\sigyo$ denotes the yield stress and $n>1$ is the power-law exponent. When $\sige < \sigyo$, the elastic term dominates, and the strain-stress relation recovers the elastic model, i.e. $\mathbf{E}\approx  \mathbf{E}^\mathrm{e}= \tilde{\mathbf{s}}$; when $\sige > \sigyo$ the nonlinear term dominates and the material yields. 

We cannot invert analytically the above equations, i.e. calculating $\tilde{\mathbf{s}}$ given the deformation $\mathbf{E}$. However, we know that $\tilde{\mathbf{s}}$ and $\mathbf{E}$ must be co-linear, i.e. $\tilde{\mathbf{s}} = c\mathbf{E}$ and hence $\tilde{\mathbf{s}}:\mathbf{E} = c\mathbf{E}:\mathbf{E}$. Therefore, we obtain
\begin{gather}
\frac{\tilde{\mathbf{s}}}{\sige}:\mathbf{E} = \frac{c\mathbf{E}:\mathbf{E}}{c\sqrt{\frac{3}{2}\mathbf{E}:\mathbf{E}}} = \sqrt{\frac{2}{3}\mathbf{E}:\mathbf{E}} \equiv \epsilon_\mathrm{E}.
\end{gather}
Substituting this relation into the expression for $\mathbf{E}$, we derive
\begin{align}
\mathbf{E}:\mathbf{E} &= \tilde{\mathbf{s}}:\mathbf{E} + \Big(\frac{\sige}{\sigyo}\Big)^n\frac{\tilde{\mathbf{s}}}{\sige}:\mathbf{E}, \text{and hence}\\
\frac{3}{2}\epsilon_\mathrm{E} &= \sige + \Big(\frac{\sige}{\sigyo}\Big)^n.
\end{align}
We can solve the equation above numerically for $\sige$ given the deformation (known $\epsilon_\mathrm{E}$). Finally, since
\begin{gather}
\frac{2}{3}\sige^2 = \tilde{\mathbf{s}}:\tilde{\mathbf{s}} = c^2\mathbf{E}:\mathbf{E} = \frac{3}{2}c^2\epsilon_\mathrm{E}^2,
\end{gather}
the co-linear factor is $c = 2\sige/3\epsilon_\mathrm{E}$, and the stress $\tilde{\mathbf{s}} = c\mathbf{E}$ follows.

\begin{figure}[t]
\centering
\includegraphics[width=.8\textwidth]{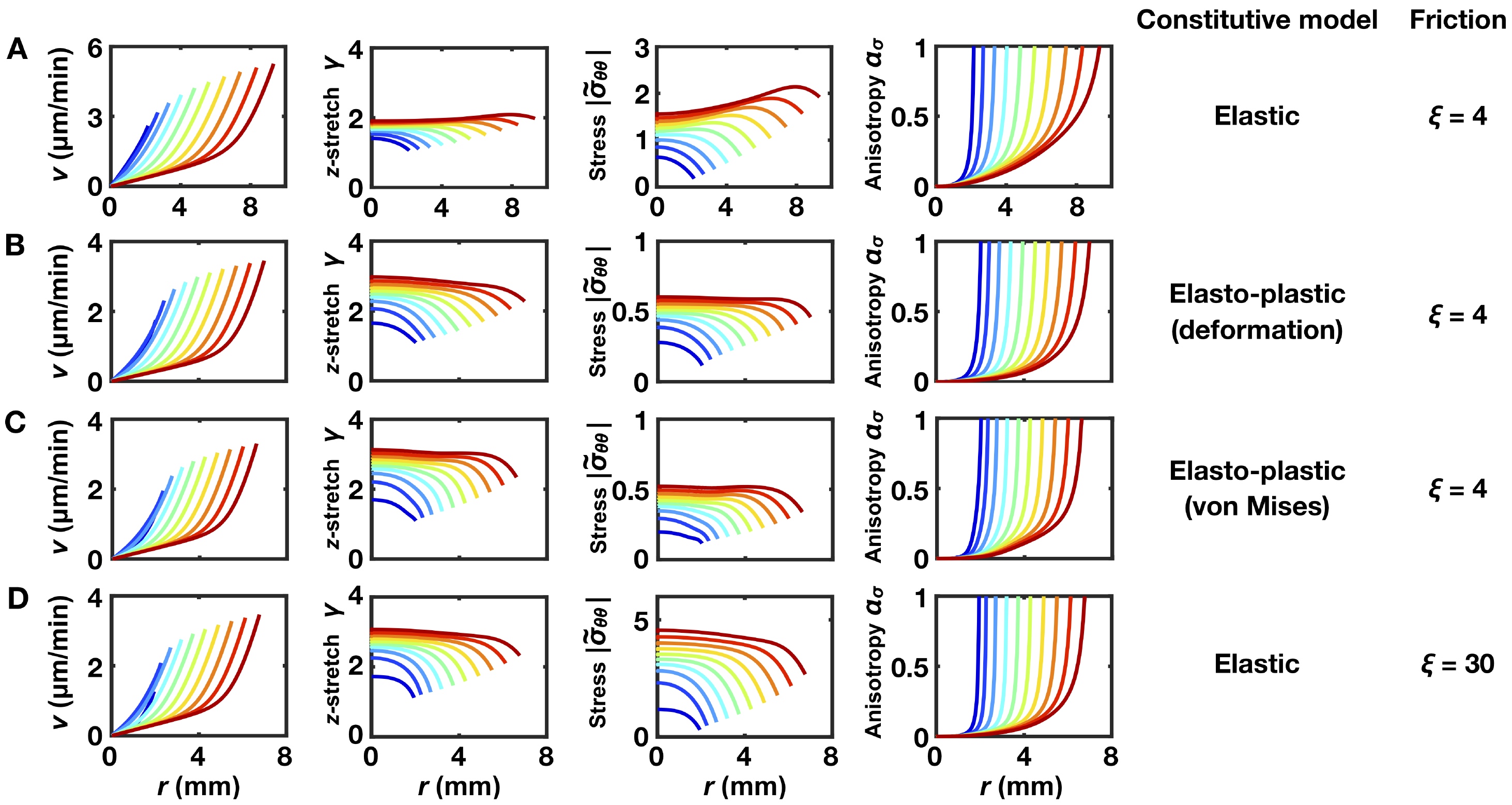}
\captionsetup{justification=justified,singlelinecheck=false,labelfont=bf}
\caption{
{\bf Comparison of the modeled biofilm expansion dynamics using an elastic constitutive model versus an elasto-plastic constitutive model.} The radial velocity field $v$ ({\it left}), the vertical stretch $\gamma$ ({\it center left}), the magnitude of the circumferential stress $\tilde{\sigma}_{\theta\theta}$ ({\it center right}), and stress anisotropy $\alpha_\sigma=(\tilde{\sigma}_{\theta\theta}-\tilde{\sigma}_{rr})/(\tilde{\sigma}_{\theta\theta}+\tilde{\sigma}_{rr})$ ({\it right}), versus radial coordinate $r$ at various times, plotted for the chemo-mechanical model using ({\bf A}) an elastic constitutive model (dimensionless friction parameter $\xi = 4$), ({\bf B}) a deformation elasto-plastic constitutive model ($\xi = 4$, power law exponent $n  = 2$, dimensionless yielding stress $\tilde{\sigma}_\mathrm{Y} = 0.2$), ({\bf C}) a von Mises elasto-plastic constitutive model ($\xi = 4$, $\tilde{\sigma}_\mathrm{Y} = 0.2$, dimensionless plastic modulus $\tilde{G}^\prime = 0.1$), and ({\bf D}) an elastic constitutive model but with larger friction ($\xi = 30$). Note that ({\bf B}-{\bf D}) have similar kinematic behaviors, but models using the elasto-plastic constitutive relation lead to stresses with smaller magnitudes, compared to the elastic counterpart.
}
\label{Fig:biofilmDynamicsPlasticity}
\end{figure}

The deformation theory of plasticity essentially describes a material with a nonlinear constitutive relation and a ``softened" tangent modulus $\Gb^\prime / \Gb \equiv \tilde{G}^\prime < 1$ after yielding. We note that the kinematic behavior of such a material with friction parameter $\xi$ resembles that of a pure elastic material with a larger $\xi^\prime$ (Fig.~\ref{Fig:biofilmDynamicsPlasticity}), but the stress (normalized by the linear elasticity modulus $\Gb$) is significantly smaller.  A simple explanation for this effect is that the biofilm enters the yielding regime in the early stage of expansion, and thus the dynamics is governed essentially by the effective friction parameter $\xi^\prime = \frac{\eta(L_0/\tau_0)}{\Gb^\prime(H/L_0)}$, which is normalized by the effective tangent modulus $\Gb^\prime$.

{\bf Linear isotropic harderning}: 
The deformation theory considers material rules without hysteresis. It appears, however, that more complicated rules are necessary especially for the case of unloading. We now discuss a von Mises-type plasticity model. The stress-strain relation after yielding is essentially given by the evolution of the yield surface, $\sige(\tilde{\mathbf{s}})-\sigy(\mathbf{E})=0$, in which $\sigy$ denotes the current yield stress, initially equal to the material parameter $\sigyo$. Our measurements show that biofilm material starts to yield at about $10\%$ strain in a simple shear experiment~\cite{yan2018bacterial}, from which we can estimate $\sigyo=0.1735\approx0.2$. Measuring the stress-strain curve under monotonic loading in the experiments also show that biofilm material is {\it strain hardening},\footnote{Not to be confused with the usage ``softened tangent modulus". See Fig.~\ref{Fig:elastoplasticModels} for illustrations.} in the sense that the yield stress $\sigy$ increases beyond the yielding point. The von Mises plasticity model also assumes that the material behaves elastically within the yield surface, which leads to hysteresis under cyclic loading (Fig.~\ref{Fig:elastoplasticModels}). 

The evolution of the yield stress $\sigy$ during plastic flow is described by an additional equation, the {\it hardening law}. In the simplest case, the hardening law is isotropic and linear: $\sigy = \sigyo + \tilde{G}^\prime\epsilon^\mathrm{p}$ where plastic modulus  $\tilde{G}^\prime < 1$ is constant and $\epsilon^\mathrm{p}$ is a scalar measure of the plastic strain. The simplest choice of $\epsilon^\mathrm{p}$ might seem to be the norm of the plastic strain tensor, i.e. $\epsilon^\mathrm{p} = (\mathbf{E}^\mathrm{p} : \mathbf{E}^\mathrm{p})^{1/2}$, but this variable does not always increase during plastic flow due to the tensorial nature of $\mathbf{E}^\mathrm{p}$. Therefore, we use the {\it cumulative
plastic strain} $\bar{\epsilon}^{\mathrm{p}}$, defined by the rate equation, $\dot{\bar{\epsilon}}^{\mathrm{p}} = (\frac{3}{2}\dot{\mathbf{E}}^\mathrm{p} :\dot{\mathbf{E}}^\mathrm{p})^{1/2}$ to characterize plastic hardening \cite{jirasek2002inelastic}.

Consider an infinitesimal (un)loading $\mathrm{d}\mathbf{E}$ when the current deviatoric stress is $\tilde{\mathbf{s}}$, then the elastic response $\tilde{\mathbf{s}}^\mathrm{e} =\tilde{\mathbf{s}}+ \mathrm{d}\mathbf{E}$ will occur if $\sige (\tilde{\mathbf{s}}^\mathrm{e} ) < \sigy$. If the von Mises stress $\sige (\tilde{\mathbf{s}}^\mathrm{e})$ surpasses the current yield stress, two things happen: (1)~the yield surface evolves; and (2)~the end stress $\tilde{\mathbf{s}}^\mathrm{p}$ falls on the new yield surface with yield stress $\sigy + \mathrm{d}\sigy$ following the hardening law. Recall the decomposition $\mathrm{d}\mathbf{E} = \mathrm{d}\mathbf{E}^\mathrm{e} + \mathrm{d}\mathbf{E}^\mathrm{p}$, the end stress is corrected by $\tilde{\mathbf{s}}^\mathrm{p}  = \tilde{\mathbf{s}}^\mathrm{e} - \mathrm{d}\mathbf{E}^\mathrm{p}$. We further assume that the plastic flow has the same direction as the deviatoric stress, i.e. $\mathrm{d}\mathbf{E}^\mathrm{p}$ is co-linear with $\tilde{\mathbf{s}}^\mathrm{p}$, and thus $\tilde{\mathbf{s}}^\mathrm{p}$ is also co-linear with $\tilde{\mathbf{s}}^\mathrm{e}$. Let $\mathrm{d}\mathbf{E}^\mathrm{p} = (\mathrm{d}\alpha)\tilde{\mathbf{s}}^\mathrm{e}$, then we obtain
\begin{gather}
\sige(\tilde{\mathbf{s}}^\mathrm{p})  = (1-\mathrm{d}\alpha)\sige(\tilde{\mathbf{s}}^\mathrm{e})  \overset{\text{hardening law}}{\longeq} \sigy + \tilde{G}^\prime \mathrm{d}\bar{\epsilon}^{\mathrm{p}} = \sigy + \mathrm{d}\alpha\tilde{G}^\prime \sige(\tilde{\mathbf{s}}^\mathrm{e}),\text{ and}\\
\mathrm{d}\alpha = \frac{1 - \sigy/\sige(\tilde{\mathbf{s}}^\mathrm{e})}{1 +  \tilde{G}^\prime}.
\end{gather} 
The stress evolution is hence given by $\tilde{\mathbf{s}}^\mathrm{p} = (1-\mathrm{d}\alpha)\tilde{\mathbf{s}}^\mathrm{e}$, or equivalently $\mathrm{d}\mathbf{s} = [\tilde{G}^\prime/(1+\tilde{G}^\prime)] \mathrm{d}\mathbf{E}$.

As expected, the magnitude of the compressive stress $\tilde{\sigma}_{\theta\theta}$ is notably smaller than that of an elastic model. Furthermore, simulations of such a material model with loading-unloading hysteresis show that the stresses quickly drop when growth is stopped, while the deformation is maintained (data not shown). Thus, plastic deformation may also explain why, in the experiments, the pattern does not seem to change after we stop biofilm growth (for example by putting biofilms in a 4$^\circ$C environment).

So far we have only discussed how elasto-plasticity of the biofilm material might change the stress state in a non-uniformly growing biofilm before mechanical instability happen.
More generally, reorganization and yielding of growing biological materials are commonly observed during morphogenesis, such as in plants~\cite{dumais2006anisotropic}, fruit flies \cite{he2014apical,fletcher2014vertex} and brain tissues \cite{Foubet021311}. Thus, the effects of viscoelasticity~\cite{matoz2019wrinkle} and elastoplasticity (Figs.~\ref{Fig:elastoplasticModels} and \ref{Fig:biofilmDynamicsPlasticity}) of biofilms on their morphological development will be an important topic for future studies.

Moreover, {\it V. cholerae} biofilms are soft ($\Gb\sim$1 kPa) and thin ($h\sim$100 $\mu$m). These features make the interfacial energies relevant in a biofilm system, which affects how mechanical instabilities happen. Note that in conventional abiotic film-substrate systems, surface energy is negligible. For example, for metal films, the surface energy is $\gamma\sim1$~J/m$^2$ but the shear modulus is $G\sim 50$~GPa, and thus the elastocapillary length, i.e. the characteristic film thickness at which surface tension could balance bending, is $l_\mathrm{ec} = \gamma/G\sim 0.01$nm; for plastics like polyethylene, $\gamma\sim$ 50 mJ/m$^2$, $G\sim 100$ MPa, and $l_\mathrm{ec}$ is about $ 0.5$nm. However, in these systems the film thickness is at least 10 nm, much larger then $l_\mathrm{ec}$.
For {\it V. cholerae} biofilms, the interfacial energy between a WT biofilm and water  $\gamma_\mathrm{bw}$ was determined to be around 50--60~mJ/m$^2$ and the biofilm-air interfacial energy is $\gamma_\mathrm{ba}\approx$30--40~mJ/m$^2$ \cite{yan2018bacterial}. Thus, biofilms are ``sticky" in the sense that the elastocapillary length (determined by its material properties) is comparable to its natural thickness ($\gamma/\Gb h\approx 0.5$ for biofilms). In fact, we previously reported that biofilms can form different types of mechanical instability patterns depending on the interfacial energies \cite{yan2019mechanical}, and researchers have also started to look at how surface energies affect other type of instabilities \cite{Liu2019elasto}. Understanding the effects of interfacial energies on the pattern formation dynamics will be an important future direction.

\bibliographystyle{ieeetr}
\bibliography{Biofilm_Ref2.bib}